\documentclass[zpreprint,zbstnp,final]{zeus_paper}
\usepackage{placeins}
\usepackage[titletoc,title]{appendix}
\usepackage{authblk}
\usepackage{lineno}
\usepackage{hyperref}
\usepackage{calc}
\usepackage[utf8]{inputenc}
\usepackage{graphicx}
\usepackage{multicol}
\usepackage{ifthen}
\usepackage{titlesec}
\usepackage{colortbl}
\usepackage{longtable}
\usepackage{ifthen}
\usepackage{import}
\usepackage{amsmath}
\usepackage{xspace}
\usepackage{ifthen}
\usepackage{mcite}
\usepackage{cite}
\usepackage{caption}
\usepackage{longtable}
\usepackage{ifthen}
\usepackage{import}
\usepackage{amsmath}
\usepackage{amssymb}
\usepackage{graphicx}
\usepackage{hyperref} 
\usepackage{bigstrut}
\usepackage{gensymb}
\usepackage{booktabs}
\usepackage{multirow}
\usepackage{adjustbox}
\usepackage{xcolor}

\usepackage{setspace}
\title{Precise determination of $\alpha_{S}(M_Z)$ from a global fit of energy-energy correlation to NNLO+NNLL predictions}

\author[a]{Adam Kardos\thanks{kardos.adam@science.unideb.hu}}
\author[b]{Stefan Kluth\thanks{stefan.kluth@mpp.mpg.de}}
\author[c]{G\'abor Somogyi\thanks{gabor.somogyi@cern.ch}}
\author[a]{Zolt\'an Tulip\'ant\thanks{tulipant.zoltan@science.unideb.hu}}
\author[b]{Andrii Verbytskyi\thanks{andrii.verbytskyi@mpp.mpg.de}}

\affil[a]{Institute of Physics, University of Debrecen,\protect\\ 4010 Debrecen, PO Box 105, Hungary}
\affil[b]{Max-Planck-Institut f\"{u}r Physik, 80805 Munich, Germany}
\affil[c]{MTA-DE Particle Physics Research Group, University of Debrecen,\protect\\ 4010 Debrecen, PO Box 105, Hungary}

\prepnum{MPP-2018-XXX}

\newcommand{\epjcbreak}[1]{}
\newcommand{\draftbreak}[1]{}
\newcommand{\arxivbreak}[1]{\\#1}
\newcommand{\FIGWONE}{0.33\linewidth}

\newcommand{\fitrangetwo}{$117-165\degree$\xspace}
\newcommand{\fitrangethree}{$60-165\degree$\xspace}
\newcommand{\fitrangefour}{$60-160\degree$\xspace}

\newcommand{\resultNNLO}{ 0.11750\pm  0.00018  {\text( exp.)}\pm 0.00102{\text(hadr.)}\pm\epjcbreak{} 0.00257{\text(ren.)}\pm 0.00078{\text(res.)}}
\newcommand{\resultNNLOabs}{ 0.11750\pm  0.00018  {\text( exp.)}\pm 0.00102{\text(hadr.)}\pm0.00257{\text(ren.)}\pm 0.00078{\text(res.)}}
\newcommand{\resultNNLOcomb}{ 0.11750\pm 0.00287{\text(comb.)}}
\newcommand{\resultNLO}{ 0.12200\pm  0.00023  {\text( exp.)}\pm 0.00113{\text(hadr.)}\pm \epjcbreak{}0.00433{\text(ren.)}\pm 0.00293{\text(res.)}}
\newcommand{\resultNLOcomb}{ 0.12200\pm 0.00535{\text(comb.)}}

\newcommand{\tabularinput}{
SLD~\protect\cite{Abe:1994mf}&$91.2(91.2)$&$ 91.2 $ &$60000 $\\
OPAL~\protect\cite{Acton:1993zh}&$91.2(91.2)$&$ 91.2 $ &$336247 $\\
OPAL~\protect\cite{Acton:1991cu}&$91.2(91.2)$&$ 91.2 $ &$128032 $\\
L3~\protect\cite{Adriani:1992gs}&$91.2(91.2)$&$ 91.2 $ &$169700 $\\
DELPHI~\protect\cite{Abreu:1992yc}&$91.2(91.2)$&$ 91.2 $ &$120600 $\\
TOPAZ~\protect\cite{Adachi:1989ej}&$59.0-60.0(59.5)$&$ 59.5 $ &$540 $\\
TOPAZ~\protect\cite{Adachi:1989ej}&$52.0-55.0(53.3)$&$ 53.3 $ &$745 $\\
TASSO~\protect\cite{Braunschweig:1987ig}&$38.4-46.8(43.5)$&$ 43.5 $ &$6434 $\\
TASSO~\protect\cite{Braunschweig:1987ig}&$32.0-35.2(34.0)$&$ 34.0 $ &$52118 $\\
PLUTO~\protect\cite{Berger:1985xq}&$34.6(34.6)$&$ 34.0 $ &$6964 $\\
JADE~\protect\cite{Bartel:1984uc}&$29.0-36.0(34.0)$&$ 34.0 $ &$12719 $\\
CELLO~\protect\cite{Behrend:1982na}&$34.0(34.0)$&$ 34.0 $ &$2600 $\\
MARKII~\protect\cite{Wood:1987uf}&$29.0(29.0)$&$ 29.0 $ &$5024 $\\
MARKII~\protect\cite{Wood:1987uf}&$29.0(29.0)$&$ 29.0 $ &$13829 $\\
MAC~\protect\cite{Fernandez:1984db}&$29.0(29.0)$&$ 29.0 $ &$65000 $\\
TASSO~\protect\cite{Braunschweig:1987ig}&$21.0-23.0(22.0)$&$ 22.0 $ &$1913 $\\
JADE~\protect\cite{Bartel:1984uc}&$22.0(22.0)$&$ 22.0 $ &$1399 $\\
CELLO~\protect\cite{Behrend:1982na}&$22.0(22.0)$&$ 22.0 $ &$2000 $\\
TASSO~\protect\cite{Braunschweig:1987ig}&$12.4-14.4(14.0)$&$ 14.0 $ &$2704 $\\
JADE~\protect\cite{Bartel:1984uc}&$14.0(14.0)$&$ 14.0 $ &$2112 $\\
}
\newcommand{\tabularresult}{
\fitrangetwo         &$0.12042\pm 0.00025$ &  $0.11760\pm 0.00020$  \\
$S^{L}$              &$765/298 = 2.57$ &  $513/298 = 1.72$  \\\hline 
\fitrangethree       &$0.12134\pm 0.00022$ &  $0.11746\pm 0.00018$  \\
$S^{L}$              &$1720/664 = 2.59$&  $1211/664 = 1.82$  \\\hline 
\fitrangefour        &$0.12200\pm 0.00023$ &  $0.11750\pm 0.00018$  \\
$S^{L}$              &$1417/623 = 2.27$&  $1022/623 = 1.64$  \\\hline 
\fitrangetwo         &$0.11796\pm 0.00022$ &  $0.11521\pm 0.00017$  \\
$S^{C}$              &$631/298 = 2.12$ &  $395/298 = 1.32$  \\\hline 
\fitrangethree       &$0.11900\pm 0.00021$ &  $0.11530\pm 0.00015$  \\
$S^{C}$              &$1557/664 = 2.34$&  $951/664 = 1.43$  \\\hline 
\fitrangefour        &$0.11973\pm 0.00022$ &  $0.11545\pm 0.00016$  \\
$S^{C}$              &$1321/623 = 2.12$&  $845/623 = 1.36$  \\\hline 
\fitrangetwo         &$0.11272\pm 0.00037$ &  $0.11044\pm 0.00029$  \\
$H^{M}$              &$1842/298 = 6.18$ &  $1201/298 = 4.03$  \\\hline 
\fitrangethree       &$0.11472\pm 0.00033$ &  $0.11180\pm 0.00023$  \\
$H^{M}$              &$3845/664 = 5.79$&  $2203/664 = 3.32$  \\\hline 
\fitrangefour        &$0.11634\pm 0.00033$ &  $0.11281\pm 0.00023$  \\
$H^{M}$              &$3091/623 = 4.96$&  $1738/623 = 2.79$  \\\hline 
\fitrangetwo         &$0.12154\pm 0.00045$ &  $0.11781\pm 0.00037$  \\
$An.^{DMW}$              &$730/295 = 2.48$ &  $558/295 = 1.89$  \\\hline 
\fitrangethree       &$0.13555\pm 0.00052$ &  $0.12937\pm 0.00039$  \\
$An.^{DMW}$              &$7525/661 = 11.38$&  $4896/661 = 7.41$  \\\hline 
\fitrangefour        &$0.13606\pm 0.00061$ &  $0.12950\pm 0.00044$  \\
$An.^{DMW}$              &$7364/620 = 11.88$&  $4827/620 = 7.78$  \\\hline 
}

\newcommand\Refr[1]     {Ref.\,\cite{#1}}
\newcommand\Refrs[1]    {Refs.\,\cite{#1}}
\newcommand\refr[1]     {ref.\,\cite{#1}}

\newcommand\eqn[1]     {eq.\,(\ref{#1})}
\newcommand\eqns[2]    {eqs.\,(\ref{#1}) and~(\ref{#2})}
\newcommand\eqnss[2]   {eqs.\,(\ref{#1})--(\ref{#2})}

\def\beq{\begin{equation}}
\def\eeq{\end{equation}}
\def\bsp#1\esp{\begin{split}#1\end{split}}
\def\bal#1\eal{\begin{align}#1\end{align}}

\newcommand\tot		  {\rm{t}}
\newcommand\tsigT      {\sigma_{\tot}}
\newcommand\tsig[2]    {\sigma^{\rm{#1}}_{#2}}
\newcommand\dsig[2]    {\rd\sigma^{{\rm #1}}_{#2}}

\newcommand{\rd}       {{\rm{d}}}

\newcommand\as  	       {\ensuremath{\alpha_{\rm{S}}}}
\newcommand\asbar[1]  	       {\ensuremath{\frac{\alpha_{\rm{S}}(#1)}{2\pi}}}
\newcommand\Oa[1]      {\ensuremath{\mathcal O(\as^{#1})}}

\newcommand{\CF}       {C_{\rm{F}}}
\newcommand{\CA}       {C_{\rm{A}}}
\newcommand{\TR}       {T_{\rm{R}}}
\newcommand{\Nc}       {N_{\rm{c}}}
\newcommand{\Nf}       {\ensuremath{n_{\rm{f}}}}

\newcommand{\xiL}      {x_{L}}
\newcommand{\xiLsq}      {x_{L}^2}
\newcommand{\xiR}      {x_{R}}
\newcommand{\xiRsq}      {(x_{R}^2)}
\newcommand{\colorfulNNLO}{{CoLoRFulNNLO}}
\newcommand\MSbar	  {\ensuremath{\overline{\rm{MS}}}}

\newcommand{\eVdist}{\kern-0.06667em}
\newcommand{\GeV}{{\,\text{Ge}\eVdist\text{V\/}}}

\begin{document}

\abstract{
We present a comparison of the computation of energy-energy correlation in $e^{+}e^{-}$ collisions 
in the back-to-back region at next-to-next-to-leading logarithmic accuracy 
matched with the next-to-next-to-leading order perturbative prediction 
 to LEP, PEP, PETRA, SLC and TRISTAN  data.  With these predictions we perform
  an extraction of the strong coupling constant taking into account 
  non-perturbative effects modelled with Monte Carlo event generators.
The final result at NNLO+NNLL precision is $\alpha_{S}(M_{Z})=\resultNNLOabs$.

}
\makezeustitle 
\newpage
\clearpage
\pagenumbering{arabic}
\pagestyle{plain}
\section{Introduction}
\label{sec:intro}

The strong interaction in the Standard Model (SM) is described by
Quantum Chromodynamics
(QCD)~\cite{Fritzsch:1973pi,Gross:1973id,Politzer:1973fx,Gross:1973ju}.
The theory successfully models the interactions between quarks and
gluons and is a source of numerous predictions.  Verifying the
predictions of QCD is instrumental for searches for physics beyond the
SM at the LHC, since the reliable prediction of SM processes as sources
of backgrounds for searches is essential.

Precision measurements of event shape distributions in $e^+e^-$ 
annihilation have provided detailed experimental tests of QCD and
remain one of the most precise tools used for extracting the strong
coupling \as\ from data~\cite{Kluth:2006bw,Dissertori:2015tfa}. Quantities
related to three-jet events are particularly well suited for this
task.

The state of the art for QCD for event shape observables currently
includes exact fixed-order next-to-next-to-leading order (NNLO)
corrections for the six standard three-jet event shapes of thrust,
heavy jet mass, total and wide jet broadening, $C$-parameter and the
two-to-three jet transition variable $y_{23}$~\cite{GehrmannDeRidder:2007hr, Weinzierl:2009ms,DelDuca:2016ily} as
well as jet cone energy fraction~\cite{DelDuca:2016ily}, oblateness
and energy-energy correlation~\cite{DelDuca:2016csb}. The numerical
matrix element integration codes described in the references allow the
straightforward computation of any suitable, i.e.\ collinear and
infrared safe event shape or jet observable.

However, fixed-order predictions have a limited kinematical range of
applicability. For small values of an event shape observable $y$
corresponding to events with two-jet like topologies the fixed-order
predictions do not converge well. This is due to terms where each power
of the strong coupling $\as^n$ is enhanced by a factor $(\ln y)^{n+1}$
(leading logs), $(\ln y)^{n}$ (next-to-leading logs) etc.  For
three-jet event shapes such logarithmically enhanced terms can be
resummed at next-to-next-to-leading logarithmic (NNLL) accuracy
\cite{deFlorian:2004mp,Becher:2008cf,Chien:2010kc,Monni:2011gb,Alioli:2012fc,
  Becher:2012qc,Banfi:2014sua}, i.e.\ up to terms $\sim(\ln y)^{n-1}$.
Resummation in next-to-next-to-next-to-leading logarithmic (N${}^3$LL)
accuracy has been achieved for the
$C$-parameter~\cite{Hoang:2014wka} and thrust~\cite{Abbate:2010xh}.  A prediction incorporating the
complete perturbative knowledge about the observable can be derived by
matching the fixed-order and resummed calculations.

For the frequently used event shapes of thrust, heavy jet mass, total and
wide jet broadening, $C$-parameter and $y_{23}$, NNLO predictions
matched to NLL resummation were presented
in~\cite{Gehrmann:2008kh}. Predictions at NNLO matched to N${}^3$LL
resummation are also known for thrust~\cite{Becher:2008cf,Abbate:2010xh} and the
 $C$-parameter~\cite{Hoang:2014wka}.

In this paper we consider the energy-energy correlation (EEC) in $e^+e^-$  annihilation 
and present NNLO predictions matched to NNLL resummation for the back-to-back 
region.  EEC was the first event shape for which a complete NNLL resummation was 
performed \cite{deFlorian:2004mp} while the fixed-order NNLO corrections to this
observable were computed recently~\cite{DelDuca:2016csb}. Moreover, 
EEC is the first event shape observable for which an analytic fixed-order NLO correction 
was computed~\cite{Dixon:2018qgp}.

The agreement between the predictions at NNLO+NNLL accuracy and the measured 
data is still not perfect. The discrepancy can be attributed mainly to non-perturbative 
hadronization corrections. We extract these corrections from data by 
comparison to state-of-the-art Monte Carlo predictions and determine the value 
of the strong coupling by comparing our results to measurements over a wide range 
of centre-of-mass energies. Our analysis allows us to target the highest precision of 
$\as$ determination and we present the first global fit of the strong coupling to EEC 
at NNLO+NNLL accuracy. Our analysis also represents the first extraction of $\as$ 
based on Monte Carlo hadronization corrections obtained from NLO Monte Carlo setups 
at NNLO+NNLL precision.

\section{EEC distribution in perturbation theory}
\label{sec:theory}
EEC is the normalized energy-weighted cross section defined in terms of the angle 
between two particles $i$ and $j$ in an event~\cite{Basham:1978bw}:
\begin{equation}
\frac{1}{\tsigT} \frac{\rd \Sigma(\chi)}{\rd \cos \chi} \equiv
        \frac{1}{\tsigT} \int \sum_{i,j} \frac{E_i E_j}{Q^2}
        \dsig{}{e^+e^- \to\, i j + X} \delta(\cos\chi - \cos\theta_{ij})\,,
\label{eq:dEEC-def}
\end{equation}
where $E_i$ and $E_j$ are the particle energies, $Q$ is the centre-of-mass energy,
$\theta_{ij} = \chi$ is the angle between the two particles and $\tsigT$ is the 
total hadronic cross section. The back-to-back region $\theta_{ij} \to  180\degree$ corresponds 
to $\chi \to \pi$, while the normalization ensures that the integral of the EEC distribution 
from $\chi = 0\degree$ to $\chi = 180\degree$ is 
unity\footnote{Refs.~\cite{deFlorian:2004mp}~and~\cite{Tulipant:2017ybb} use the opposite 
convention of $\theta_{ij} = 180\degree-\chi$ such that the back-to-back region corresponds to 
$\chi\to 0\degree$. Here we use $\theta_{ij} = \chi$ throughout which agrees with the experimental 
convention.}.

\subsection{Fixed-order and resummed calculations}
\label{ssec:fo-res}

The differential EEC distribution has been computed numerically at NLO accuracy 
in perturbation theory some time ago~\cite{Richards:1983sr,Ellis:1983fg,
Richards:1982te,Chao:1982wb,Schneider:1983iu,Ali:1982ub,Kunszt:1989km,
Falck:1988gb,Glover:1994vz,Clay:1995sd,Kramer:1996qr} and efforts towards 
obtaining an analytic result at this order~\cite{Belitsky:2013ofa,Gituliar:2017umx} have 
culminated in a complete calculation very recently~\cite{Dixon:2018qgp}. 
The NNLO prediction has also been obtained in \refr{DelDuca:2016csb} using 
the \colorfulNNLO{} method \cite{Somogyi:2006da,Somogyi:2006db,DelDuca:2016ily}. 
At the default renormalization scale\footnote{We use the \MSbar{} renormalization 
scheme throughout the paper.} of $\mu=Q$ the fixed-order prediction reads
\begin{multline}
\bigg[\frac{1}{\tsig{}{0}} \frac{\rd \Sigma(\chi,Q)}{\rd \cos \chi}\bigg]_{\mathrm{f.o.}} = 
	\asbar{Q} \frac{\rd A(\chi)}{\rd \cos \chi } 
	+
	\left(\asbar{Q}\right)^2 \frac{\rd B(\chi) }{\rd \cos \chi }
	\epjcbreak{}+
	\left(\asbar{Q}\right)^3 \frac{\rd C(\chi)}{\rd \cos \chi }
	+
	\Oa{4}\,,
\label{eq:dEEC-fo-sigma0}
\end{multline}
where $A$, $B$ and $C$ are the perturbative coefficients at LO, NLO and NNLO, normalized 
to the LO cross section for $e^+e^- \to \mbox{hadrons}$, $\tsig{}{0}$. In massless QCD 
this normalization cancels all electroweak coupling factors, and the dependence on the 
collision energy enters only through $\as(Q)$. However, experiments measure the distribution 
normalized to the total hadronic cross section, so physical predictions must be normalized to $\tsigT$. 
The distribution normalized to the total hadronic cross section can be 
obtained from the expansion in \eqn{eq:dEEC-fo-sigma0} through multiplying by $\tsig{}{0}/\tsigT$. 
For massless quarks, this ratio is independent of all electroweak couplings and reads
\begin{equation*}
\frac{\tsig{}{0}}{\tsigT} = 
	1 - \asbar{Q} A_{\tot} + \left(\asbar{Q}\right)^2 
	\left(A_{\tot}^2 - B_{\tot}\right) + \Oa{3}\,,
\end{equation*}
with
\begin{equation*}
A_{\tot} = \frac{3}{2}\CF 
\end{equation*}
and
\begin{equation*}
B_{\tot} = \CF \left[\left(\frac{123}{8} - 11 \zeta_3\right) \CA
	- \frac{3}{8} \CF - \left(\frac{11}{2} - 4 \zeta_3\right) \Nf \TR \right]\,.
\end{equation*}
The colour factors which appear above are given by
\begin{equation*}
\CA = 2\Nc\TR\,,
\qquad
\CF = \frac{\Nc^2-1}{\Nc}\TR
\qquad\mbox{and}\qquad 
\TR = \frac{1}{2}\,,
\end{equation*}
while $\Nf$ denotes the number of light quark flavours.

The renormalization scale dependence of the fixed-order prediction can be restored 
using the renormalization group equation for $\as$ and one finds 
\begin{align}\begin{split}
\bigg[\frac{1}{\tsigT} \frac{\rd \Sigma(\chi,\mu)}{\rd \cos \chi}\bigg]_{\mathrm{f.o.}} &= 
	\asbar{\mu} \frac{\rd \bar{A}(\chi,\xiR)}{\rd \cos \chi } 
	+\epjcbreak{&+}
	\left(\asbar{\mu}\right)^2 \frac{\rd \bar{B}(\chi,\xiR) }{\rd \cos \chi }+\\
	&+\left(\asbar{\mu}\right)^3 \frac{\rd \bar{C}(\chi,\xiR)}{\rd \cos \chi }
	+
	\Oa{4}\,,
\label{eq:dEEC-fo}
\end{split}\end{align}
where
\begin{align*}\begin{split}
\bar{A}(\chi,\xiR) &= A(\chi)\,,
\\
\bar{B}(\chi,\xiR) &= B(\chi) + \left(\frac{1}{2}\beta_0 \ln\xiRsq 
	- A_{\tot}\right) A(\chi)\,,
\\
\bar{C}(\chi,\xiR) &= C(\chi) + \left(\beta_0 \ln\xiRsq - A_{\tot}\right) B(\chi)+\\
	&+ \bigg(\frac{1}{4} \beta_1 \ln\xiRsq 
	+ \frac{1}{4}\beta_0^2 \ln^2\xiRsq
	-  A_{\tot} \beta_0 \ln\xiRsq 
	+\epjcbreak{&+} A_{\tot}^2 - B_{\tot} \bigg) A(\chi)\,,
\end{split}\end{align*}
with $\xiR = \mu/Q$. Finally, using three-loop running the scale dependence of 
the strong coupling is given by 
\begin{align*}\begin{split}
\as(\mu) &= \frac{4\pi}{\beta_0 t}
\bigg[1 - \frac{\beta_1}{\beta_0^2 t} \ln t + \epjcbreak{&+}   \left(\frac{\beta_1}{\beta_0^2 t}\right)^2
\left(\ln^2 t - \ln t - 1 + \frac{\beta_0 \beta_2}{\beta_1^2}\right)\bigg]\,.
\end{split}\end{align*}
Here, $t = \ln(\mu^2/\Lambda_{\rm{QCD}}^2)$ and the $\beta_i$ are the \MSbar{}-scheme 
coefficients of the QCD beta function,
\begin{equation*}\begin{split}
\beta_0 &= \frac{11\CA}{3} - \frac{4\Nf \TR}{3}\,,
\\
\beta_1 &= \frac{34}{3} \CA^2 - \frac{20}{3} \CA \TR \Nf - 4 \CF \TR \Nf\,,
\\
\beta_2 &= \frac{2857}{54} \CA^3
	- \left(\frac{1415}{27} \CA^2 + \frac{205}{9} \CA \CF - 2 \CF^2\right) \TR \Nf
	+\epjcbreak{&+} \left(\frac{158}{27} \CA + \frac{44}{9} \CF\right) \TR^2 \Nf^2\,.
\end{split}\end{equation*}

The fixed-order perturbative predictions diverge for both small and large values of $\chi$, 
due to the presence of large logarithmic contributions of infrared origin. Concentrating
on the back-to-back region $\chi \to 180\degree$, these contributions take the form 
$\as^n \log^{2n-1} y$, where 
\begin{equation*}
y = \cos^2 \frac{\chi}{2}\,.
\end{equation*}
As $y$ decreases, the logarithms become large and invalidate the use of the fixed-order
perturbative expansion. In order to obtain a description of EEC in this limit, the 
logarithmic contributions must be resummed to all orders. This resummation has been computed 
at NNLL accuracy in \Refr{deFlorian:2004mp}\footnote{Note that the NNLL $A^{(3)}$ coefficient in 
\Refr{deFlorian:2004mp} is incomplete. The full coefficient has been derived in \Refr{Becher:2010tm}.} 
while in \Refr{Moult:2018jzp} a factorization theorem for EEC was derived based on soft-collinear 
effective theory which will allow to preform the resummation at N$^3$LL accuracy once the 
corresponding NNLO jet function is computed. Since the complete jet function is currently 
not available, we use the NNLL results and formalism of \Refr{deFlorian:2004mp} in the following. 
The resummed prediction at the default scale of $\mu=Q$ can be written as
\beq
\bigg[\frac{1}{\tsigT}\frac{\rd \Sigma(\chi,Q)}{\rd \cos \chi}\bigg]_{\mathrm{res.}} =
	\frac{Q^2}{8} H(\as(Q)) \int_0^\infty \rd b\, b\, J_0(b\, Q \sqrt{y}) S(Q,b)\,.
\label{eq:dEEC-res}
\eeq
The large logarithmic corrections are exponentiated in the Sudakov form factor,
\beq
S(Q,b) = \exp\Bigg\{-\int_{b_0^2/b^2}^{Q^2} \frac{\rd q^2}{q^2} 
	\left[A(\as(q^2)) \ln\frac{Q^2}{q^2} + B(\as(q^2))\right]\Bigg\}\,.
\label{eq:dEEC-Sudakov}
\eeq
The zeroth order Bessel function $J_0$ in \eqn{eq:dEEC-res} and $b_0 = 2 e^{-\gamma_{\rm E}}$ 
in \eqn{eq:dEEC-Sudakov} have a kinematic origin. The functions $A$, $B$ (not to be confused with 
the fixed-order expansion coefficients appearing in \eqn{eq:dEEC-fo-sigma0}) and $H$ in 
\eqns{eq:dEEC-res}{eq:dEEC-Sudakov} are free of logarithmic corrections and can be 
computed as perturbative expansions in $\as$,
\bal
A(\as) &= \sum_{n=1}^{\infty} \left(\frac{\as}{4\pi}\right)^n A^{(n)}\,,
\label{eq:A-func}
\\
B(\as) &= \sum_{n=1}^{\infty} \left(\frac{\as}{4\pi}\right)^n B^{(n)}\,,
\label{eq:B-func}
\\
H(\as) &= 1+ \sum_{n=1}^{\infty} \left(\frac{\as}{4\pi}\right)^n H^{(n)}\,.
\label{eq:H-func}
\eal
Explicit expressions for the expansion coefficients (up to NNLL accuracy) in our normalization 
conventions can be found in \Refr{Tulipant:2017ybb}.

It is possible to perform the $q^2$ integration in \eqn{eq:dEEC-Sudakov} analytically and 
the Sudakov form factor can be written as

\begin{align}\begin{split}
S(Q,b) = \exp[& 
	L g_1(a_{\rm{S}} \beta_0 L) 
	+ g_2(a_{\rm{S}} \beta_0 L)
	+\epjcbreak{&+} a_{\rm{S}} g_3(a_{\rm{S}} \beta_0 L) + \ldots]\,,
\label{eq:dEEC-Sudakov-gi}
\end{split}\end{align}
where $a_{\rm{S}} = \as(Q)/(4\pi)$ and $L = \ln (Q^2 b^2/b_0^2)$ corresponds to $\ln y$ at 
large $b$ (the $y \ll 0$ limit corresponds to $Q b \gg 1$ through a Fourier transformation). 
Writing the Sudakov form factor this way clearly shows that $S(Q,b)$ depends on its variables 
only through the dimensionless combination $b\,Q$. The functions $g_1$, $g_2$ and $g_3$ 
correspond to the LL, NLL and NNLL contributions. Their explicit expressions can be found 
in \Refrs{deFlorian:2004mp,Tulipant:2017ybb}.

So far, we have not considered the dependence of the resummed prediction on the 
renormalization scale. Besides the replacement of $\as(Q)$ by $\as(\mu)$ in 
\eqns{eq:dEEC-res}{eq:dEEC-Sudakov-gi}, the resummation functions $g_i(\lambda)$ also 
acquire renormalization scale dependence,
\begin{equation*}\begin{split}
g_1(\lambda,\xiR) &= g_1(\lambda)\,,\\
g_2(\lambda,\xiR) &= g_2(\lambda) + \lambda^2 g_1'(\lambda) \ln \xiRsq\,,\\
g_3(\lambda,\xiR) &= g_3(\lambda) 
	+ \left[\frac{\beta_1}{\beta_0} \lambda^2 g_1'(\lambda) 
		+ \beta_0 \lambda g_2'(\lambda)\right] \ln \xiRsq+\epjcbreak{&+}\left[\frac{\beta_0}{2} \lambda^3 g_1''(\lambda) 
		+ \beta_0 \lambda^2 g_1'(\lambda)\right] \ln^2 \xiRsq\,,
\end{split}\end{equation*}
where the prime denotes differentiation with respect to $\lambda$.

The factorization between the constant and  logarithmic terms $H(\as)$ and $S(Q,b)$ in \eqn{eq:dEEC-res} 
also involves some arbitrariness, since the argument of the large logarithm $L$ can always be rescaled as 
\begin{equation*}
L = \ln(Q^2 b^2/b_0^2) = \ln (\xiLsq Q^2 b^2/b_0^2) - \ln(\xiLsq)\,, 
\end{equation*}
provided that $\xiL$ is independent of $b$ and that $\xiL = {\mathcal O}(1)$ when $Q b \gg 1$. 
This arbitrariness is parametrized by $\xiL$, which plays a role in the resummed computation 
which is analogous to the role played by the renormalization scale in the fixed-order 
calculation. This rescaling of the logarithm introduces some modifications of the resummed formulae 
and the expansion coefficients in \eqnss{eq:A-func}{eq:H-func}. We find
\begin{align*}
\tilde{A}^{(n)}(\xiL) & = A^{(n)}\,,
\\
\tilde{B}^{(n)}(\xiL) &= B^{(n)} - A^{(n)} \ln(\xiL^2)\,,
\\
\tilde{H}^{(1)}(\xiL) &= H^{(1)} - \beta_0 g_2'(0) \ln(\xiL^2) + \beta_0 g_1'(0) \ln^2(\xiL^2)\,,
\end{align*}
while the Sudakov form factor in \eqn{eq:dEEC-Sudakov-gi} is also modified as follows
\beq
\bsp
S(Q,b,\xiR,\xiL) =
	\exp\bigg[& 
	\tilde{L} g_1\left(a_{\rm{S}} \beta_0 \tilde{L}, \frac{\xiR}{\xiL}\right) 
	+\epjcbreak{&+} g_2\left(a_{\rm{S}} \beta_0 \tilde{L}, \frac{\xiR}{\xiL}\right) +
\\&
	+ \as g_3\left(a_{\rm{S}} \beta_0 \tilde{L}, \frac{\xiR}{\xiL}\right) + \ldots\bigg]\,,
\label{eq:dEEC-Sudakov-gi-xiR-xiL}
\esp
\eeq
where $\tilde{L} = \ln(\xiL^2 Q^2 b^2/b_0^2)$. 

\subsection{Matching the fixed-order and resummed predictions}
\label{ssec:fo-res}
%
%
In order to obtain a prediction which is valid over a wide kinematical range\footnote{We note 
that another resummation in the forward limit would be required to describe EEC over the full 
angular range.} the fixed-order and resummed calculations must be matched. Here we employ the 
log-$R$ matching scheme as worked out for EEC in \Refr{Tulipant:2017ybb}, and limit ourselves 
to recalling the final results.

In the log-$R$ matching scheme for EEC we consider the cumulative distribution
\beq
\bsp
\frac{1}{\tsigT} \widetilde{\Sigma}(\chi,\mu) &\equiv
        \frac{1}{\tsigT} \int_0^\chi \rd \chi'\, (1-\cos\chi')
                \frac{\rd \Sigma(\chi',\mu)}{\rd \chi'}
=\epjcbreak{&=}
	\frac{1}{\tsigT} \int_0^{y(\chi)} \rd y'\, 2(1-y')  
		\frac{\rd \Sigma(y',\mu)}{\rd y'}\,.
\esp
\label{eq:EECcos-def}
\eeq
The differential EEC distribution is easily recovered from $\widetilde{\Sigma}(\chi,\mu)$,
\begin{equation*}
\frac{1}{\tsigT} \frac{\rd \Sigma(\chi,\mu)}{\rd \chi} = \frac{1}{1-\cos\chi}\frac{\rd}{\rd \chi}
\left[\frac{1}{\tsigT} \widetilde{\Sigma}(\chi,\mu)\right]\,.
\end{equation*}
The particular linear combination of moments introduced in \eqn{eq:EECcos-def} has the property 
that the divergence of the differential EEC distribution in the forward region ($\chi \to 0$) is 
suppressed by the factor of $(1-\cos\chi)$. Hence, in contrast to EEC itself, the fixed-order 
cumulative coefficients of $\widetilde{\Sigma}(\chi,\mu)$ can be computed reliably. Furthermore, 
one can show that in massless QCD this cumulative distribution is unity when $\chi = 180\degree$. 
Hence, we can integrate the fixed-order differential distribution in \eqn{eq:dEEC-fo} and 
use the unitarity constraint $\widetilde{\Sigma}(\pi,\mu)/\tsigT = 1$ to all orders in $\as$ to 
fix the constants of integration,
\begin{multline}
\bigg[\frac{1}{\tsigT} \widetilde{\Sigma}(\chi,\mu)\bigg]_{\mathrm{f.o.}} = 
	1 
        + 
        \asbar{\mu} \bar{{\mathcal A}}(\chi,\xiR)
	+\epjcbreak{+}
	\left(\asbar{\mu}\right)^2 \bar{{\mathcal B}}(\chi,\xiR)
	+
	\left(\asbar{\mu}\right)^3 \bar{{\mathcal C}}(\chi,\xiR)
	+
	\Oa{4}\,.
\label{eq:EECcos-fo}
\end{multline}
Moreover, starting from \eqn{eq:dEEC-res} and using the definition of $\widetilde{\Sigma}$, 
\eqn{eq:EECcos-def}, we obtain the following expression for the resummed prediction:\footnote{Note 
a misprint in eq.~(3.12) of \Refr{Tulipant:2017ybb} where an overall factor of $1/2$ appears erroneously.}
\begin{multline}
\bigg[\frac{1}{\tsigT} \widetilde{\Sigma}(\chi,\mu)\bigg]_{\mathrm{res.}} =
        H(\as(\mu)) \int_0^\infty  
        \bigg[ Q\sqrt{y}(1-y) J_1(b\, Q\sqrt{y})
                +\epjcbreak{+} \frac{2 y}{b} J_2(b\, Q\sqrt{y}) \bigg] S(Q,b) \rd b\,\,,
\label{eq:EECcos-res}
\end{multline}
where the Sudakov form factor $S(Q,b)$ is the one given in \eqn{eq:dEEC-Sudakov-gi}. 

The final expression for the matched prediction was derived in \Refr{Tulipant:2017ybb} and reads
\begin{multline}
\ln\bigg[\frac{1}{\tsigT} \widetilde{\Sigma}(\chi,\mu)\bigg] =
\ln\bigg\{ \frac{1}{H(\as(\mu))}
\bigg[ \frac{1}{\tsigT} \widetilde{\Sigma}(\chi,\mu) \bigg]_{\mathrm{res.}} \bigg\}
-\epjcbreak{-}\ln\bigg\{ \frac{1}{H(\as(\mu))}
\bigg[ \frac{1}{\tsigT} \widetilde{\Sigma}(\chi,\mu) \bigg]_{\mathrm{res.}} \bigg\}_{\mathrm{f.o.}}
\draftbreak{}\arxivbreak{}+
    \frac{\as(\mu)}{2\pi}
        \bar{\mathcal{A}}(\chi,\mu)\,+\,
    \epjcbreak{+}\left(\frac{\as(\mu)}{2\pi}\right)^2
        \bigg[\bar{\mathcal{B}}(\chi,\mu) - \frac{1}{2}\bar{\mathcal{A}}^2(\chi,\mu)\bigg]+
\\+
    \left(\frac{\as(\mu)}{2\pi}\right)^3
        \bigg[\bar{\mathcal{C}}(\chi,\mu) - \bar{\mathcal{A}}(\chi,\mu)\bar{\mathcal{B}}(\chi,\mu) 
          + \frac{1}{3}\bar{\mathcal{A}}^3(\chi,\mu) \bigg]\,.
\label{eq:EECcos-log-R-match}
\end{multline}
Here $\bigg\{ \frac{1}{H(\as(\mu))} \bigg[ \frac{1}{\tsigT} \widetilde{\Sigma}(\chi,\mu) \bigg]_{\mathrm{res.}} 
\bigg\}_{\mathrm{f.o.}}$ is the fixed-order expansion of the resummed result,
\begin{multline}
\frac{1}{H(\as(\mu))}
	\bigg[\frac{1}{\tsigT} \widetilde{\Sigma}(\chi,\mu)\bigg]_{\mathrm{res.}} =
	1
	+
	\frac{\as(\mu)}{2\pi} \bar{\mathcal A}_{\mathrm{res.}}(\chi,\mu) 
	+\epjcbreak{+}
	\left(\frac{\as(\mu)}{2\pi}\right)^2 \bar{\mathcal B}_{\mathrm{res.}}(\chi,\mu) 
+\arxivbreak{+}\draftbreak{+}
	\left(\frac{\as(\mu)}{2\pi}\right)^3 \bar{\mathcal C}_{\mathrm{res.}}(\chi,\mu)
	+
	\Oa{4}\,.
\label{eq:EECcos-res-exp}
\end{multline}%
The expansion coefficients $\bar{\mathcal A}_{\mathrm{res.}}$, $\bar{\mathcal B}_{\mathrm{res.}}$ 
and $\bar{\mathcal C}_{\mathrm{res.}}$ can be found in \Refr{Tulipant:2017ybb}.

Notice that the function $H(\as)$ does not appear in \eqn{eq:EECcos-log-R-match} at all. 
In the log-$R$ matching scheme such non-logarithmically enhanced contributions should not be 
exponentiated, instead these terms, as well as subdominant logarithmic contributions, are all 
implicit in the unsubtracted parts of the fixed-order coefficients $\bar{\mathcal{A}}$, 
$\bar{\mathcal{B}}$ and $\bar{\mathcal{C}}$ \cite{Catani:1992ua}. Thus the log-$R$ matched 
prediction can be computed without the explicit knowledge of $H^{(n)}$.

Finally, we comment on our implementation of the unitarity constraint 
$\widetilde{\Sigma}(\pi,\mu)/\tsigT = 1$. It can be shown that this constraint can be satisfied 
by modifying the resummation formula in \eqn{eq:dEEC-res} such that in the kinematical limit $y=1$ 
the Sudakov form factor is unity. This may be achieved in several ways and here we choose a very 
simple solution and modify the resummation coefficients $\tilde{A}^{(n)}$ and $\tilde{B}^{(n)}$ 
according to
\beq
\bsp
\tilde{A}^{(n)}(\xiL) &\to \tilde{A}^{(n)}(y,\xiL) = \tilde{A}^{(n)}(\xiL)(1-y)^p\,,
\\
\tilde{B}^{(n)}(\xiL) &\to \tilde{B}^{(n)}(y,\xiL) = \tilde{B}^{(n)}(\xiL)(1-y)^p\,,
\esp
\label{eq:unitarity-p}
\eeq
where $p$ is a positive number.\footnote{A modification similar in spirit was employed in 
\Refr{Bizon:2017rah} although in the context of matching the fixed-order and resummed 
predictions for transverse observables in Higgs hadroproduction.} This modification is fully 
legitimate since it does not modify the logarithmic structure of the result and introduces only 
power-suppressed terms. In practice, we set $p=1$ and quantify the impact of this modification 
by comparing the results to those obtained with $p=2$.

\subsection{Finite $b$-quark mass corrections}
\label{ssec:mb}
The theoretical prediction presented above was computed in massless QCD. However, the 
assumption of vanishing quark masses is not fully justified, especially at lower energies, where 
$b$-quark mass effects are relevant at the percent level~\cite{Gehrmann:2012sc}. 
In order to take $b$-quark mass corrections into account, we subtract the fraction of $b$-quark events, 
$r_b(Q)$ from the massless result and add back the corresponding massive contribution. Hence, we 
include mass effects directly at the level of matched distributions,
\begin{multline}
\frac{1}{\tsigT} \frac{\rd \Sigma(\chi,Q)}{\rd \cos \chi} = 
	(1 - r_b(Q))\bigg[\frac{1}{\tsigT} \frac{\rd \Sigma(\chi,Q)}{\rd \cos \chi}\bigg]_{\mathrm{massless}} 
	+\epjcbreak{+} r_b(Q)\bigg[\frac{1}{\tsigT} \frac{\rd \Sigma(\chi,Q)}{\rd \cos \chi}\bigg]^{NNLO^{*}}_{\mathrm{massive}} \,.
\label{eq:mb-corr-base}
\end{multline}
Here $\Big[\frac{1}{\tsigT} \frac{\rd \Sigma(\chi,Q)}{\rd \cos \chi}\Big]_{\mathrm{massless}}$ is the 
NNLO+NNLL matched distribution, computed in the log-$R$ matching scheme in massless QCD as 
outlined above, while $\Big[\frac{1}{\tsigT} \frac{\rd \Sigma(\chi,Q)}{\rd \cos \chi}\Big]^{NNLO^{*}}_{\mathrm{massive}}$ 
is the fixed-order massive distribution. As the complete NNLO correction to this distribution is currently 
unknown, we model it by supplementing the massive NLO prediction of the parton level Monte Carlo 
generator {\tt Zbb4}~\cite{Nason:1997tz}, with the NNLO coefficient of the massless fixed-order result. 

We define the fraction of $b$-quark events as the ratio of the total $b$-quark production cross section 
divided by the total hadronic cross section,
\begin{equation*}
r_b(Q) \equiv \frac{\tsig{}{\mathrm{massive}}(e^+e^- \to b\bar{b})}
	{\tsig{}{\mathrm{massive}}(e^+e^- \to \mathrm{hadrons})}\,.
\end{equation*}
We evaluate the ratio of these cross sections at NNLO accuracy (\Oa{2} in the strong coupling) including 
the exact $b$-quark mass corrections at $\Oa{}$ and the leading mass terms up to $(m_b^2/Q^2)^2$ at 
$\Oa{2}$~\cite{Chetyrkin:2000zk}. We note that the electroweak coupling factors do not cancel in this ratio 
and the summation over quark flavours has to be carried out explicitly when computing 
$\tsig{}{\mathrm{massive}}(e^+e^- \to \mathrm{hadrons})$.

Distributions for $\Big[\frac{1}{\tsigT} \frac{\rd \Sigma(\chi,Q)}{\rd \cos \chi}\Big]_{\mathrm{massive}}$ were 
generated for each of the considered energies using a pole $b$-quark mass of $m_b = 4.75\GeV$, which is 
consistent with world average estimations of pole mass $4.78\pm0.06\GeV$~\cite{Olive:2016xmw}.

In order to assess the uncertainty associated to the modelling of $b$-quark mass corrections, we have 
investigated two alternative approaches for including them in our predictions. In approach $A$ \eqn{eq:mb-corr-base} 
is modified to 
\begin{multline*}
\frac{1}{\tsigT} \frac{\rd \Sigma(\chi,Q)}{\rd \cos \chi} = 
	\bigg[\frac{1}{\tsigT} \frac{\rd \Sigma(\chi,Q)}{\rd \cos \chi}\bigg]_{\mathrm{massless}} 
	+\epjcbreak{+} r_b(Q)\bigg[\frac{1}{\tsigT} \frac{\rd \Sigma(\chi,Q)}{\rd \cos \chi}\bigg]^{NLO}_{\mathrm{massive}}
	-r_b(Q)\bigg[\frac{1}{\tsigT} \frac{\rd \Sigma(\chi,Q)}{\rd \cos \chi}\bigg]^{NLO}_{\mathrm{massless}} \,,
\end{multline*}
i.e., we simply subtract the massless fixed-order NLO prediction multiplied by the fraction $r_b(Q)$ 
of $b$-quark events and add back the corresponding massive NLO distribution. Approach $B$ is defined in a way 
very similar to our baseline, \eqn{eq:mb-corr-base}, but we do not include any NNLO corrections to the massive 
distribution,
\begin{multline*}
\frac{1}{\tsigT} \frac{\rd \Sigma(\chi,Q)}{\rd \cos \chi} = 
	(1 - r_b(Q))\bigg[\frac{1}{\tsigT} \frac{\rd \Sigma(\chi,Q)}{\rd \cos \chi}\bigg]_{\mathrm{massless}} 
	+\epjcbreak{+} r_b(Q)\bigg[\frac{1}{\tsigT} \frac{\rd \Sigma(\chi,Q)}{\rd \cos \chi}\bigg]^{NLO}_{\mathrm{massive}} \,.
\end{multline*}
Hence, $\Big[\frac{1}{\tsigT} \frac{\rd \Sigma(\chi,Q)}{\rd \cos \chi}\Big]^{NLO}_{\mathrm{massive}}$ is simply 
the prediction obtained with {\tt Zbb4}.

\section{Extraction  procedure}
\label{sec:fit}
To extract the strong coupling the predictions described above were 
confronted with the available data sets. Namely, the data  obtained 
in SLD~\cite{Abe:1994mf},  L3~\cite{Adriani:1992gs}, 
 DELPHI~\cite{Abreu:1992yc}, OPAL~\cite{Acton:1993zh,Acton:1991cu}, TOPAZ~\cite{Adachi:1989ej}, TASSO~\cite{Braunschweig:1987ig}, 
 JADE\cite{Bartel:1984uc}, \epjcbreak{} MAC~\cite{Fernandez:1984db}, MARKII~\cite{Wood:1987uf}, CELLO~\cite{Behrend:1982na}  
and PLUTO~\cite{Berger:1985xq}
 experiments were included. 
The information on used data is summarised in Tab.~\ref{tab:input}.

\begin{table}[!htbp]\centering
\begin{tabular}{|c|c|c|c|}\hline
Experiment                                      & $\sqrt{s},\GeV$, data &$\sqrt{s},\GeV$, MC  & Events  \\\hline\hline
\tabularinput\hline
\end{tabular}
\caption{
Data used in the extraction procedure. The average of $\sqrt{s}$ is given in the brackets.}
\label{tab:input}
\end{table}

The criteria to include the data were high precision of differential 
distributions obtained with charged and neutral final state particles
 in the full $\chi$ range,
presence of corrections for detector effects, correction for  initial state photon radiation
 and sufficient amount of supplementary information.
Therefore, data sets without  supplementary information~\cite{Berger:1980yh}, with  large uncertainties~\cite{Adeva:1991vw}, 
superseded datasets~\cite{Akrawy:1990hy,Abe:1994wv} and measurements 
unfolded only to charged particles in the final state~\cite{Abreu:2000ck} are not included in the analysis.

The  data sets selected for the extraction procedure have high precision and 
the measurements from different experiments performed at close energy points are consistent\footnote{Some observed differences between the
measurements performed at $\sqrt{s}=91.2\GeV$ are not statistically significant once the systematical uncertainties and 
correlations are taken into account.}.
 This justifies their use in the extraction procedure in a wide centre-of-mass energy interval,
 similarly to studies of thrust~\cite{Gehrmann:2012sc} and $C$ parameter~\cite{Hoang:2015hka} 
 and allows us to target the highest precision of $\as$ determination with available theoretical 
 predictions.

\subsection{Monte Carlo generation setup}
\label{sec:mcsetup}

In a previous study~\cite{Tulipant:2017ybb} the non-perturbative effects for the EEC distribution 
were modelled with an analytic approach. In this paper the non-perturbative effects in 
the $e^{+}e^{-}\rightarrow hadrons$ process are modelled using state-of-the-art particle-level 
Monte Carlo (MC) generators. The non-perturbative corrections of the energy-energy correlation 
distributions were extracted as ratios of energy-energy correlation distributions at hadron and parton 
level in the simulated samples.

In this study the  MC generators 
{\tt SHERPA2.2.4}~\cite{Gleisberg:2008ta}\footnote{Partially updated to version 2.2.5.}
and {\tt Herwig7.1.1}~\cite{Bellm:2015jjp}
were used.

The $e^{+}e^{-}\rightarrow hadrons $ MC  samples were generated  
at centre-of-mass energies 
$\sqrt{s}=14.0$, $22.0$, $29.0$, $34.0$, $53.2$, $59.5$ and  $91.2\GeV$.
In all cases, the simulation of initial state radiation was disabled and generator settings 
were defaults if the opposite is not  stated explicitly.
The value of the strong coupling used for the hard process  was set 
to $\as(M_{Z})=0.1181$~\cite{Olive:2016xmw}.

The {\tt SHERPA2.2.4} samples were generated with the MENLOPS method using the matrix 
element generators {\tt AMEGIC}~\cite{Krauss:2001iv}, {\tt COMIX}~\cite{Duhr:2006iq} and 
the {\tt GoSam}~\cite{Cullen:2014yla} one-loop library  to produce matrix elements for 
 $e^{+}e^{-}\rightarrow Z/\gamma\rightarrow {\rm 2,3,4,5\ partons}$ processes. 
The $2-$parton final state processes had NLO accuracy in perturbative QCD.
The  QCD matrix  elements  were calculated assuming massive $b$-quarks.
 The merging parameter $Y_{\mathrm{cut}}$ was  set to $10^{-2.75} \simeq 1.778\times 10^{-3}$.
  
To test the fragmentation and hadronization model dependence, the events 
 generated with {\tt SHERPA2.2.4}  were hadronized using the  Lund string fragmentation 
model~\cite{Sjostrand:2006za} or the cluster fragmentation  model~\cite{Winter:2003tt}.
The first setup is labelled below as $S^{L}$  and the second as $S^{C}$.  

To assure proper fragmentation of heavy quarks and heavy hadron decays the cluster fragmentation model was adjusted. 
The value of  {\tt SPLIT\_LEADEXPONENT} parameter was set to $1.0$,
the parameter {\tt M\_DIQUARK\_OFFSET} was set to $0.55$,
the production of charm  and beauty baryons was enhanced by factors $0.8$ and $1.7$.

For the cross-check of {\tt SHERPA2.2.4} samples, the \epjcbreak{} {\tt Herwig7.1.1} generator was used.
The {\tt Herwig7.1.1} samples were generated with the MENLOPS
method using the {\tt MadGraph5}~\cite{Alwall:2011uj} matrix element generator and 
the {\tt GoSam}~\cite{Cullen:2014yla} one-loop library to produce matrix elements of 
the  $e^{+}e^{-}\rightarrow Z/\gamma\rightarrow {\rm 2,3,4,5\ partons}$ processes. 
The $2-$parton final state processes again had NLO accuracy in perturbative QCD and 
the matrix elements  were calculated assuming massive $b$-quarks.
The merging parameter was set to $\sqrt{s}\times10^{-1.25} \simeq \sqrt{s}\times 5.623\times 10^{-2}$.
For the modelling of the hadronization process the default implementation of 
the cluster fragmentation model~\cite{Webber:1983if} was used.
To improve the modelling of beauty production at the lowest energies, 
the $b$-quark nominal mass was changed from the default value of 
$5.3\GeV$ to $5.1\GeV$. This setup  is labelled below as $H^{M}$.

\subsection{Estimation of hadronization effects from MC models}
\label{sec:mchad}
Estimation of hadronization corrections is an integral part of comparing 
the parton-level QCD predictions 
to the data measured on hadron (particle) level.
Despite the fact that under certain conditions the 
local parton-hadron duality leads to close  values of quantities on 
parton and hadron level, the difference between 
them is not negligible and should be taken into account in precise analyses.
One way to do so is to apply correction factors 
estimated from MC simulations to the perturbative QCD prediction.
The factors, called hadronization corrections $H/P$, are defined as ratios
of the corresponding quantities at parton level to the same quantities at 
hadron level at every point of the considered distribution.

To obtain the EEC distributions, the generated MC samples were 
processed in the same way as data (see e.g.\ Ref.~\cite{Acton:1993zh}), using 
partons before hadronization for parton-level calculations and  
undecayed/stable particles for hadron-level calculations.
For the parton-level calculations the parton energies were used as provided by 
the MC generators.

The predictions obtained with all 
setups  describe the data well for all 
ranges of $\chi$ with the exception of the regions near $\chi=0\degree$ and $\chi=90\degree$, 
for all values of $\sqrt{s}$. 
For $\sqrt{s}<29\GeV$ the $H^{M}$ setup is sensitive to the 
$b$-quark mass and the corresponding predictions are not reliable. 

\begin{figure}\centering
\includegraphics[width=\FIGWONE]{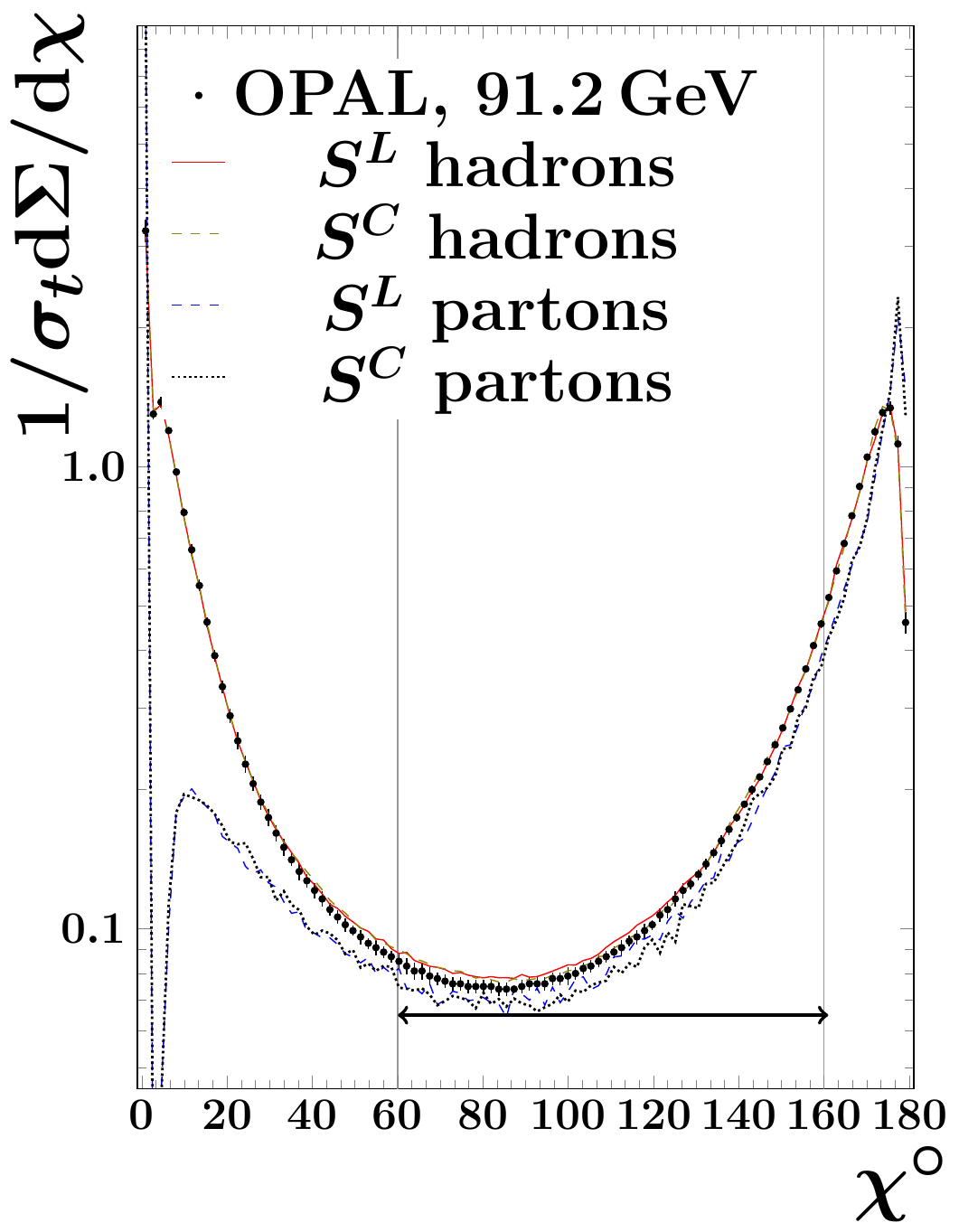}\includegraphics[width=\FIGWONE]{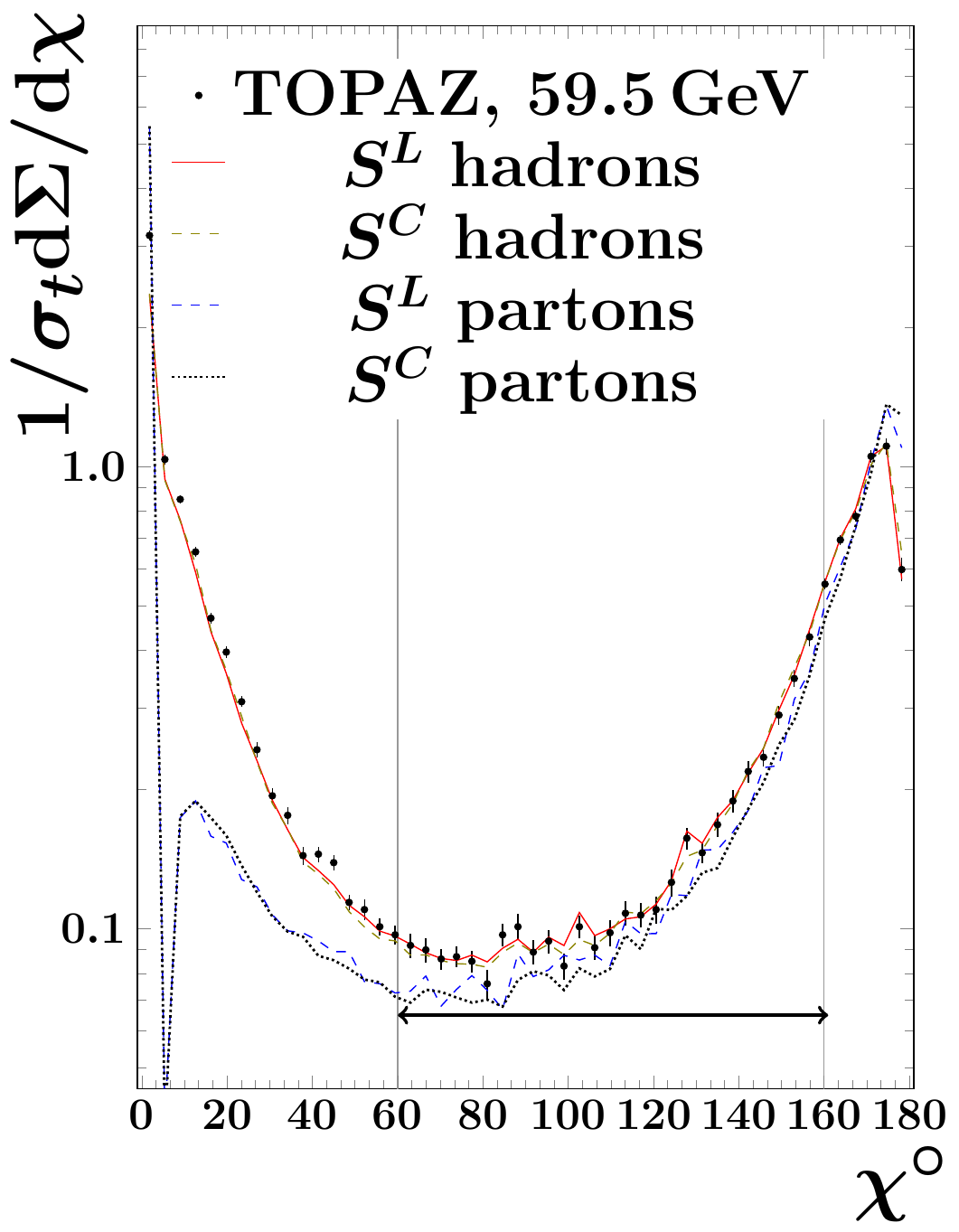}\epjcbreak{}\includegraphics[width=\FIGWONE]{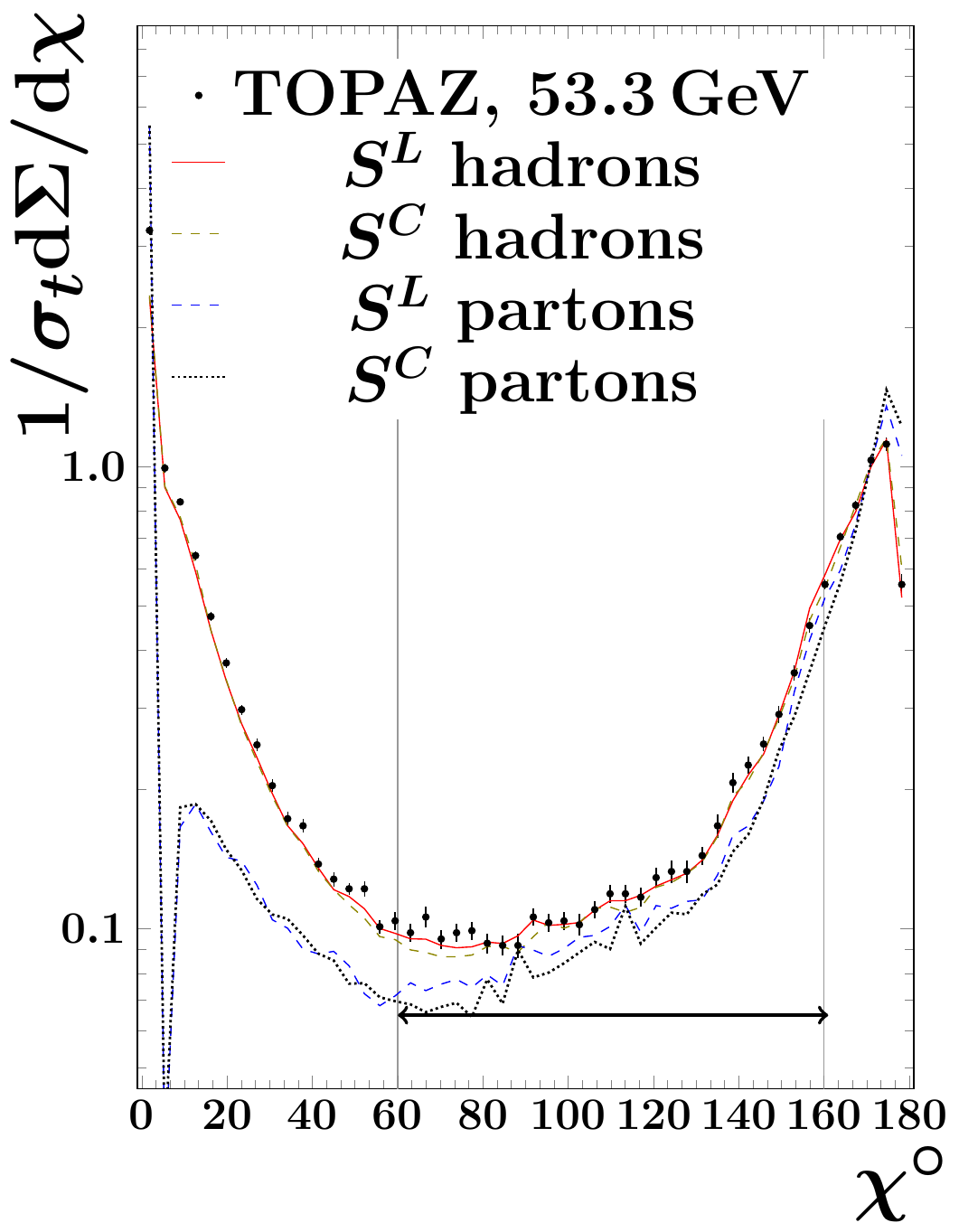}\draftbreak{}\arxivbreak{}\includegraphics[width=\FIGWONE]{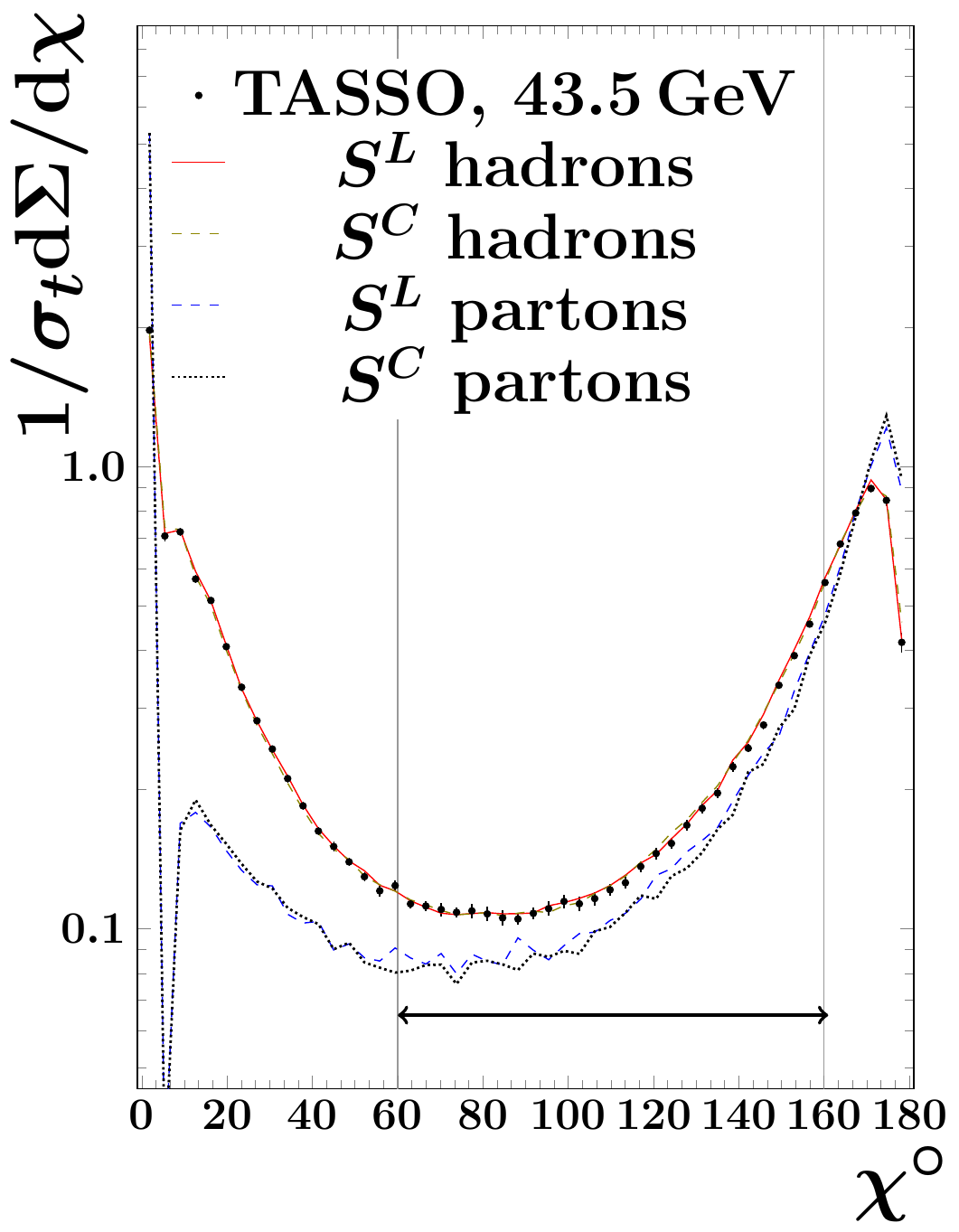}\epjcbreak{}\includegraphics[width=\FIGWONE]{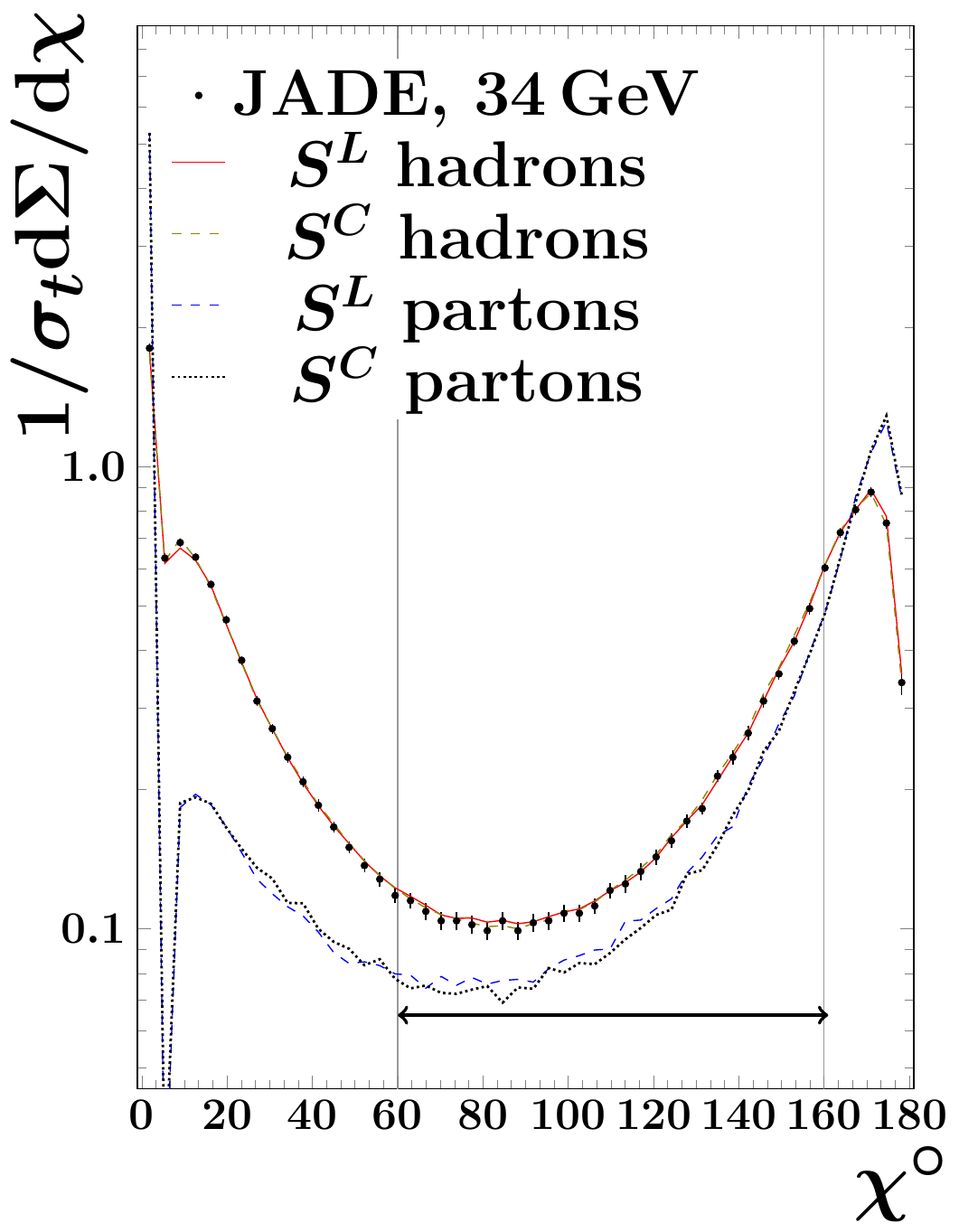}\includegraphics[width=\FIGWONE]{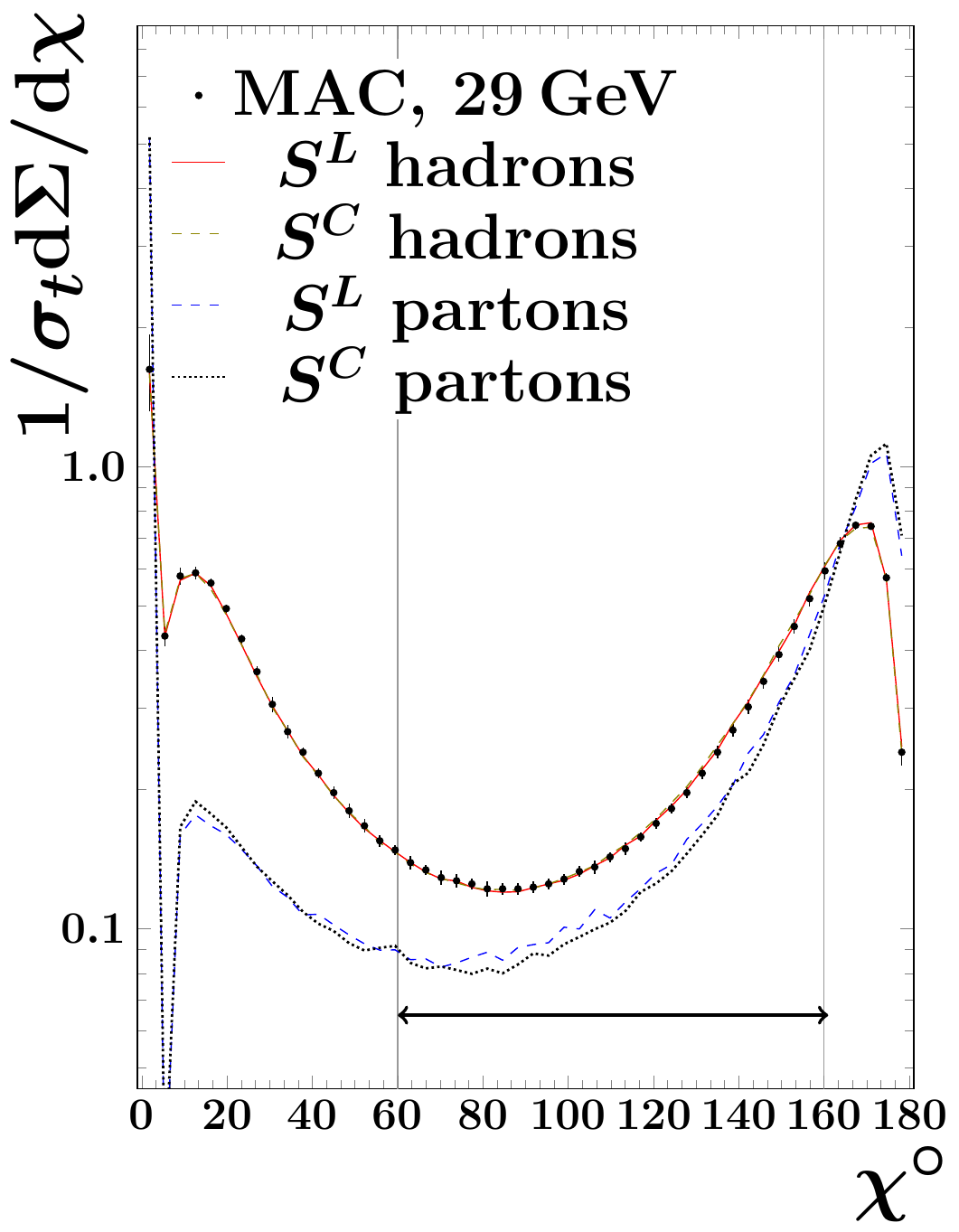}\\
\includegraphics[width=\FIGWONE]{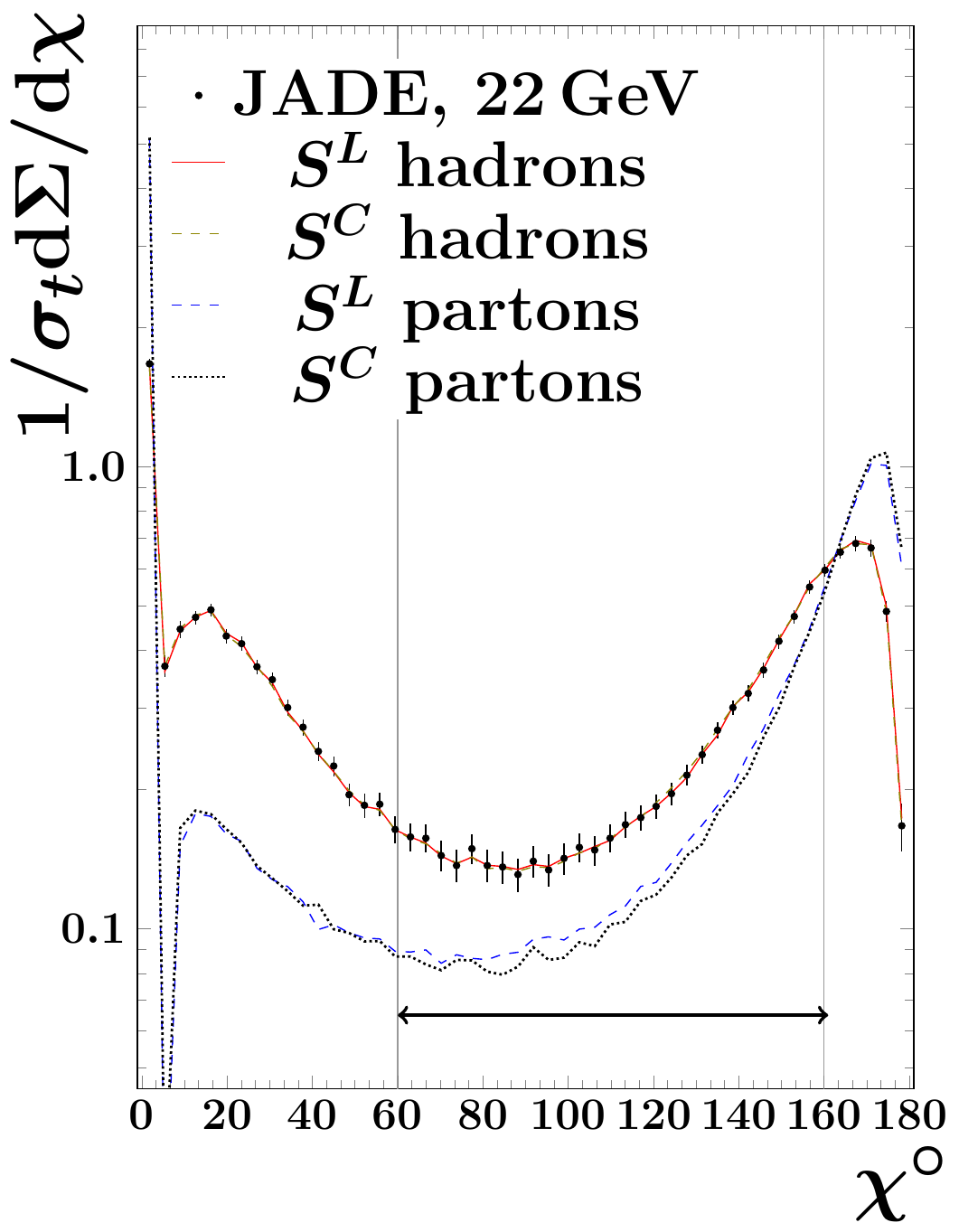}\includegraphics[width=\FIGWONE]{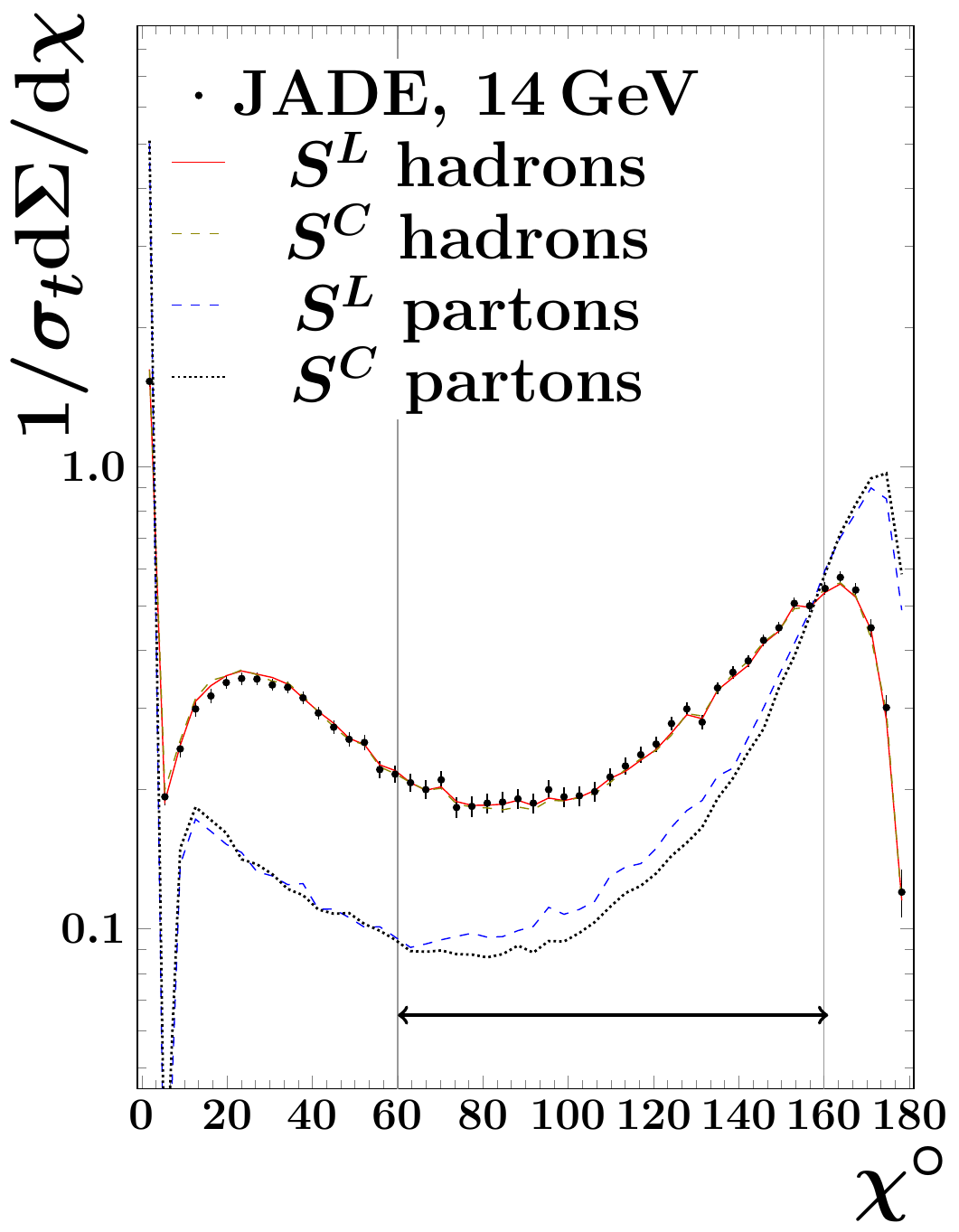}\\
\caption{
Data and Monte Carlo predictions  obtained  with the $S^{L}$  and  $S^{C}$ setups
at parton and hadron level. Reweighting was applied.
}
\label{fig:mcsherparew}
\end{figure}

\begin{figure}\centering
\includegraphics[width=\FIGWONE]{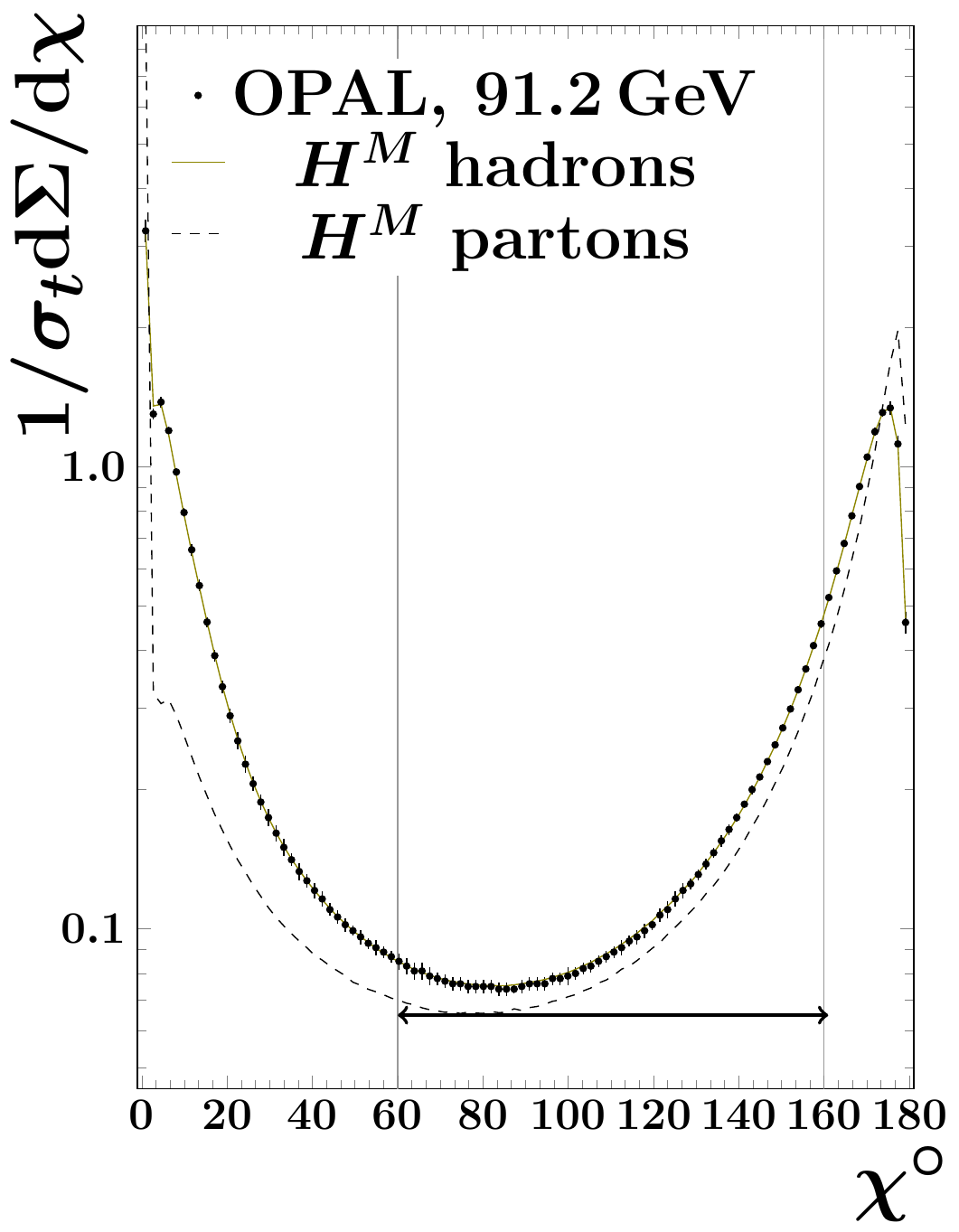}\includegraphics[width=\FIGWONE]{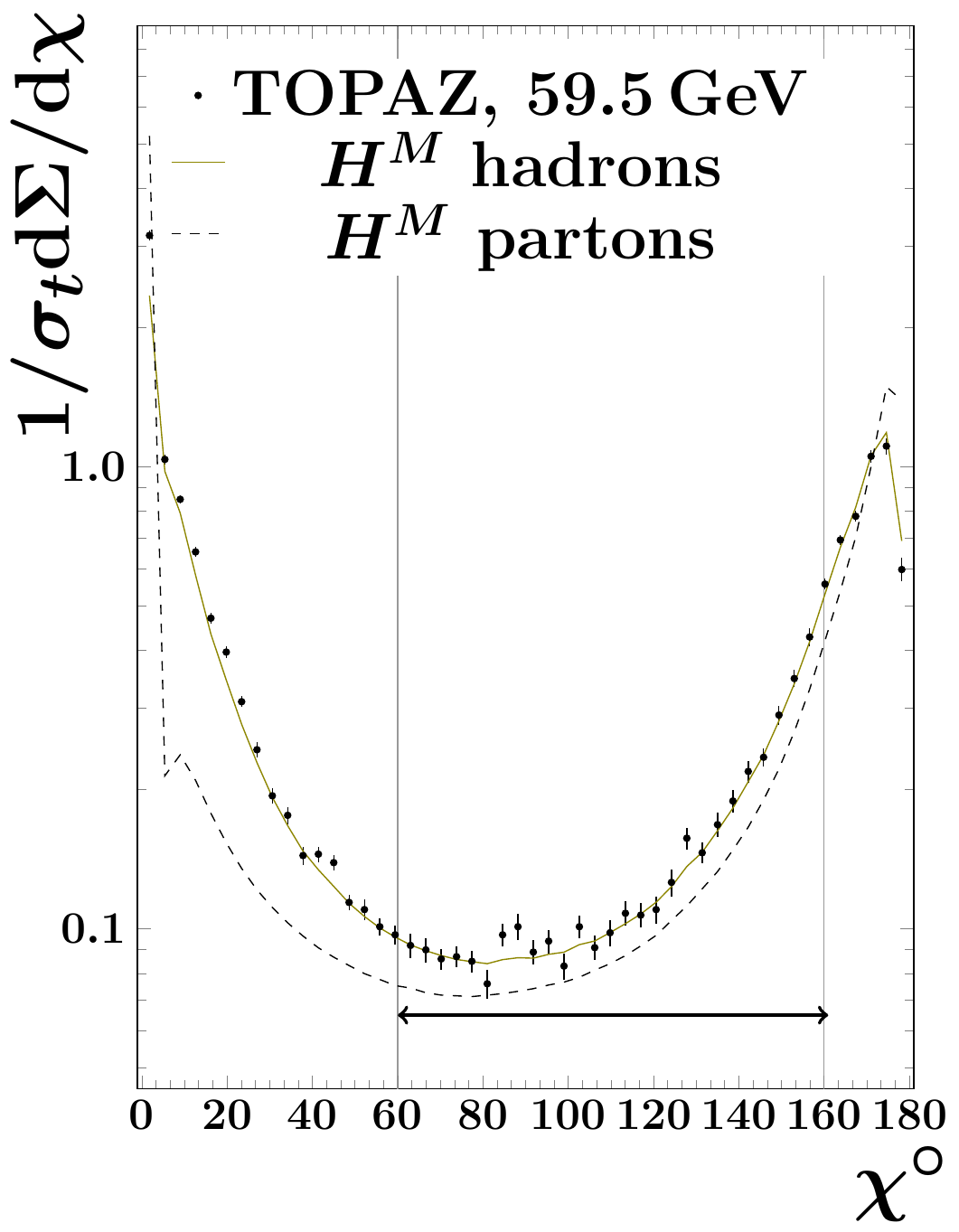}\epjcbreak{}\includegraphics[width=\FIGWONE]{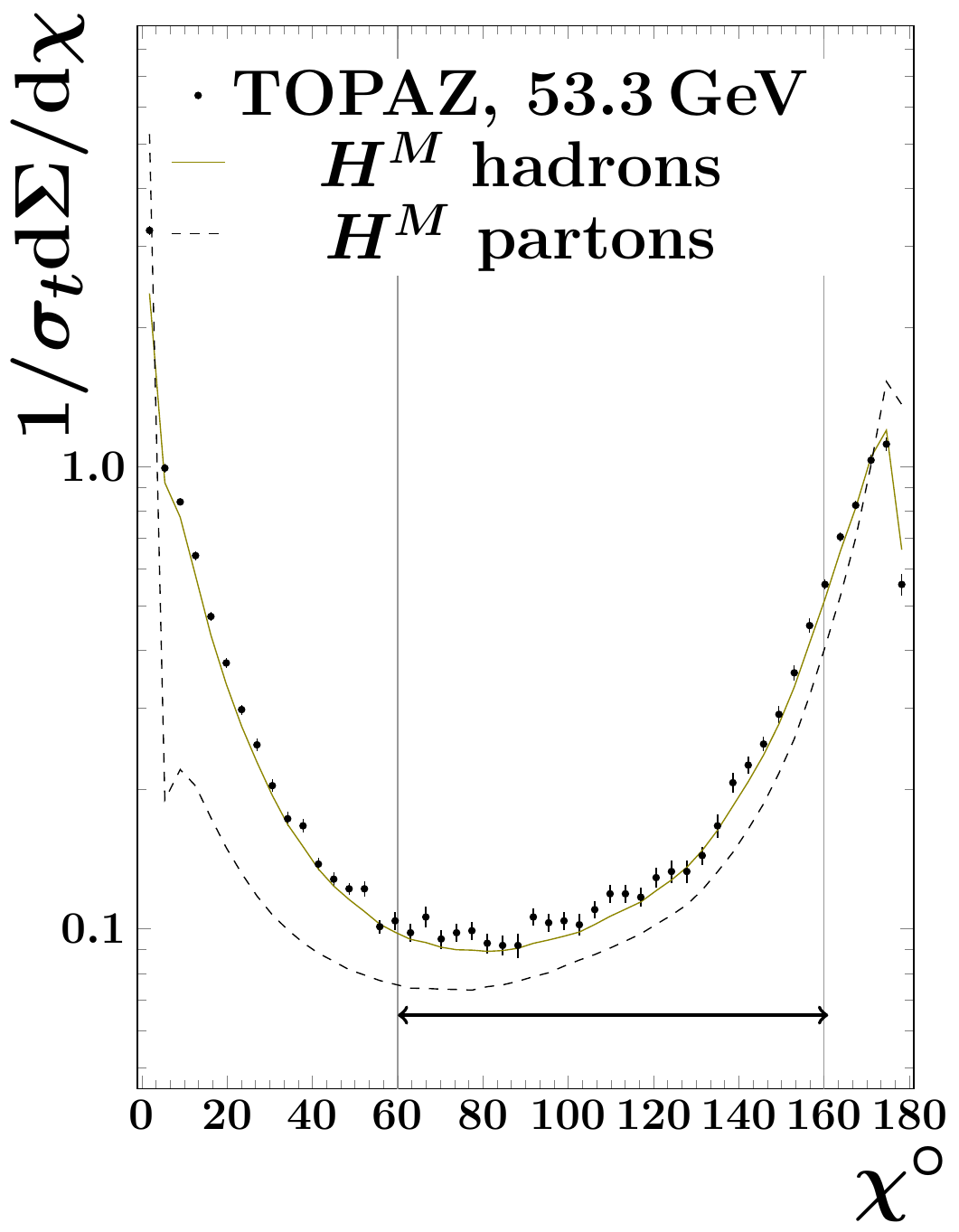}\draftbreak{}\arxivbreak{}\includegraphics[width=\FIGWONE]{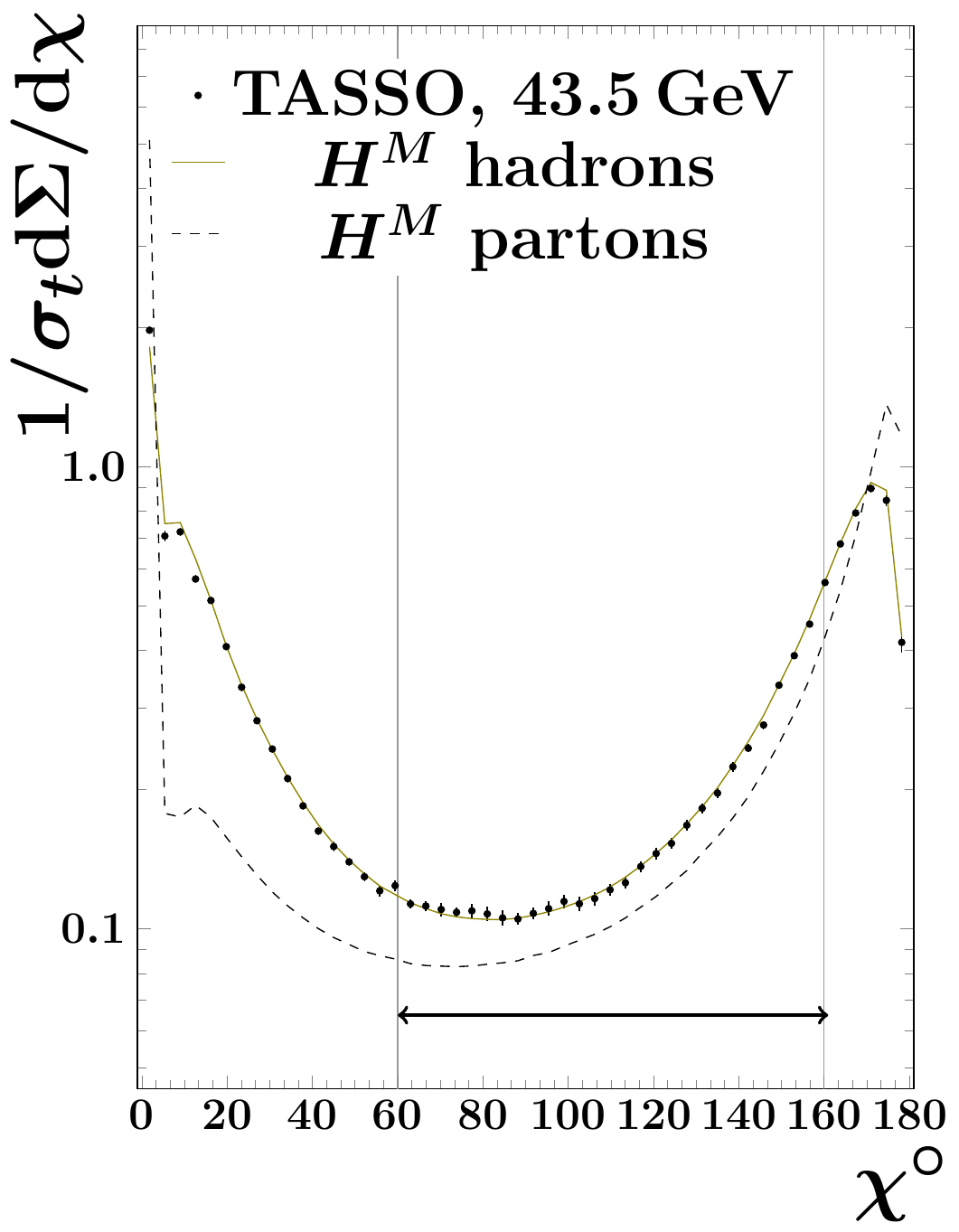}\epjcbreak{}\includegraphics[width=\FIGWONE]{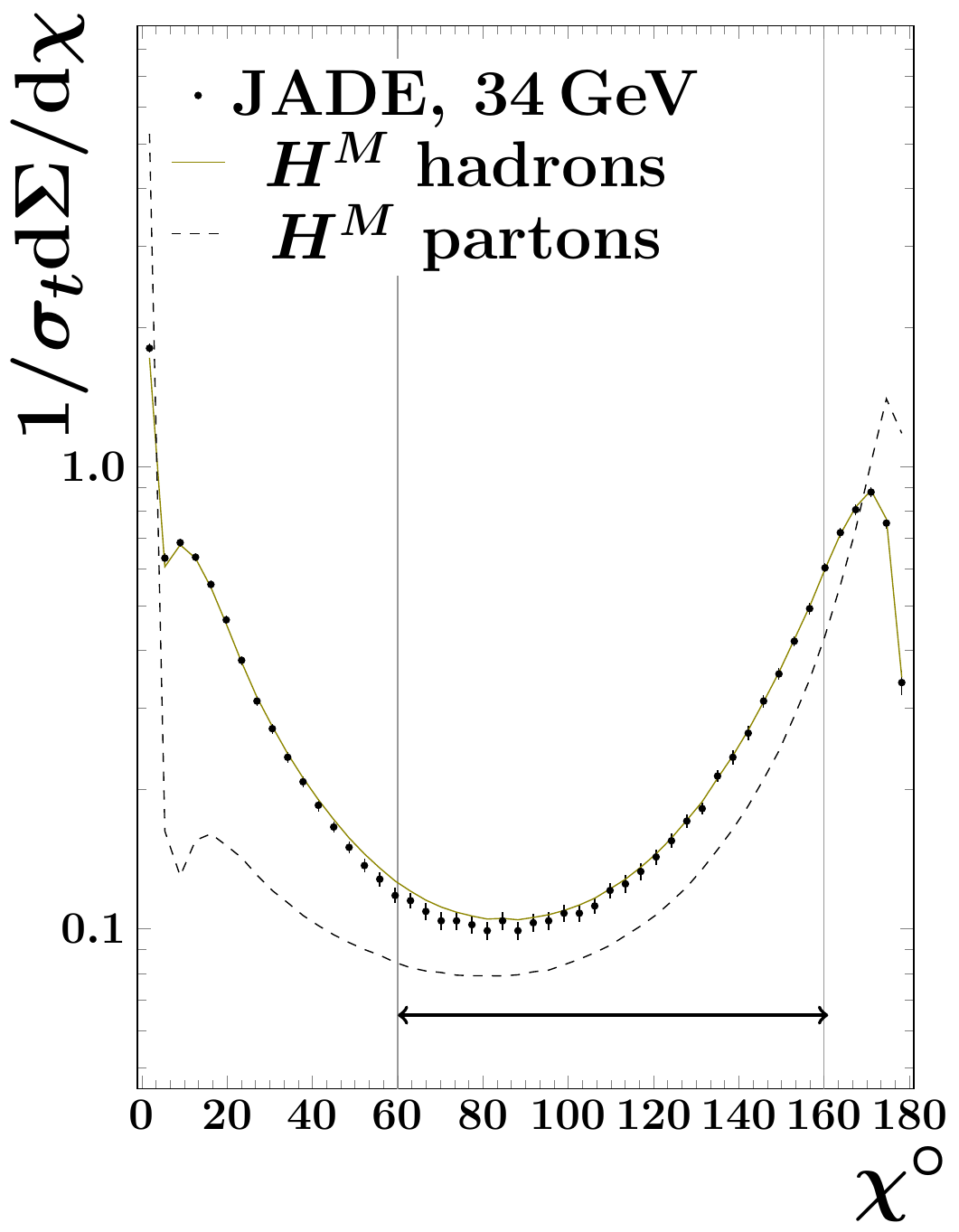}\includegraphics[width=\FIGWONE]{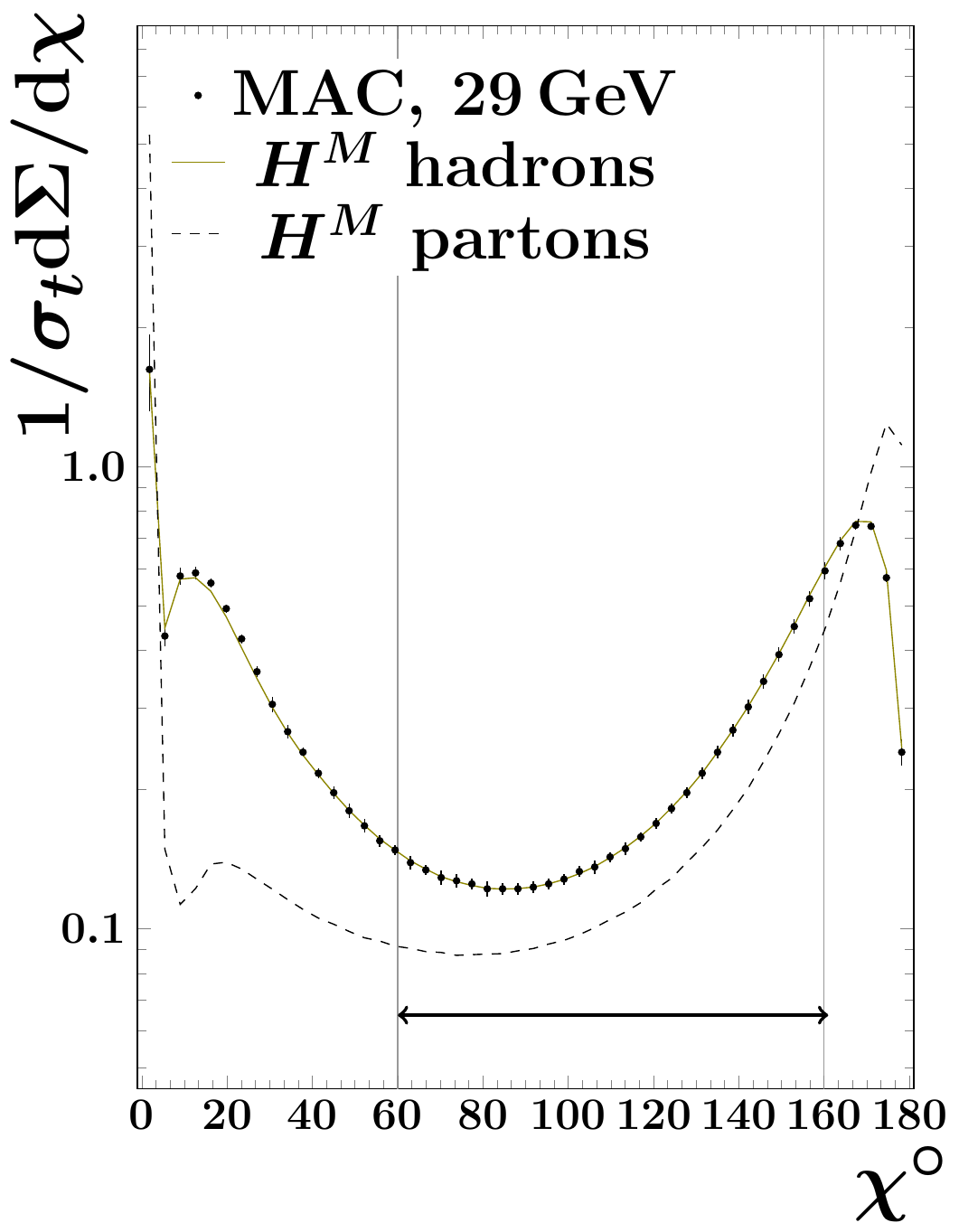}\\
\includegraphics[width=\FIGWONE]{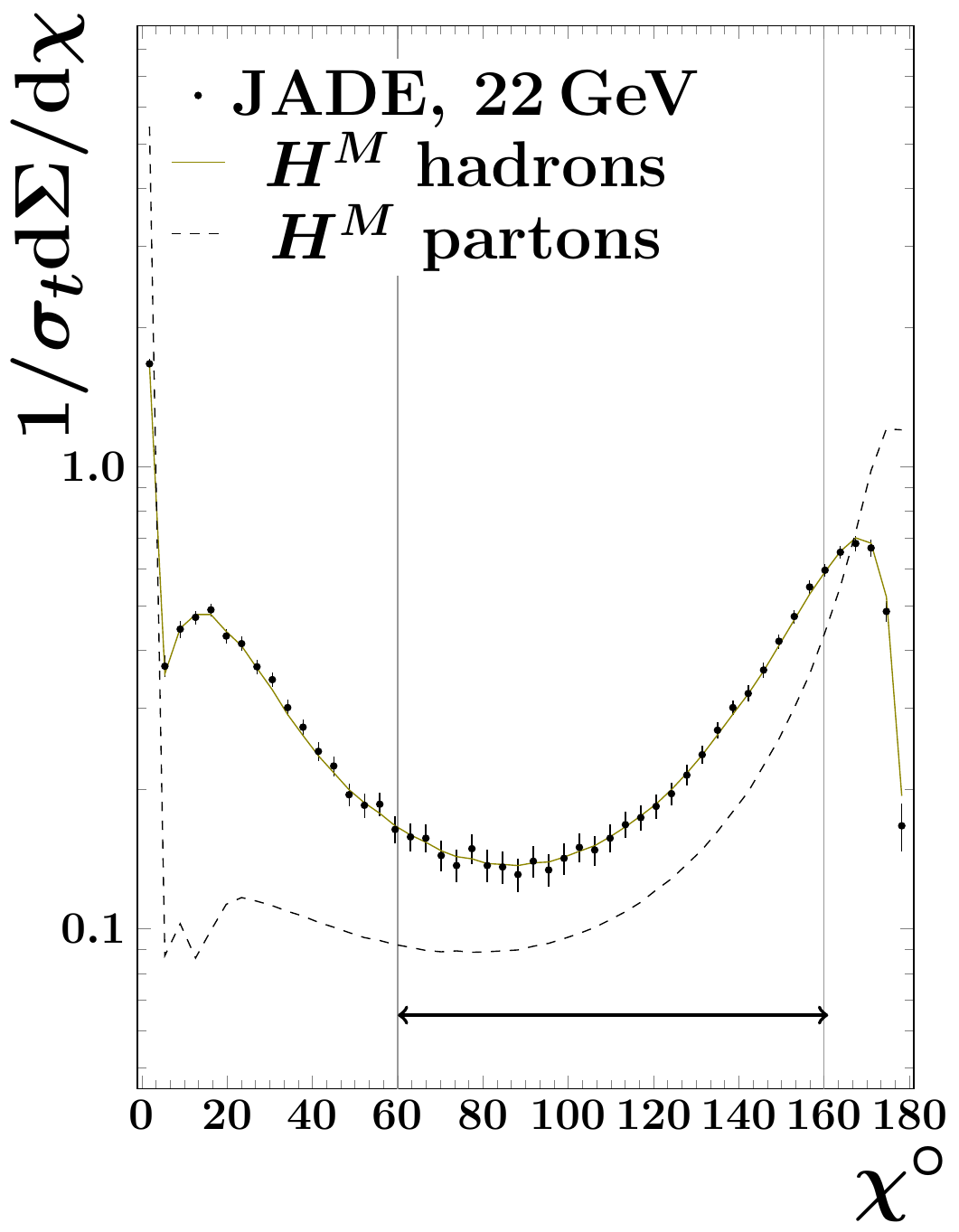}\includegraphics[width=\FIGWONE]{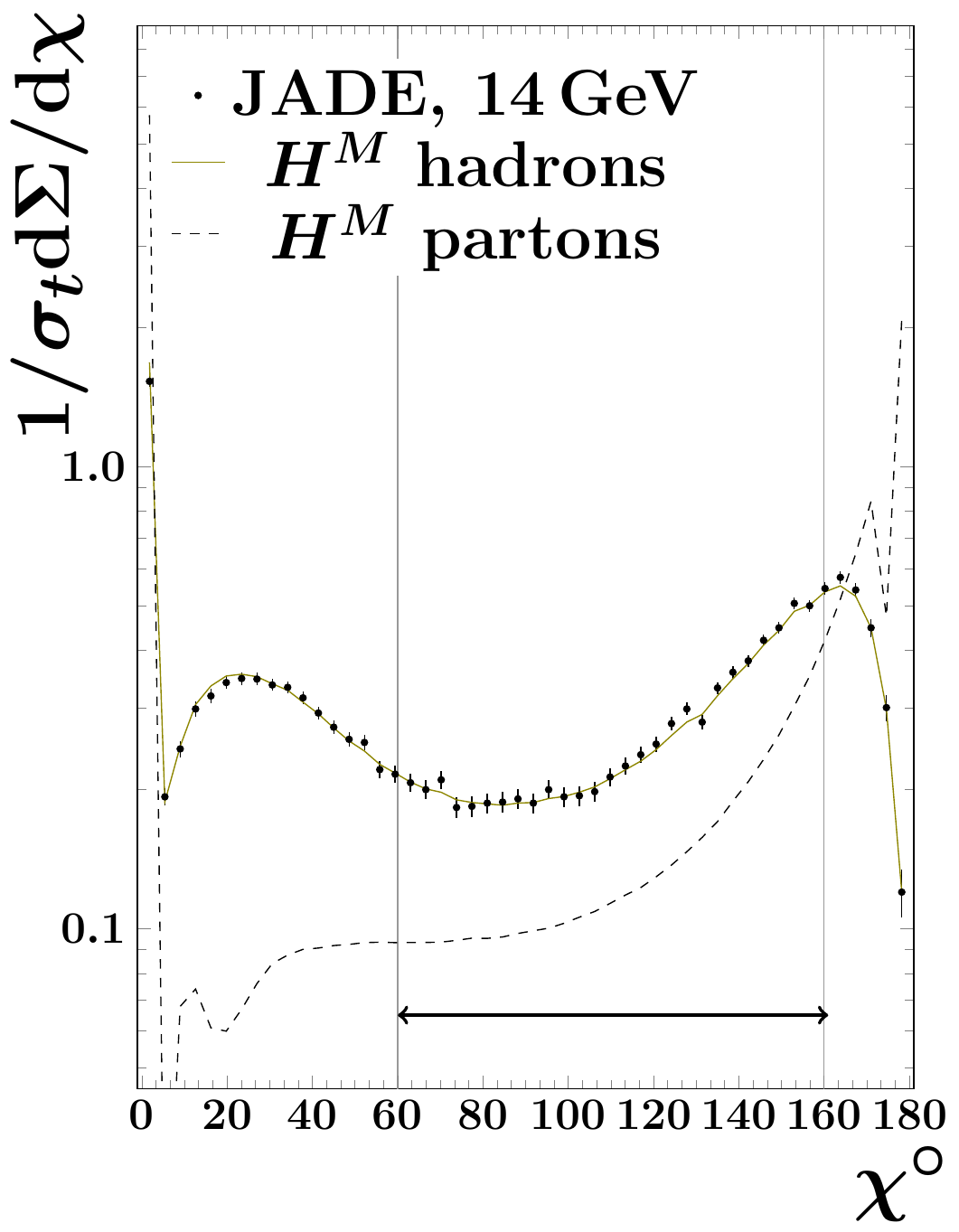}\\
\caption{
Data and Monte Carlo predictions  obtained  with the $H^{M}$  setup  
at parton and hadron level. Reweighting was applied.
}
\label{fig:mcherwigrew}
\end{figure}

However, to assure an even better description of data, a reweighting procedure was applied to the 
simulated samples. 
The samples were reweighted at hadron level on an event-by-event basis to describe the data 
and the corresponding event weights were propagated to the parton level. 
The resulting distributions are shown in Fig.~\ref{fig:mcsherparew} for the {\tt SHERPA2.2.4} 
setups and in Fig.~\ref{fig:mcherwigrew} for the  {\tt Herwig7.1.1} setup.

\begin{figure}\centering
\includegraphics[width=\FIGWONE]{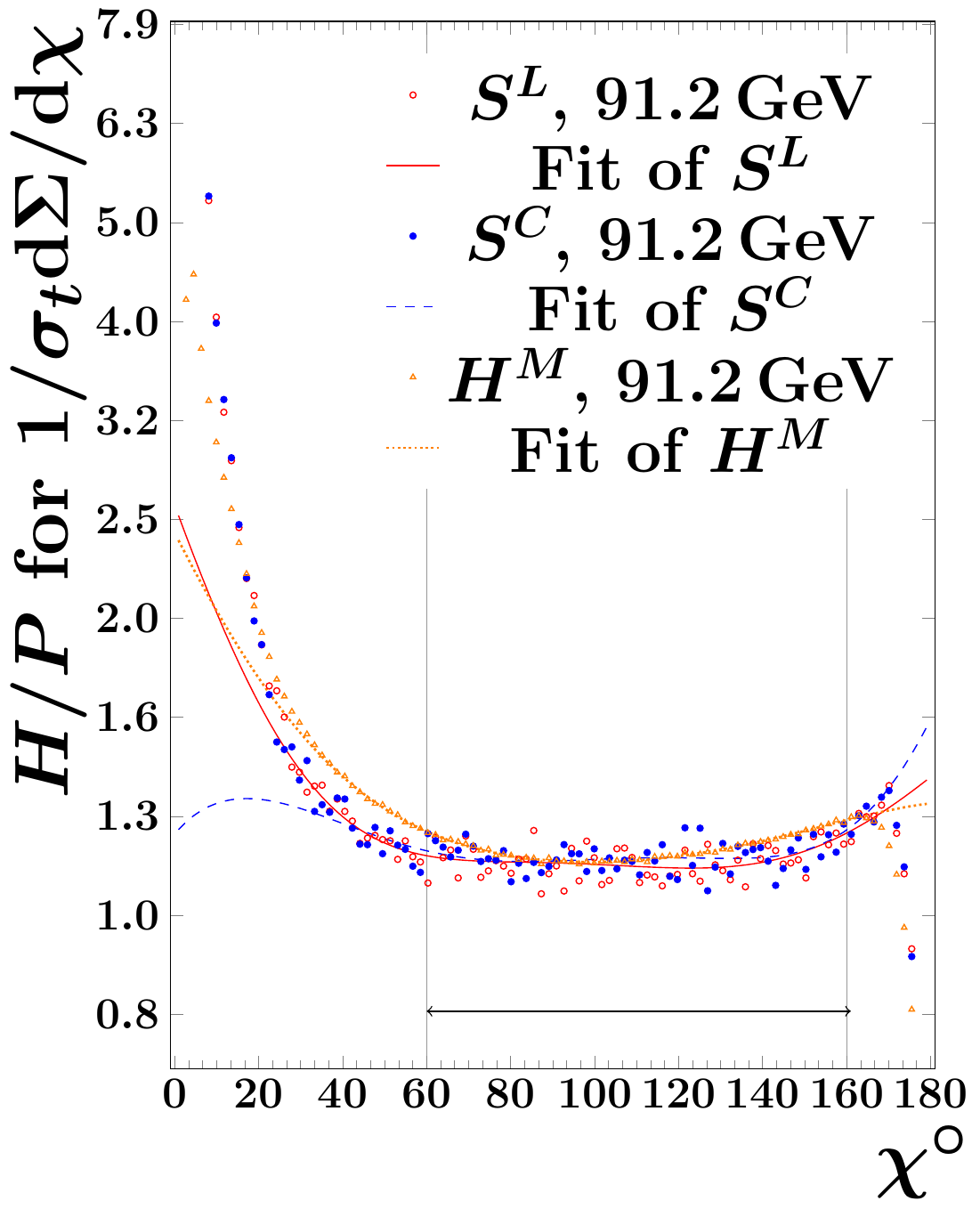}\includegraphics[width=\FIGWONE]{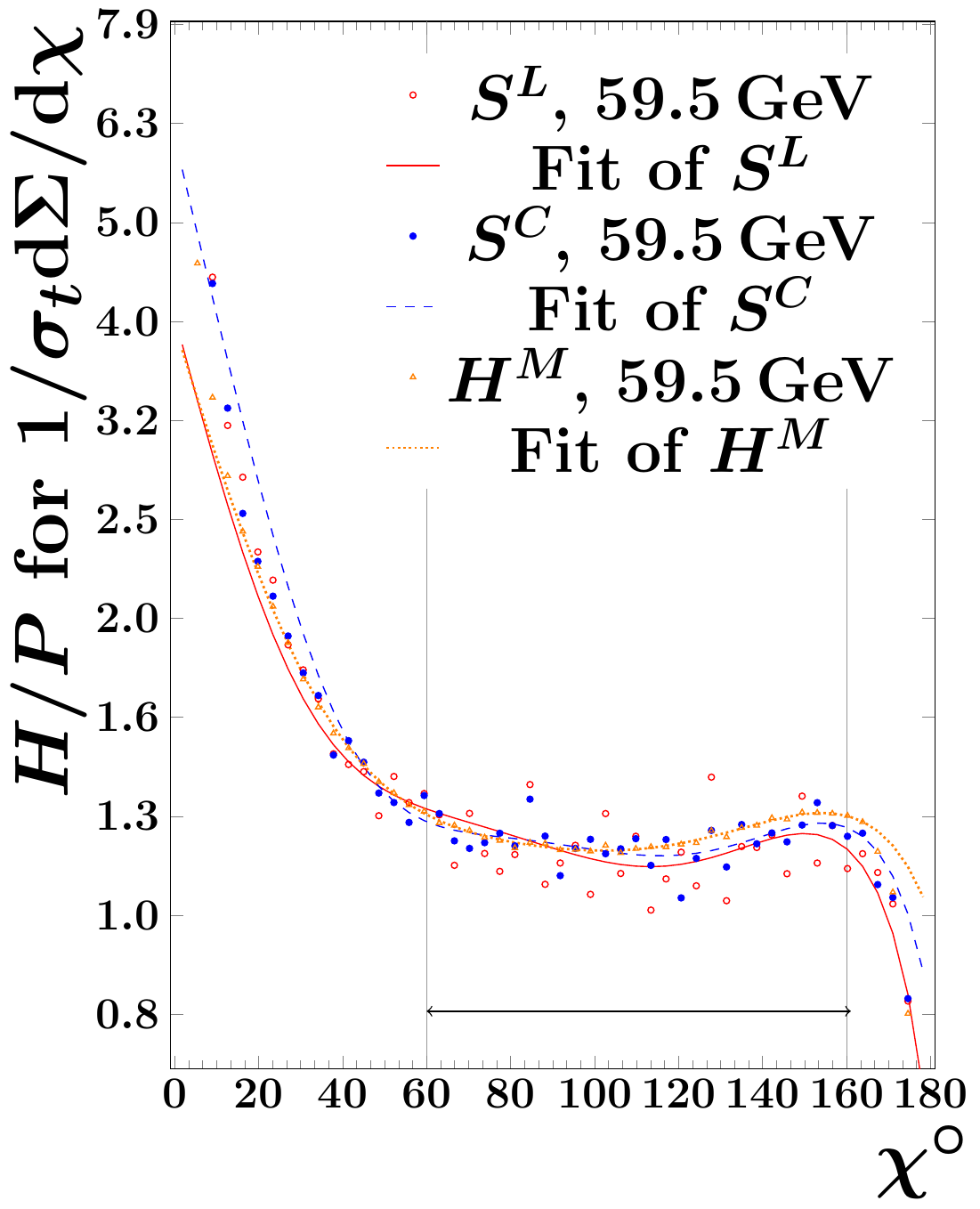}\epjcbreak{}\includegraphics[width=\FIGWONE]{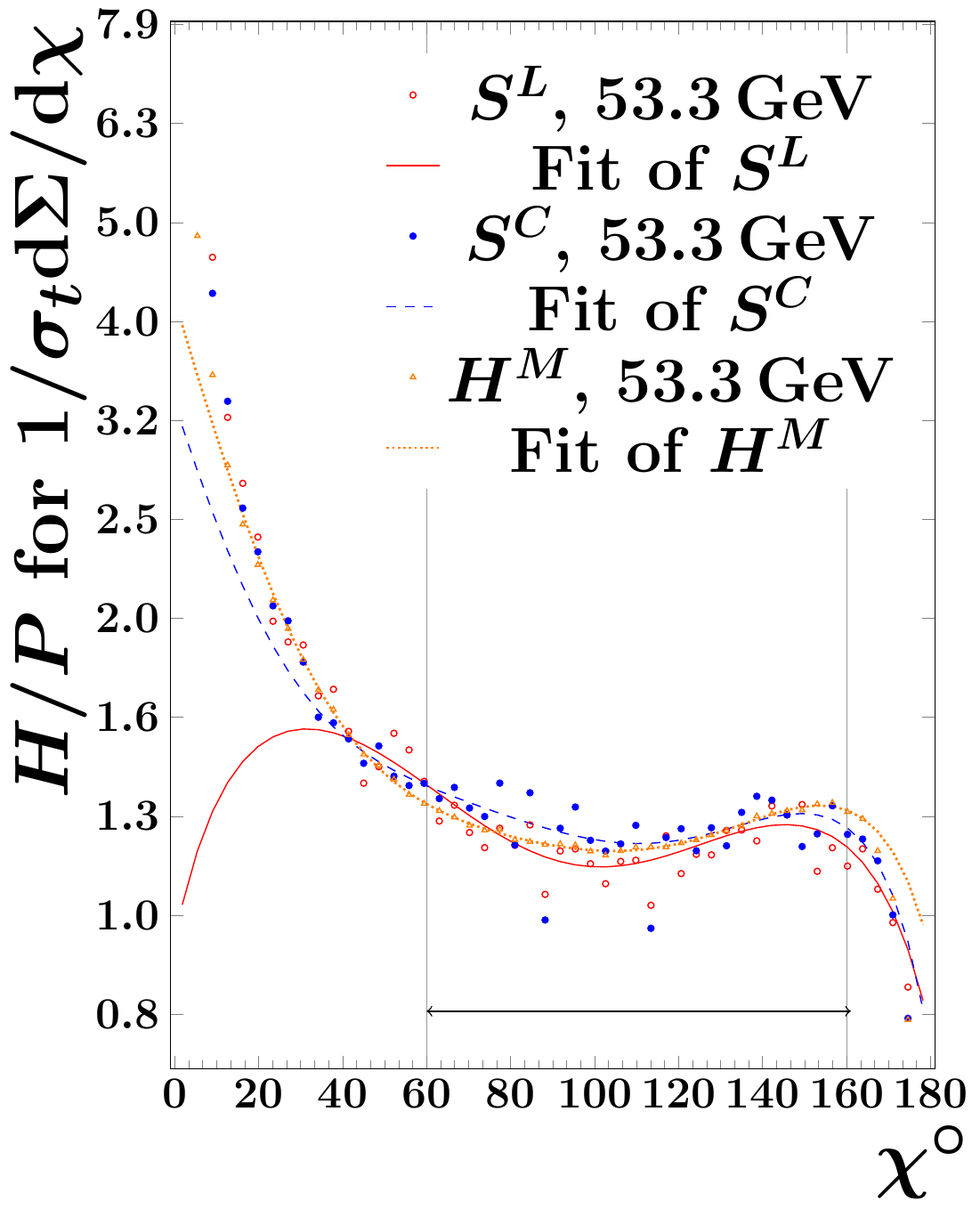}\arxivbreak{}\draftbreak{}\includegraphics[width=\FIGWONE]{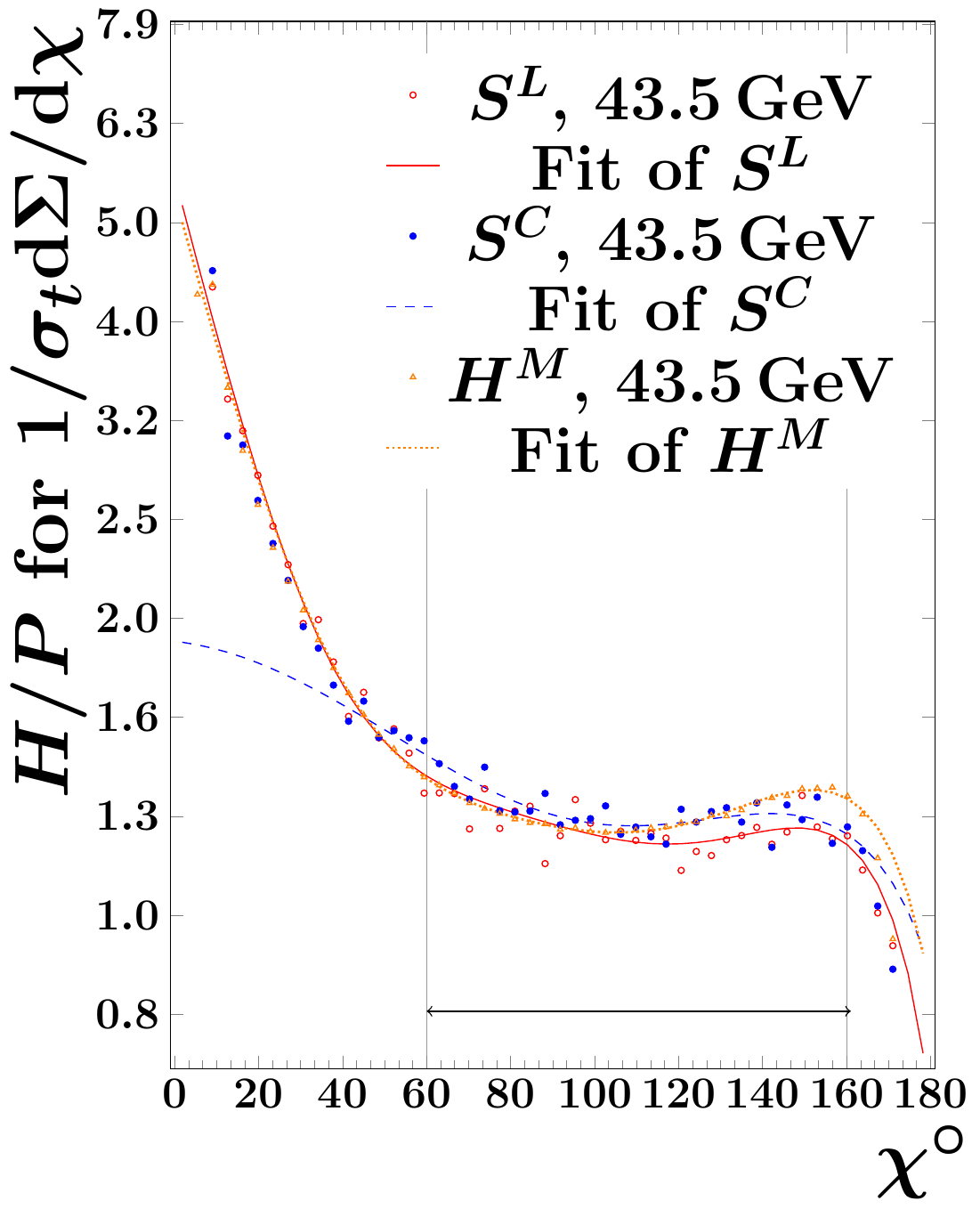}\epjcbreak{}\includegraphics[width=\FIGWONE]{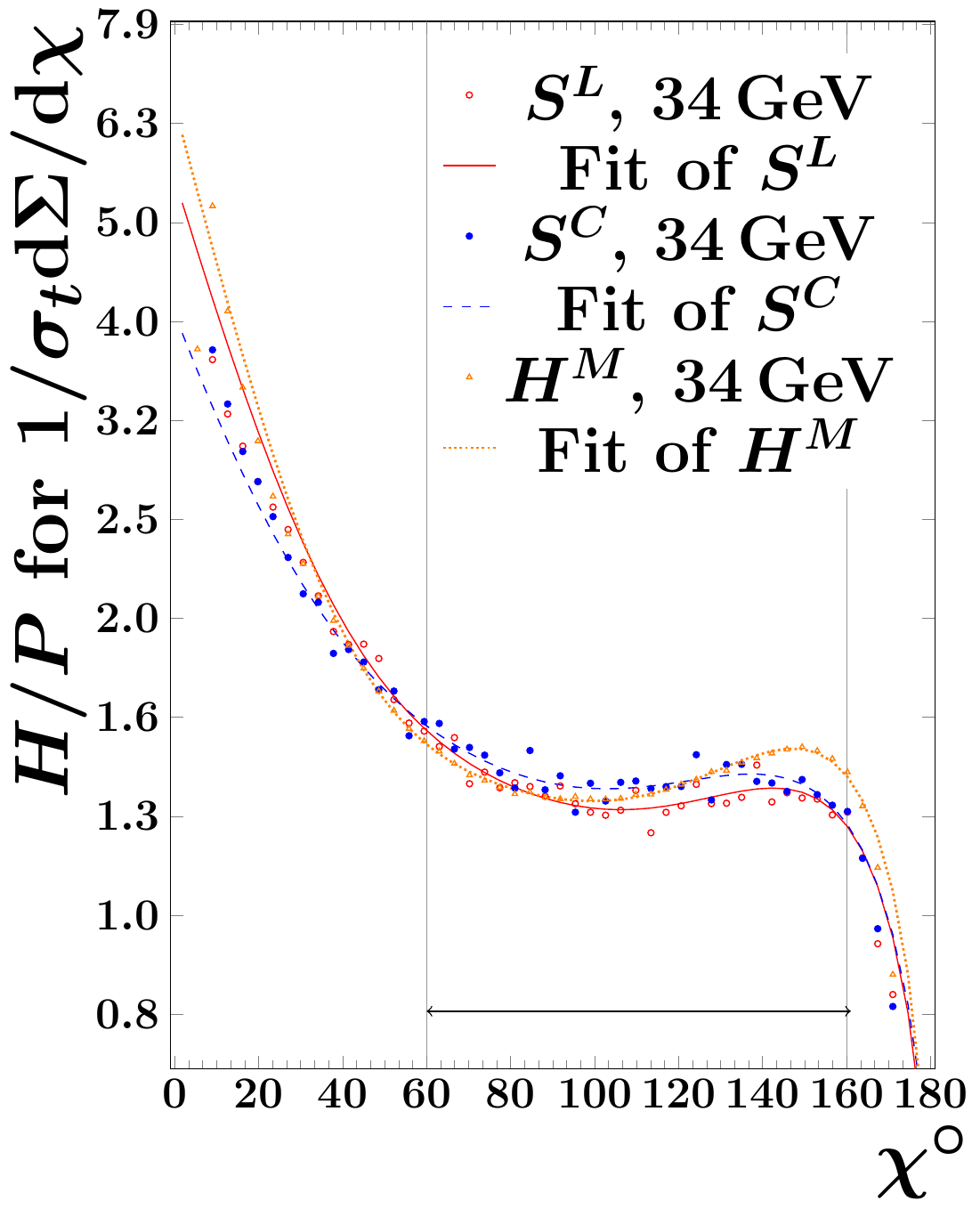}\includegraphics[width=\FIGWONE]{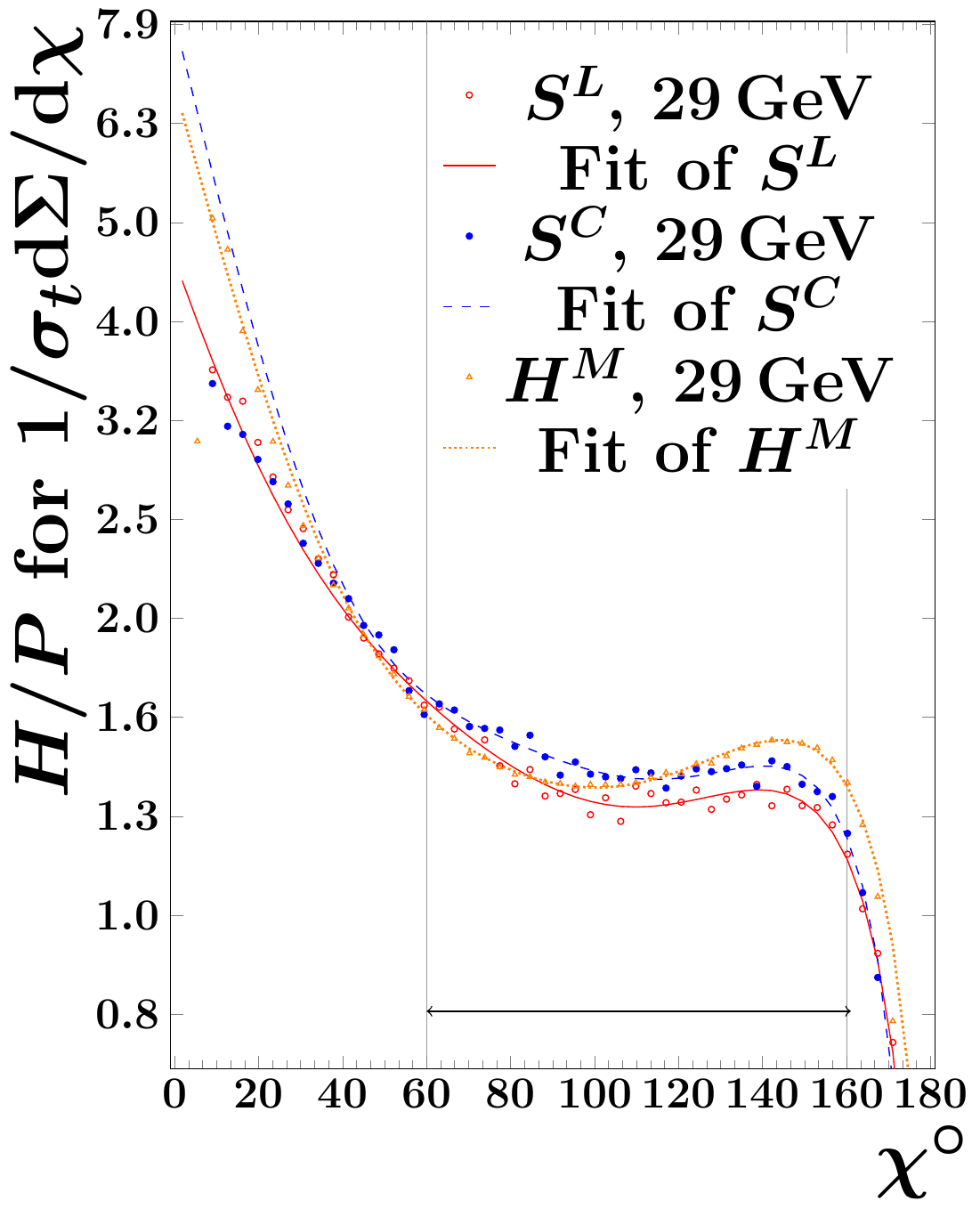}\\
\includegraphics[width=\FIGWONE]{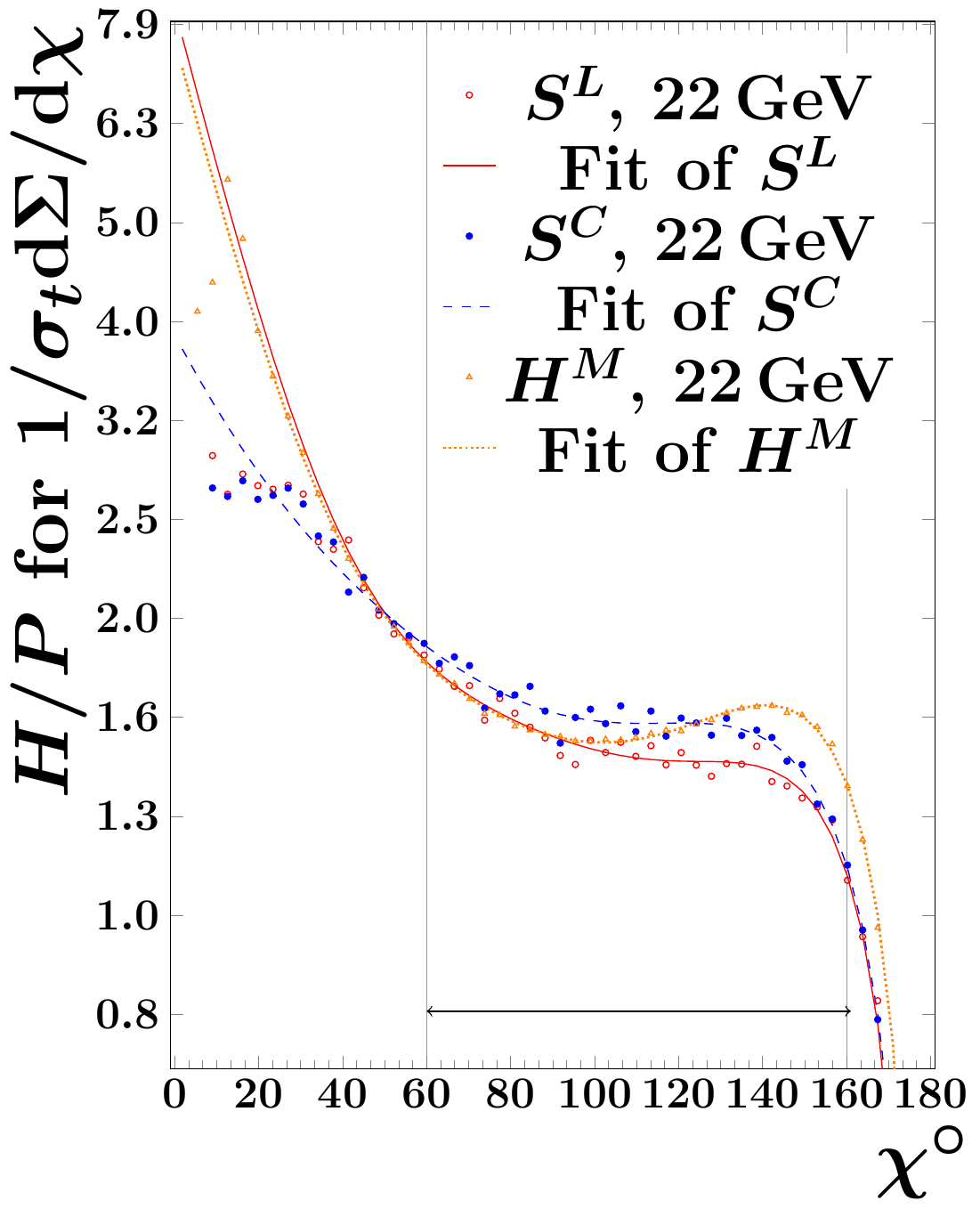}\includegraphics[width=\FIGWONE]{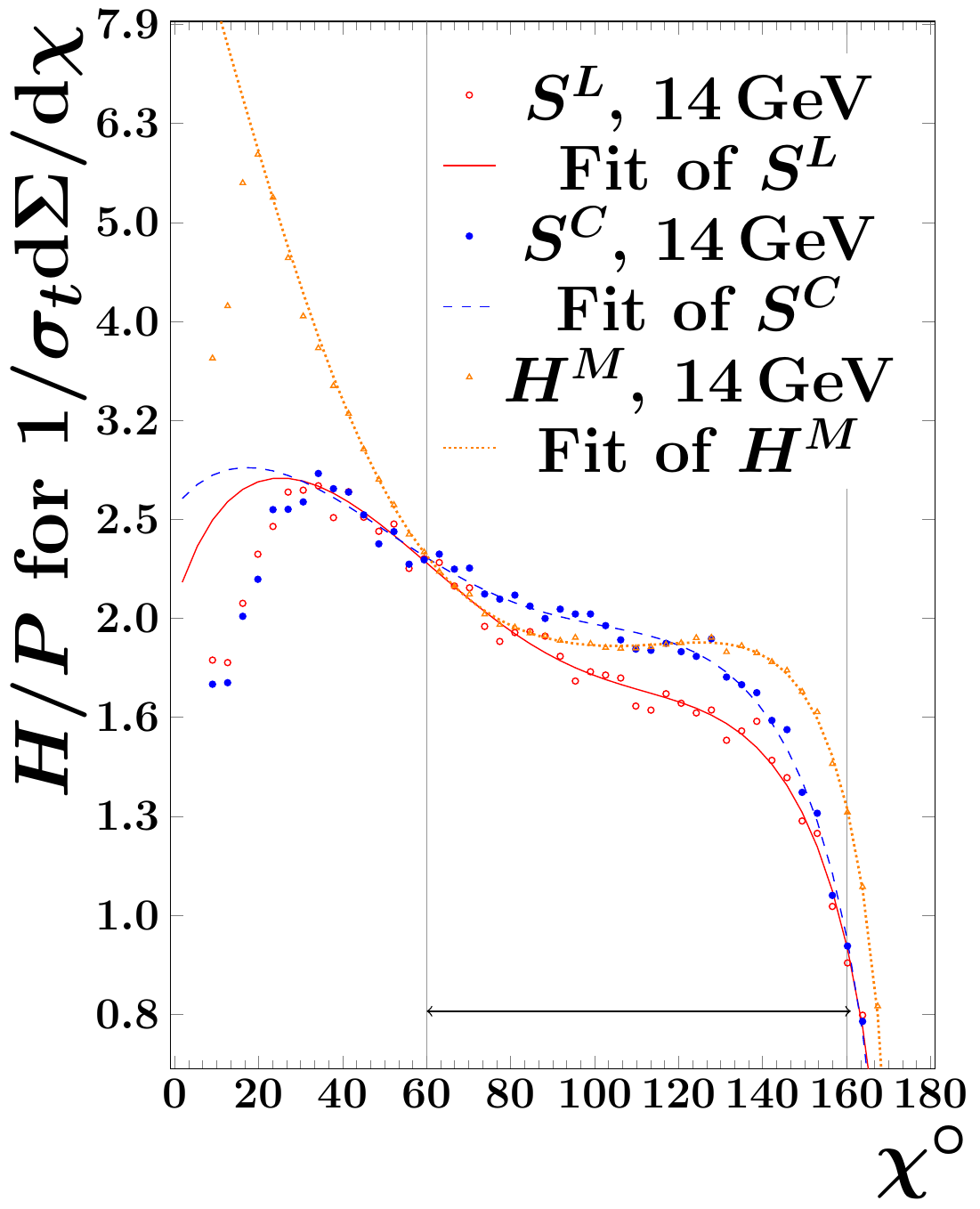}\\
\caption{
Hadronization corrections obtained with different setups of Monte Carlo 
event simulations for every energy point and corresponding parametrizations.
Event-by-event reweighting was applied.
The used fit range is indicated with a thick line. 
}
\label{fig:hadrrew}
\end{figure}

As the {\tt SHERPA2.2.4} setups give the most stable and physically reliable 
predictions these are used in the analysis 
for reference hadronization corrections ($S^{L}$) and for systematic studies ($S^{C}$).
The corresponding hadronization corrections together with parametrizations 
are shown 
in Fig.~\ref{fig:hadrrew} for reweighted samples.

\subsection{Estimation of statistical correlations between measurements from MC models}

To perform an accurate extraction procedure, the available data and uncertainties were examined 
and for every measured set of data a covariance matrix was built. The procedure 
consisted of multiple steps.

In the first step the systematic uncertainties were recalculated and 
separated from statistical uncertainties when this was possible.
For the measurements with the uncertainties  rounded to one
significant digit~\cite{Acton:1993zh,Braunschweig:1987ig,Bartel:1984uc}  
the uncertainties were expanded 
assuming maximal uncertainty before rounding.
The measurements of TASSO~\cite{Braunschweig:1987ig} were converted from
the $d\Sigma/d\cos\chi$ form to $d\Sigma/d\chi$  using values of 
$\cos\chi$ on the bin edges.

For the  data from TOPAZ~\cite{Adachi:1989ej}  the systematic uncertainty 
was calculated   
from an estimated  relative 
systematic uncertainty of $\pm4\%$~\cite{Adachi:1989ej}.

Taking into account the uncertainties of $\as$ extraction 
analysis~\cite{Berger:1985xq} the same was done for PLUTO~\cite{Berger:1985xq} data.
The systematic uncertainty estimation of $\pm5\%$ for TASSO~\cite{Braunschweig:1987ig}
is based on the upper limit of $10\%$ for the total uncertainty mentioned 
in the paper~\cite{Braunschweig:1987ig}.
The systematic uncertainties from 
DELPHI~\cite{Abreu:1992yc} and 
SLD~\cite{Abe:1994mf}   were used as provided in the original papers.

For all remaining data sets  the published 
combined uncertainty was treated as statistical.

\begin{figure}\centering
\includegraphics[width=0.32\linewidth]{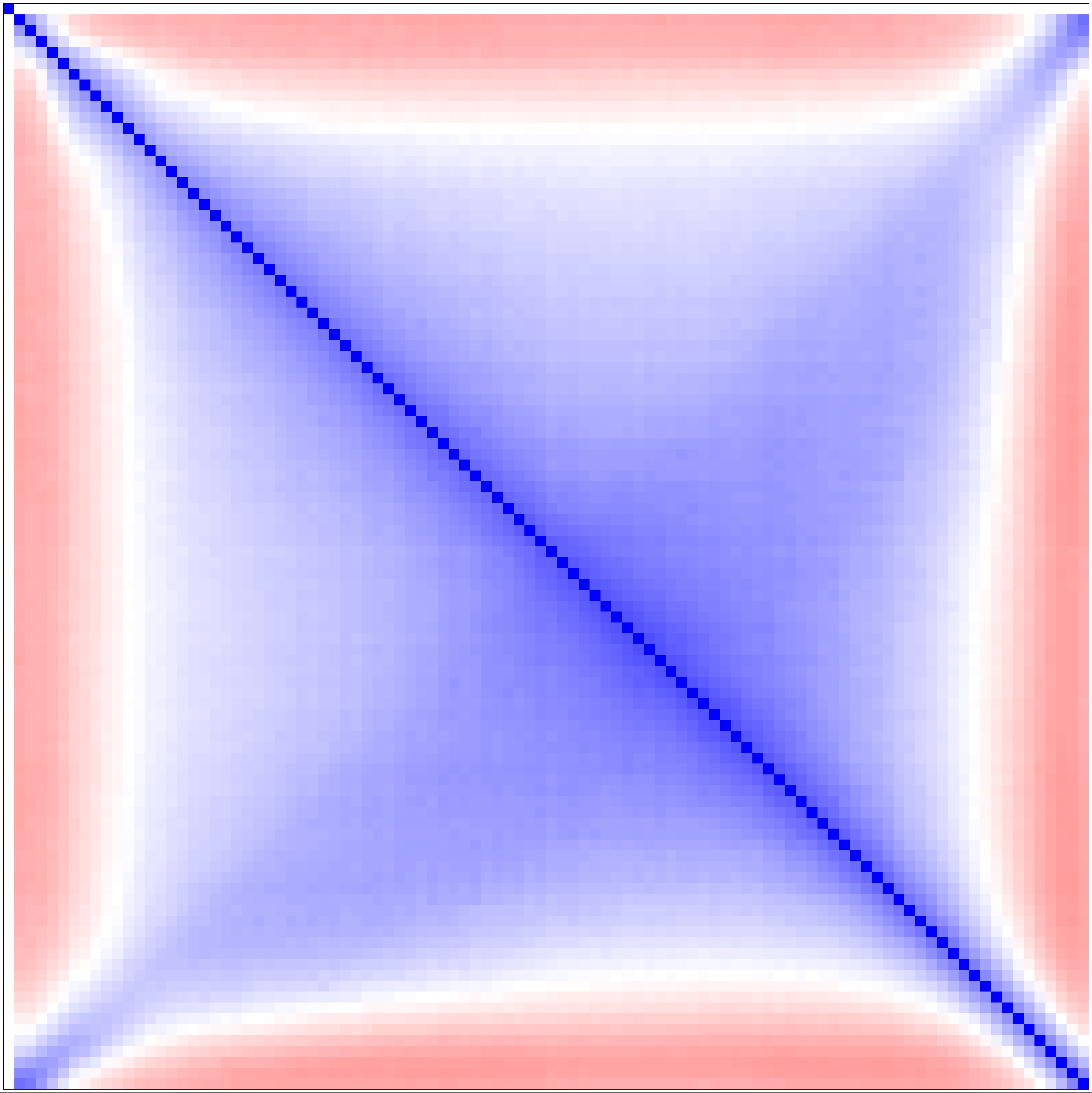}\includegraphics[width=0.32\linewidth]{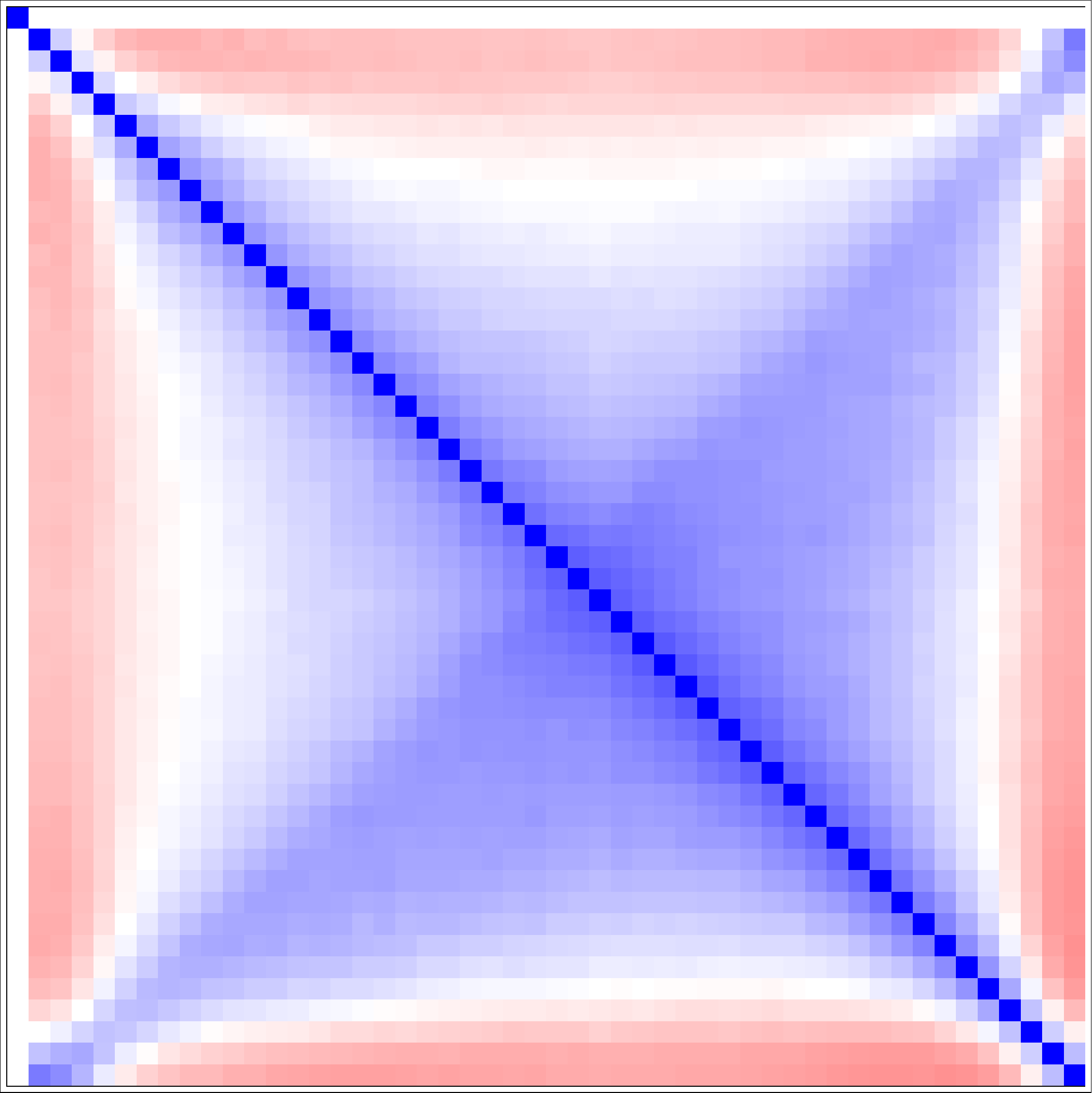}\includegraphics[width=0.32\linewidth]{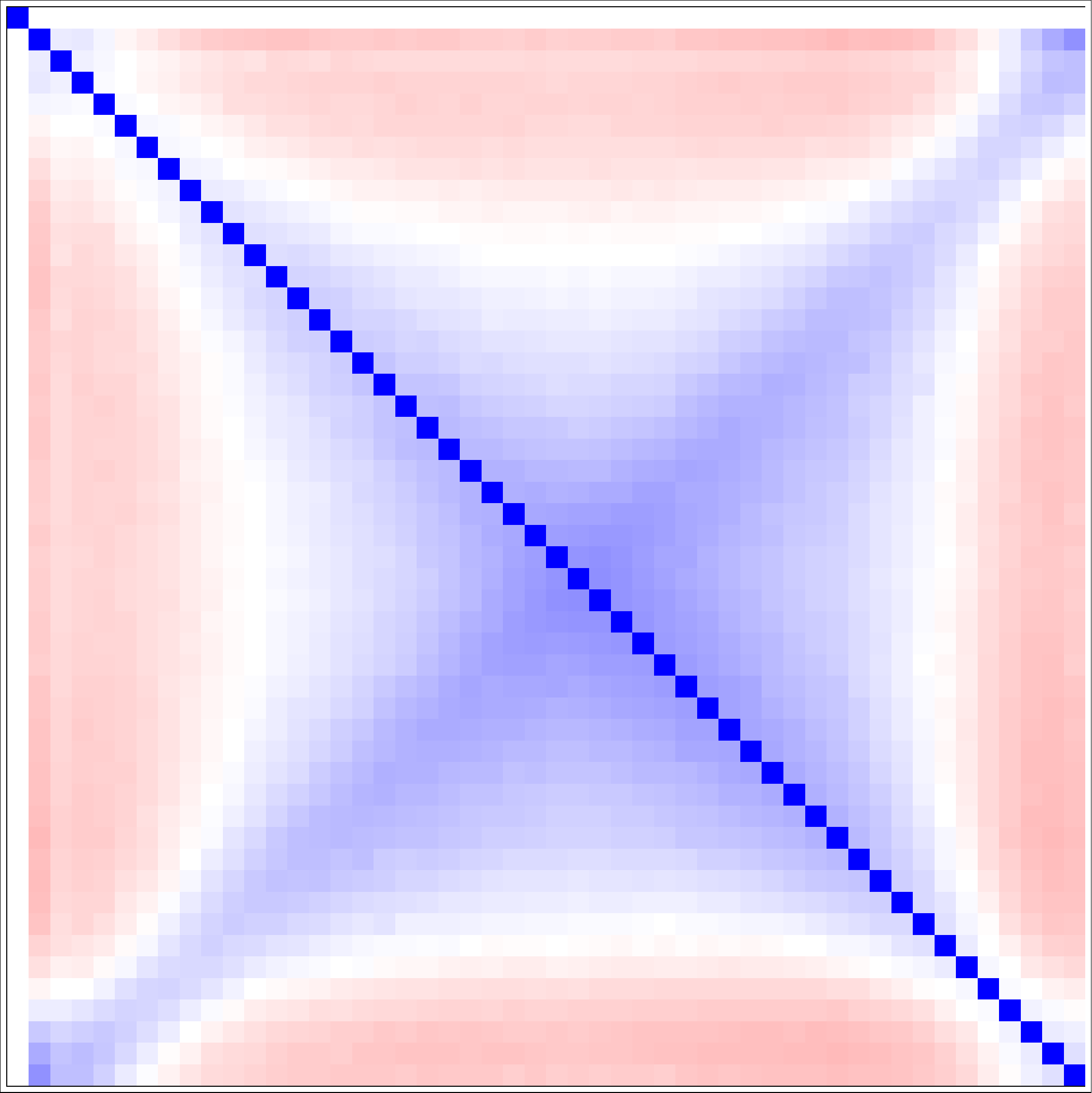}\\
\includegraphics[width=0.96\linewidth]{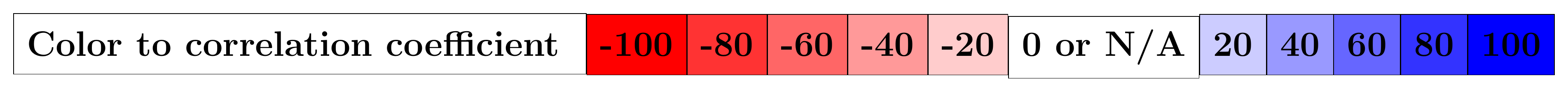}\\
\caption{A graphical representation of statistical correlation matrices for $\Sigma(\chi)$ obtained 
from Monte Carlo simulated events with the $S^{L}$ setup for 
 OPAL~\protect\cite{Acton:1991cu} data at $\sqrt{s}=91.2\GeV$ (left),
 TOPAZ~\protect\cite{Adachi:1989ej} data at $\sqrt{s}=53.3\GeV$ (centre),
 and JADE~\protect\cite{Bartel:1984uc} data at $\sqrt{s}=22\GeV$ (right). 
 The bottom left corner for each figure corresponds to $\chi=0\degree$ and the 
 bottom right to $\chi=180\degree$.}
\label{fig:corr}
\end{figure}

The measurements of $\Sigma$ are provided in the original publications
  without correlations
between the individual points. 
The correlation matrix was estimated from the Monte Carlo samples 
in terms of Fisher correlation coefficients~\cite{Fisher:1915,Fisher:1921}.
Some of the obtained correlation matrices are shown in Fig.~\ref{fig:corr}.
The obtained correlation coefficients are sizeable, up to $0.5$ for the 
closest points, which highlights the importance of properly taking into account 
the correlations between measured points in the fits.  
The obtained correlation matrix together with statistical uncertainties 
was used to build a statistical covariance matrix for every data set.

To construct the systematic covariance matrix, the 
 systematic uncertainties from the original publications were used 
 with an assumption that these are 
 positively correlated with correlation coefficient $\rho=0.5$ between closest points. 
The correlations between the uncertainties of data from different experiments or 
different beam energies were neglected.
The final covariance matrix used in the fit for every data set was a sum 
of statistical and systematic covariance matrices.

\subsection{Fit procedure and estimation of uncertainties}
\label{sec:fit}
The strong coupling extraction procedure is based on the comparison of data 
to the perturbative QCD prediction combined with non-perturbative (hadronization) 
corrections. The perturbative part of the predictions was calculated in 
every bin as described in previous sections.
To tame the statistical fluctuations present in the obtained binned 
hadronization correction distributions, these were parametrized with 
analytic functions, expressed as a sum of polynomials of $\chi-90\degree$. 
The value of the fitted function at the bin centre was used as the correction factor.

To find the optimal value of $\as$, the {\tt MINUIT2}~\cite{minuit,James:2004xla} 
program was used to minimize 
$$\chi^2(\as)=\sum_{\rm data\ sets}\chi^2(\as)_{\rm data\ set},$$ 
where the $\chi^2(\as)$ value  was calculated  for each data set  as
\begin{equation*}
\chi^2(\as)=(\vec{D}-\vec{P}(\as))V^{-1}(\vec{D}-\vec{P}(\as))^{T},
\end{equation*}
with $\vec{D}$ standing for the vector of data points, $\vec{P}(\as)$ for the 
vector of calculated predictions and $V$ for the covariance matrix for $\vec{D}$.
The default scale used in the fit procedure was $\mu=Q=\sqrt{s}$.

\begin{figure}[htbp]\centering
\includegraphics[width=\FIGWONE]{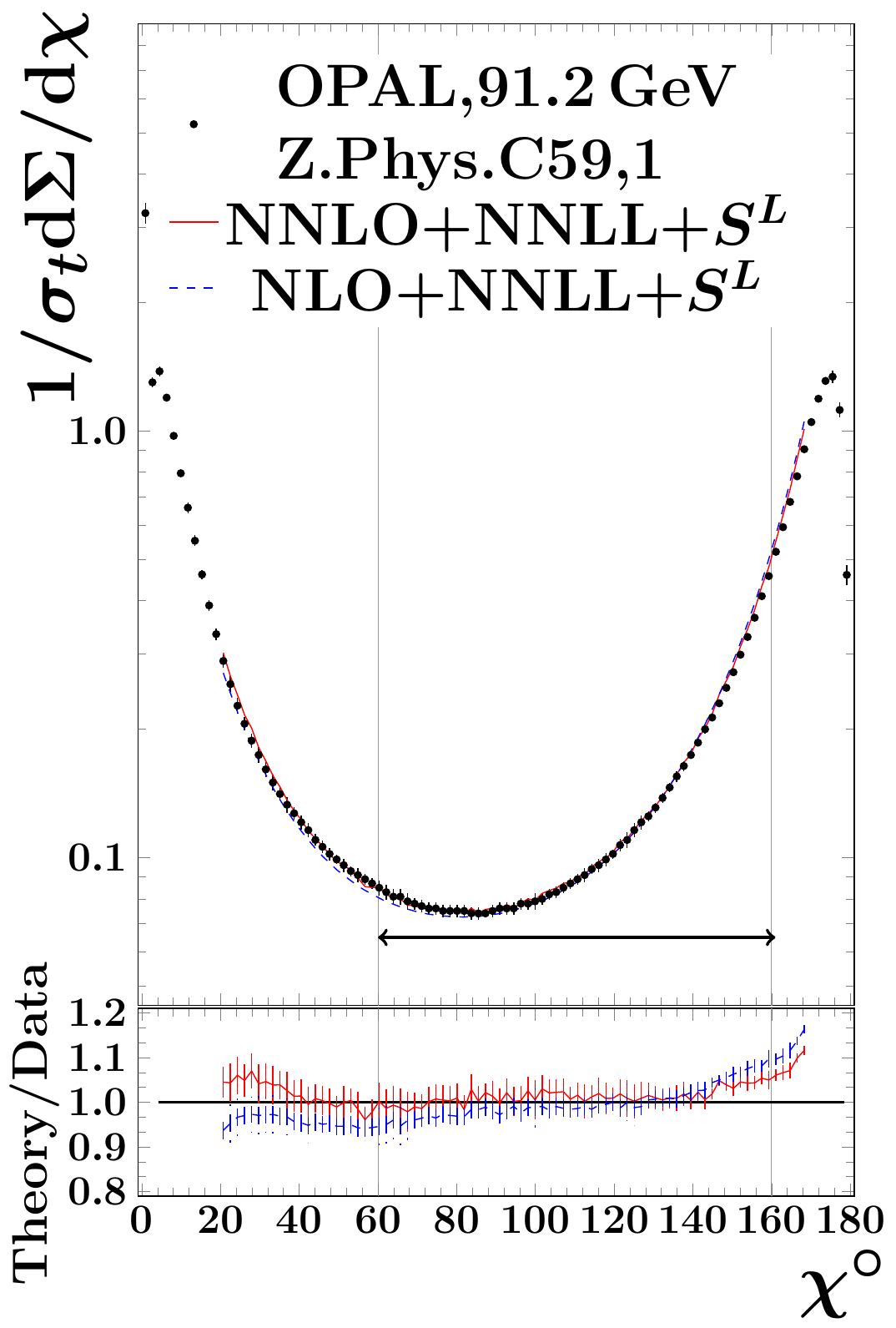}\includegraphics[width=\FIGWONE]{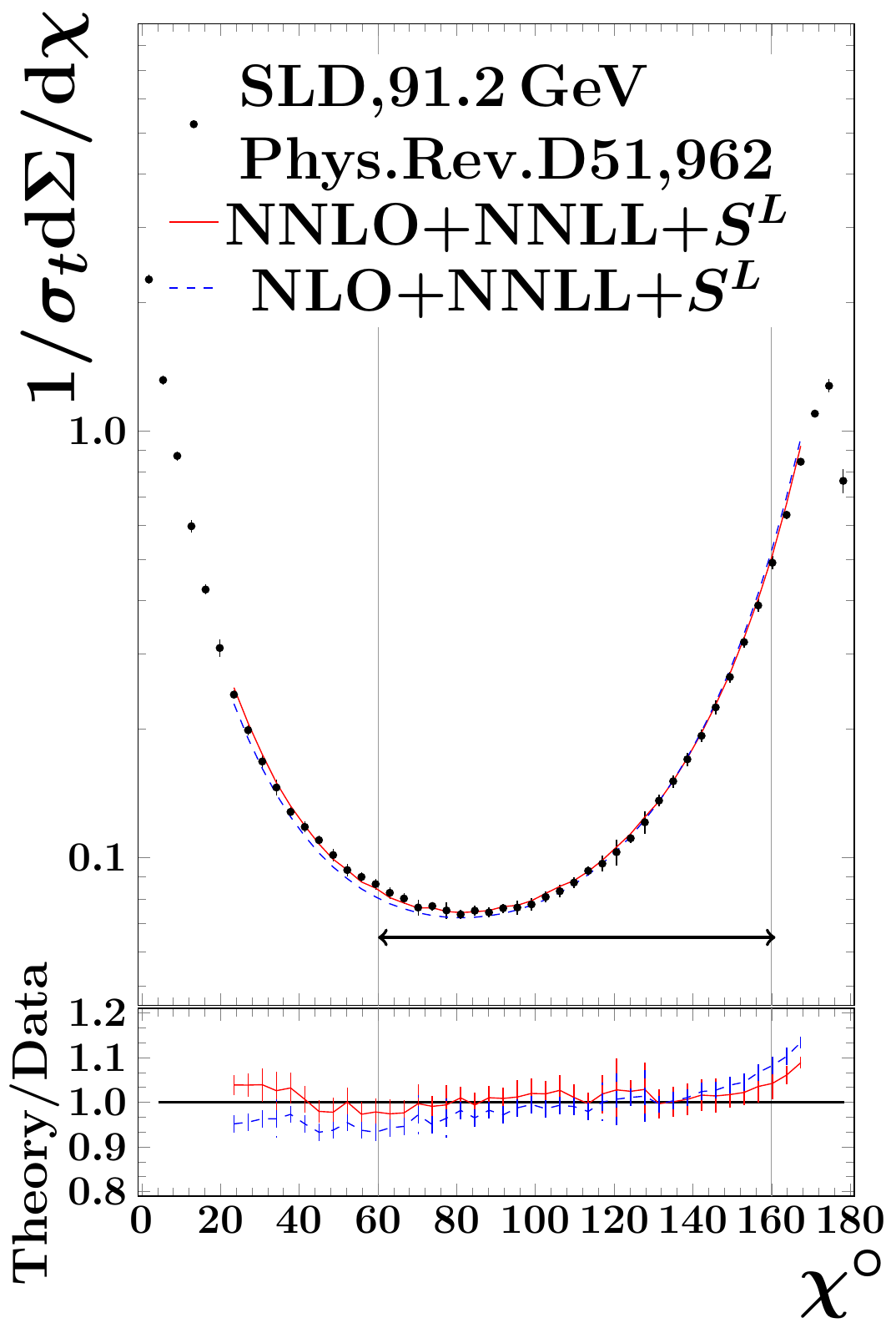}\epjcbreak{}\includegraphics[width=\FIGWONE]{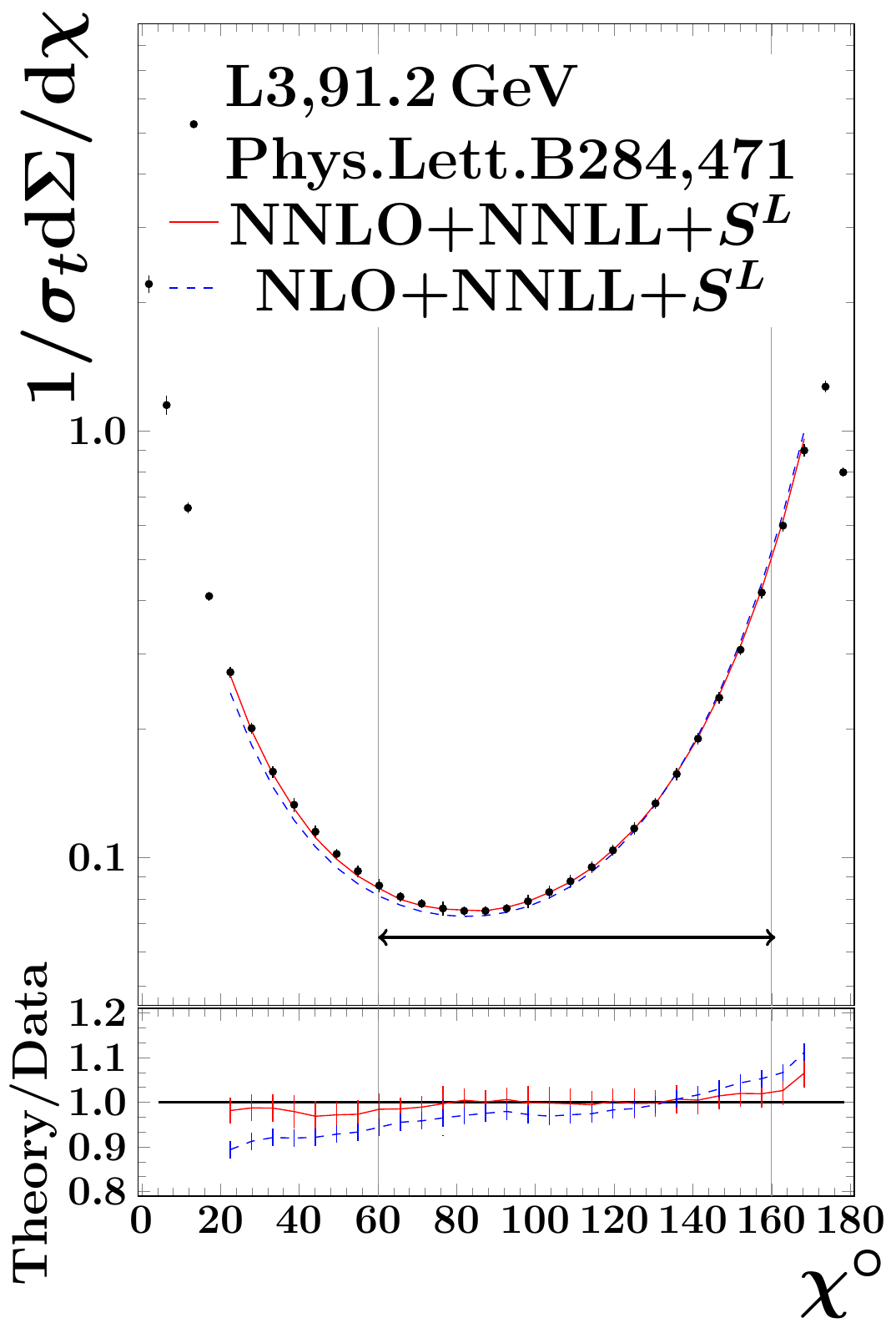}\arxivbreak{}\draftbreak{}\includegraphics[width=\FIGWONE]{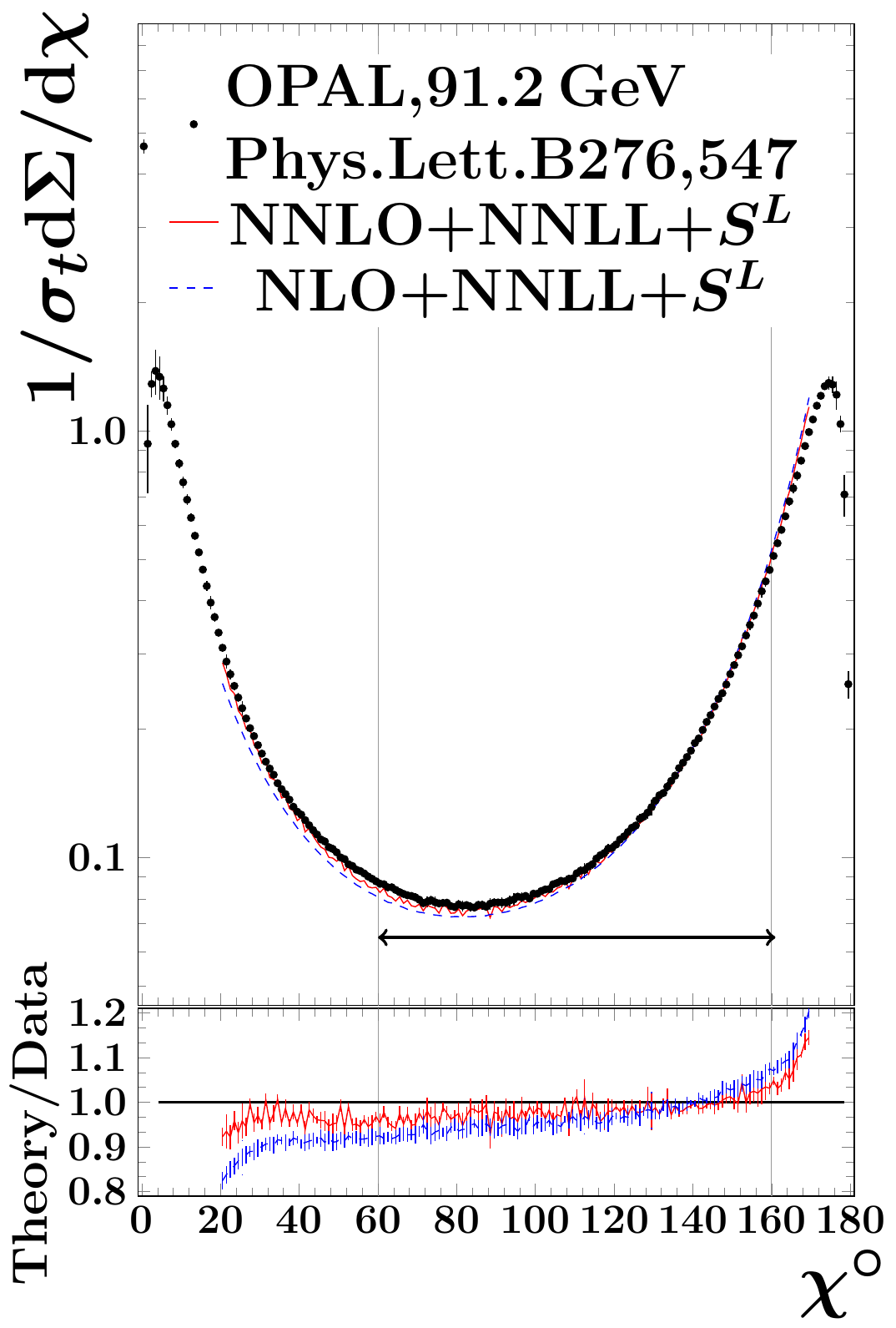}\epjcbreak{}\includegraphics[width=\FIGWONE]{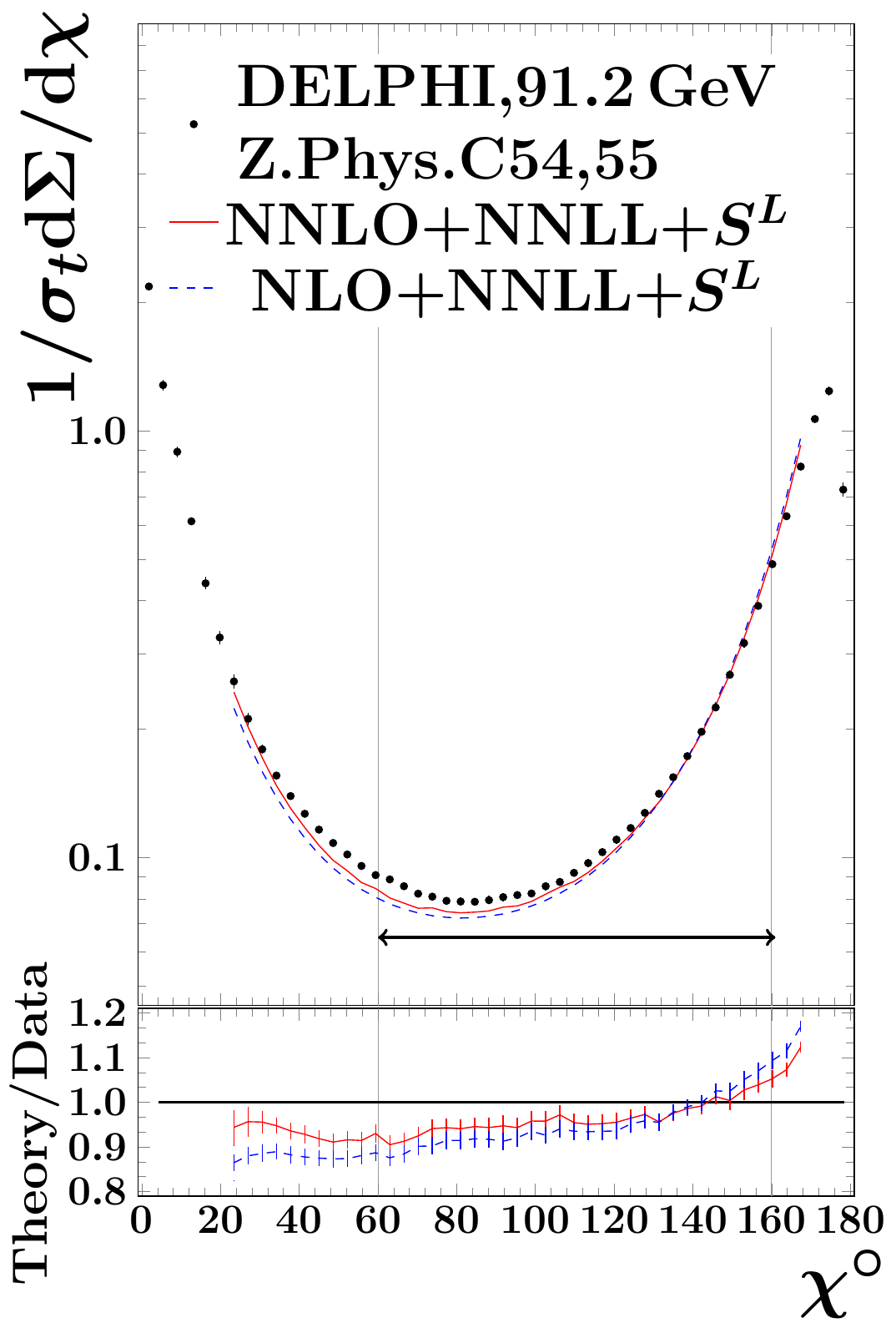}\includegraphics[width=\FIGWONE]{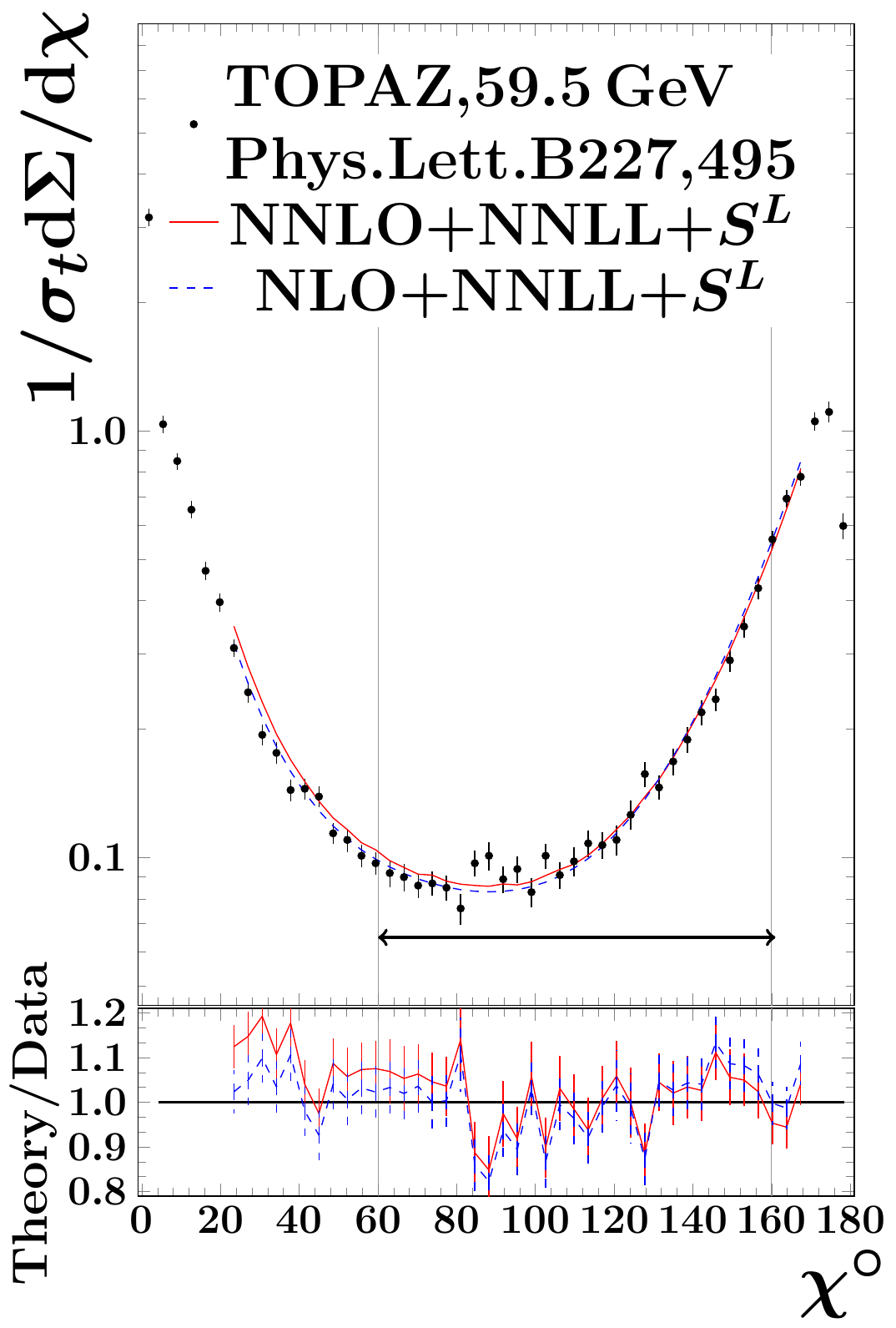}\\
\caption{
Fits of theory predictions to the data at $\sqrt{s}=59.5-91.2\GeV$. 
The used fit range is shown with thick line. For the ratio plot
 only the uncertainties of the data are taken into account.
}
\label{fig:result:nineone}
\end{figure}

\begin{figure}[htbp]\centering
\includegraphics[width=\FIGWONE]{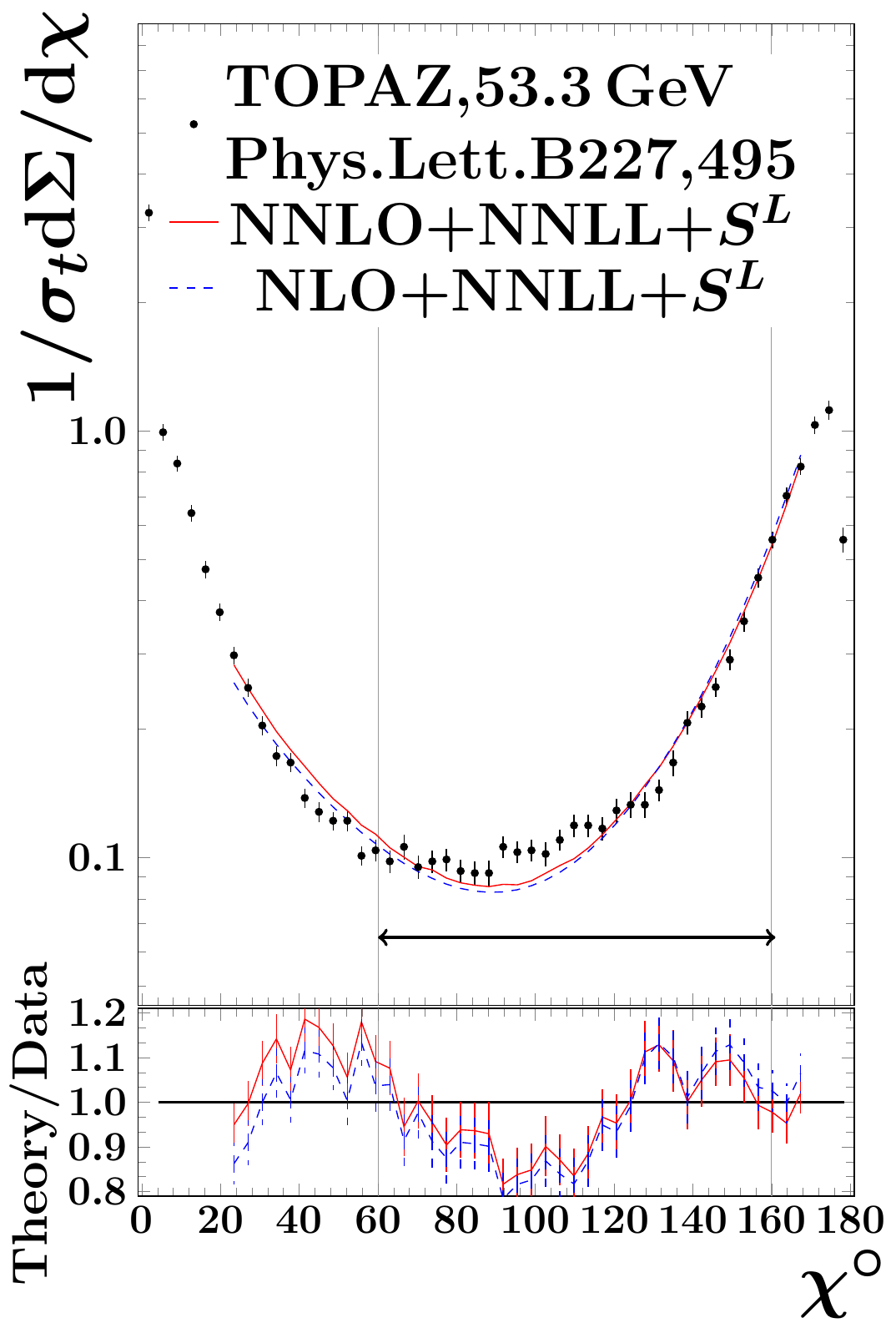}\includegraphics[width=\FIGWONE]{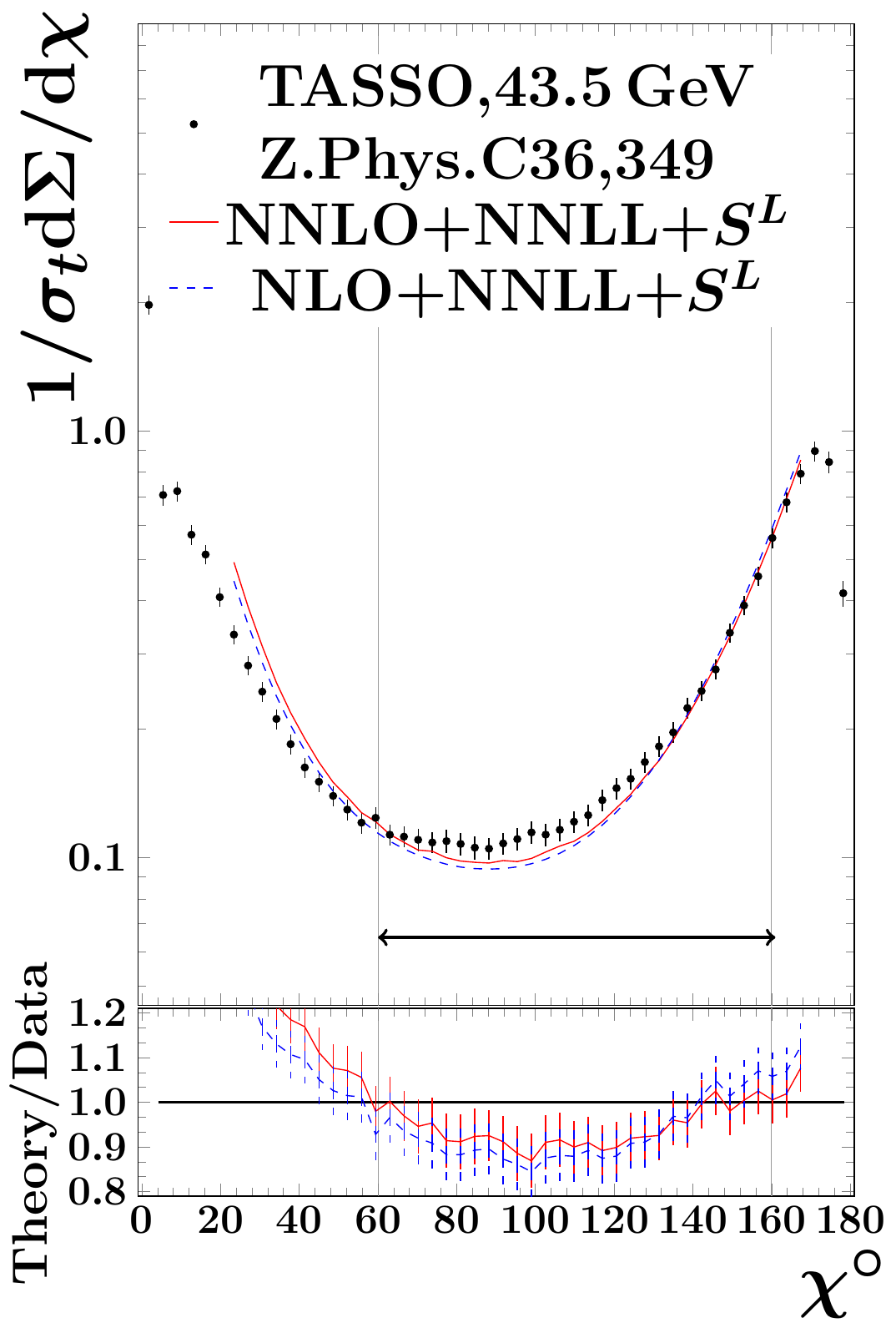}\epjcbreak{}\includegraphics[width=\FIGWONE]{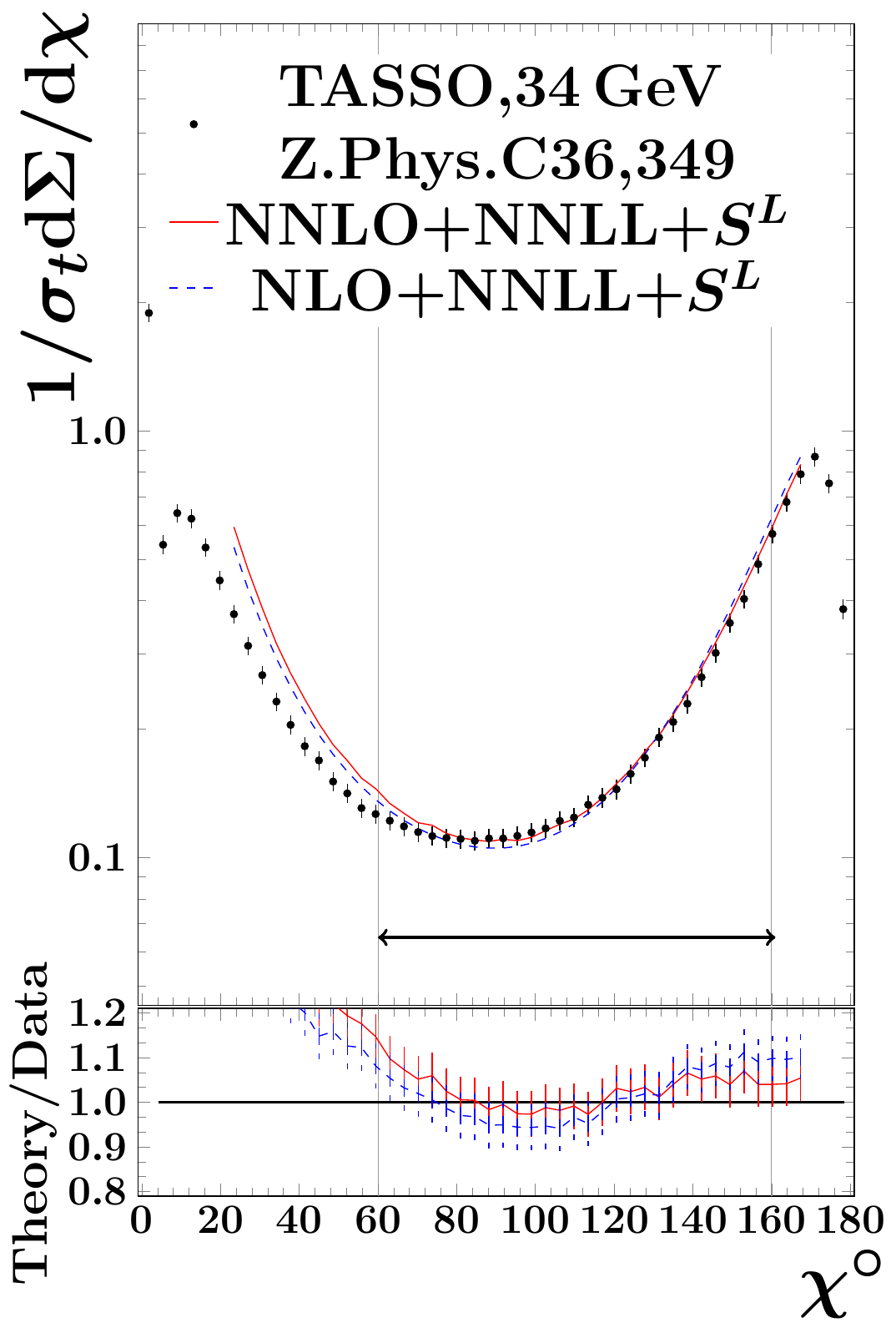}\arxivbreak{}\draftbreak{}\includegraphics[width=\FIGWONE]{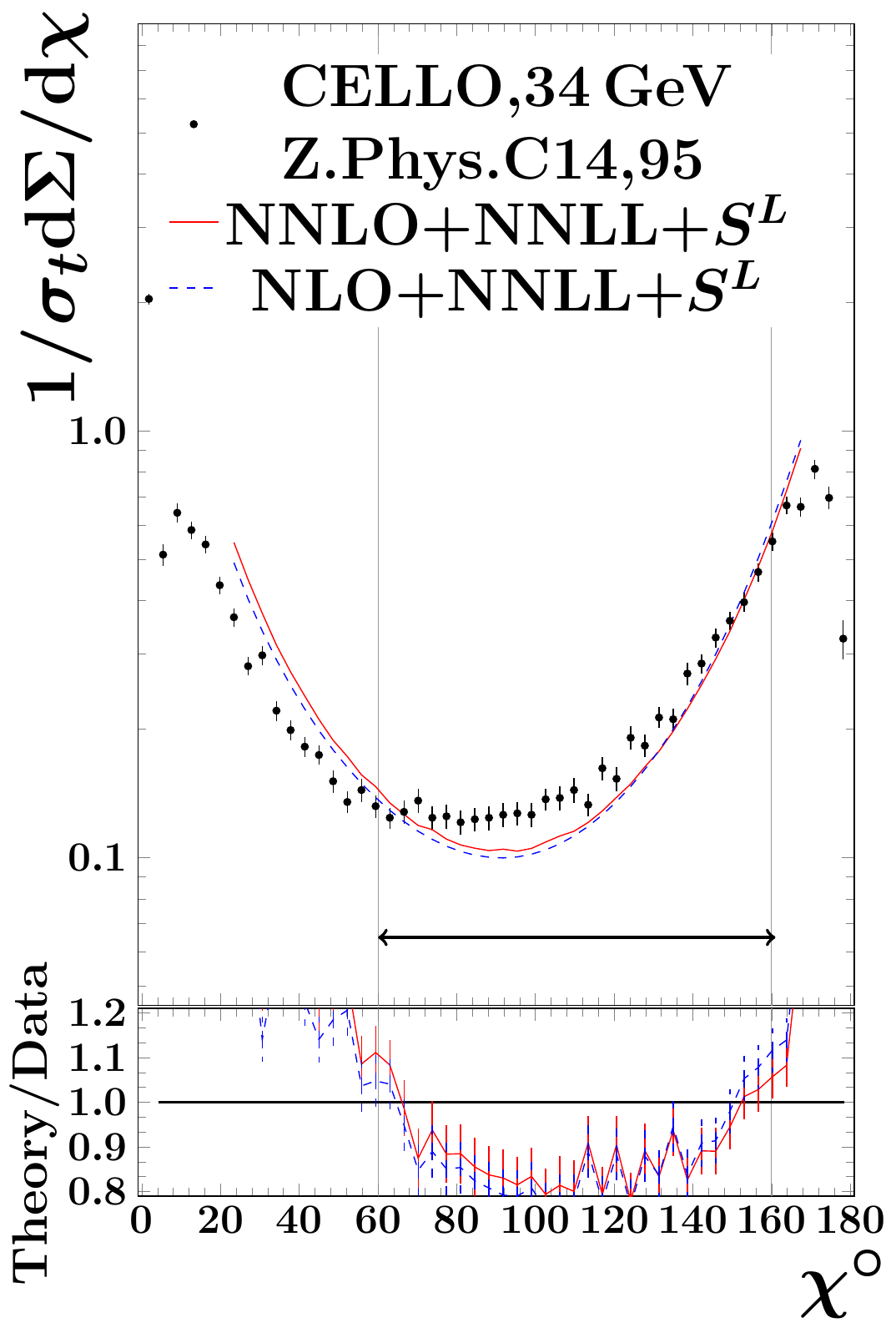}\epjcbreak{}\includegraphics[width=\FIGWONE]{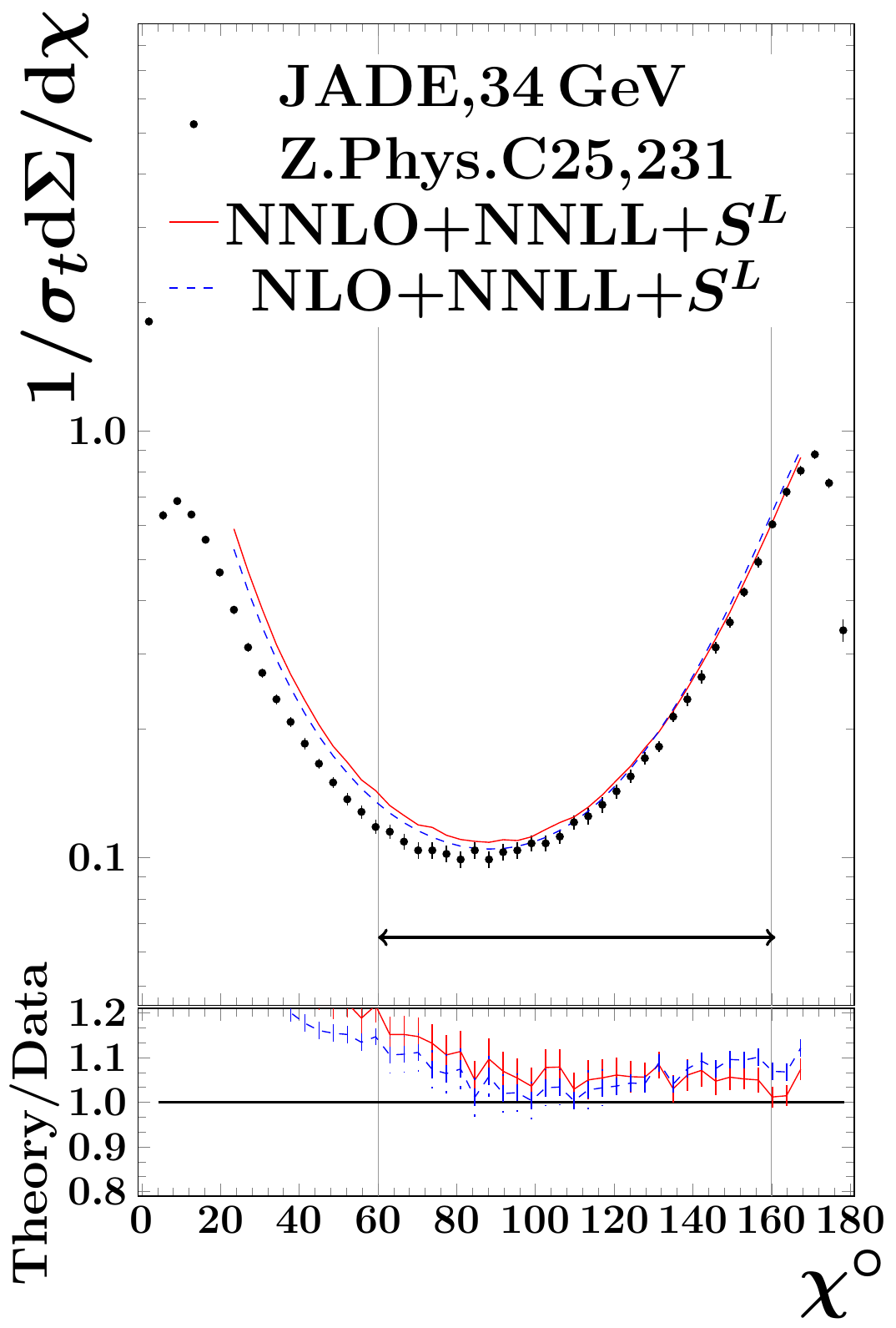}\includegraphics[width=\FIGWONE]{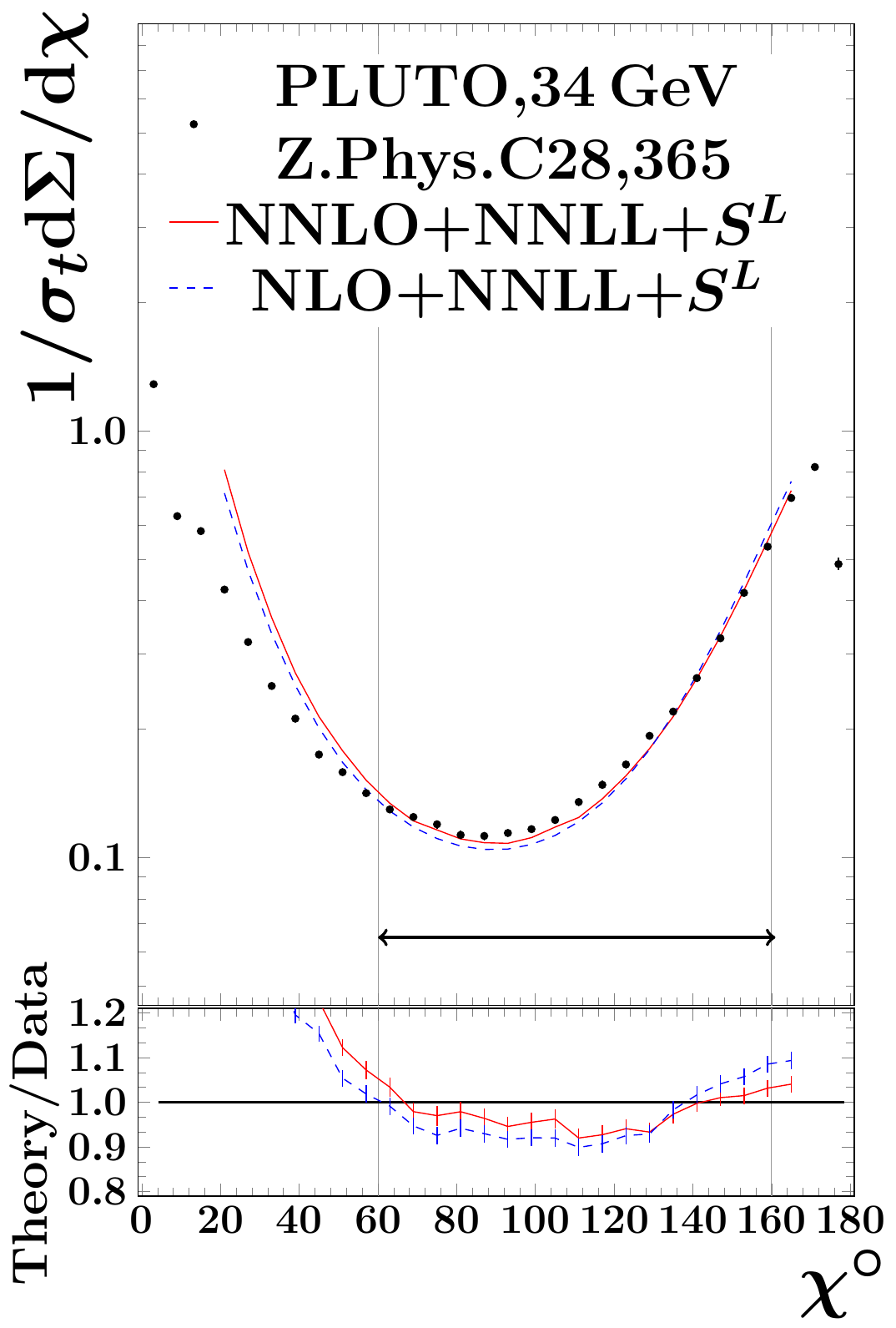}\\
\caption{
Fits of theory predictions to the data for $\sqrt{s}=34-53.3\GeV$. 
The used fit range is shown with thick line. For the ratio plot 
only the uncertainties of the data are taken into account.
}
\label{fig:result:threefour}
\end{figure}

\begin{figure}[htbp]\centering
\includegraphics[width=\FIGWONE]{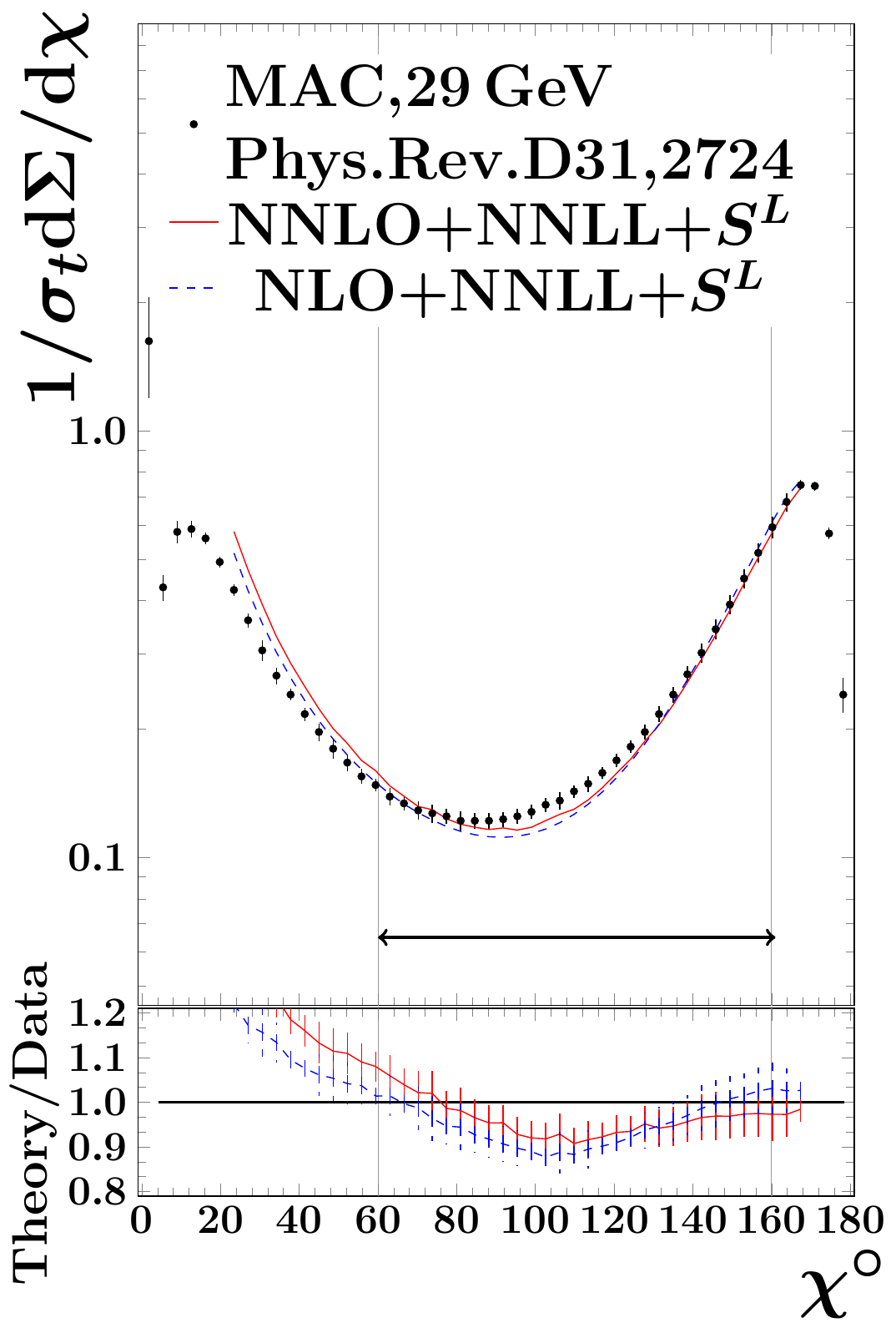}\includegraphics[width=\FIGWONE]{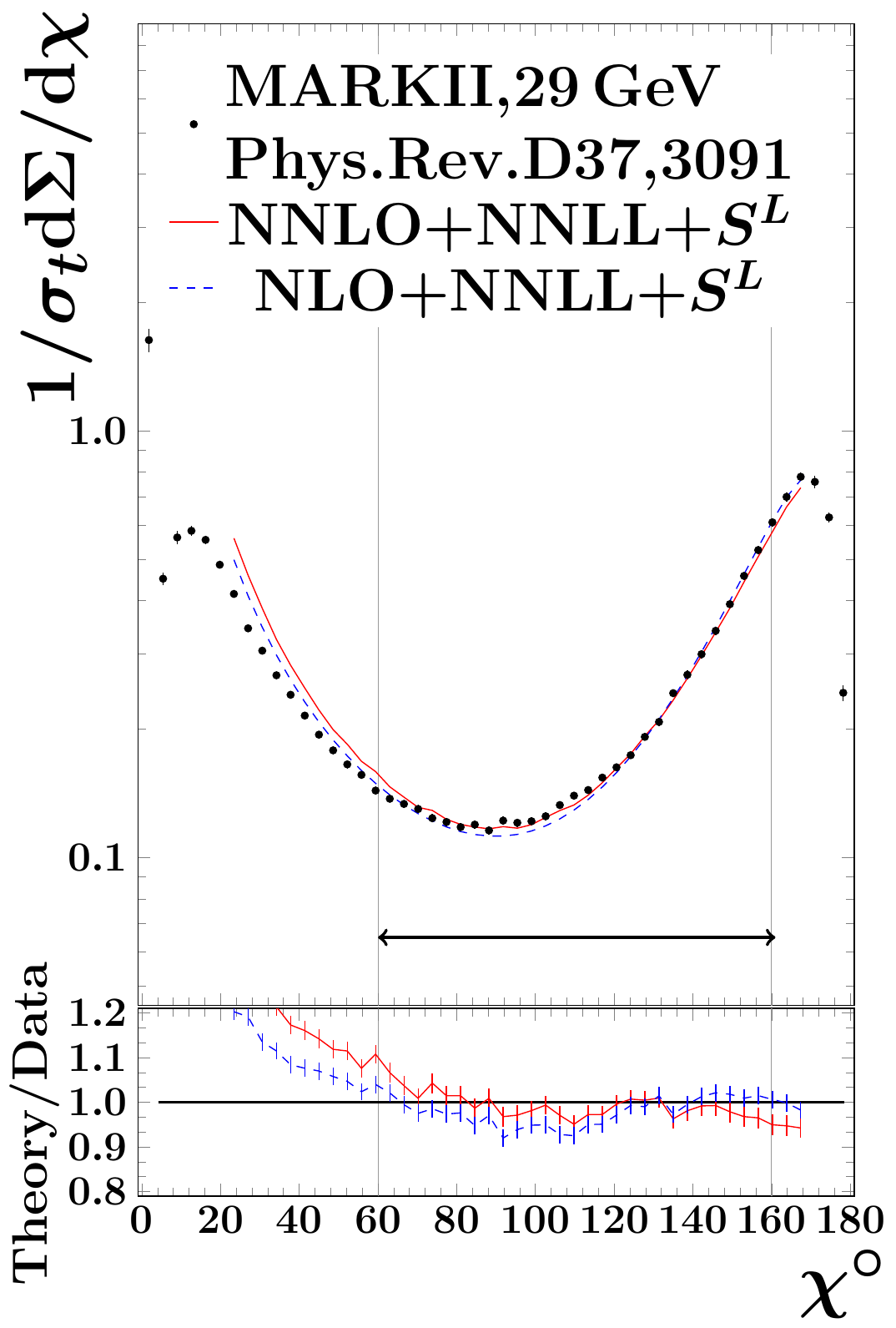}\epjcbreak{}\includegraphics[width=\FIGWONE]{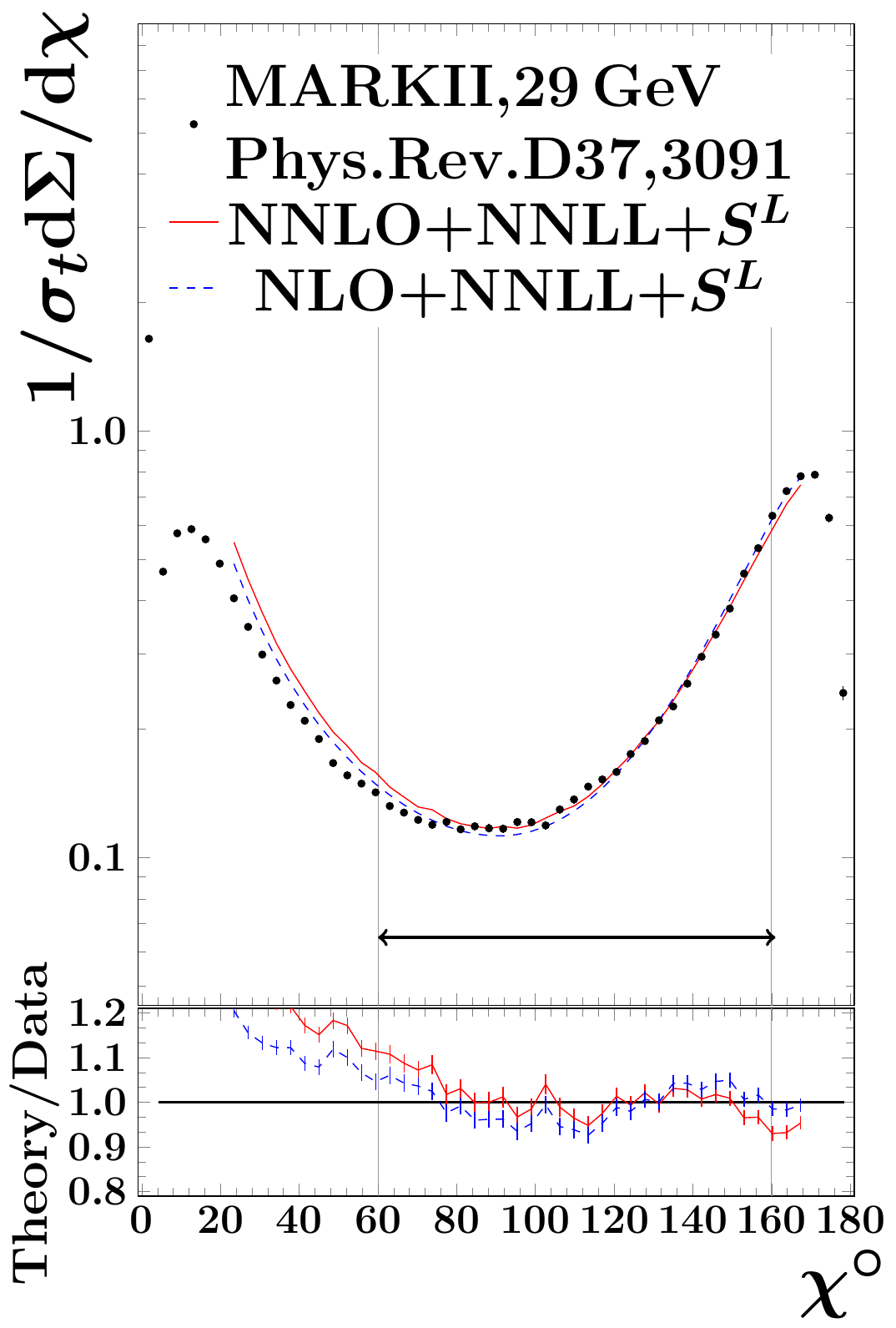}\arxivbreak{}\draftbreak{}\includegraphics[width=\FIGWONE]{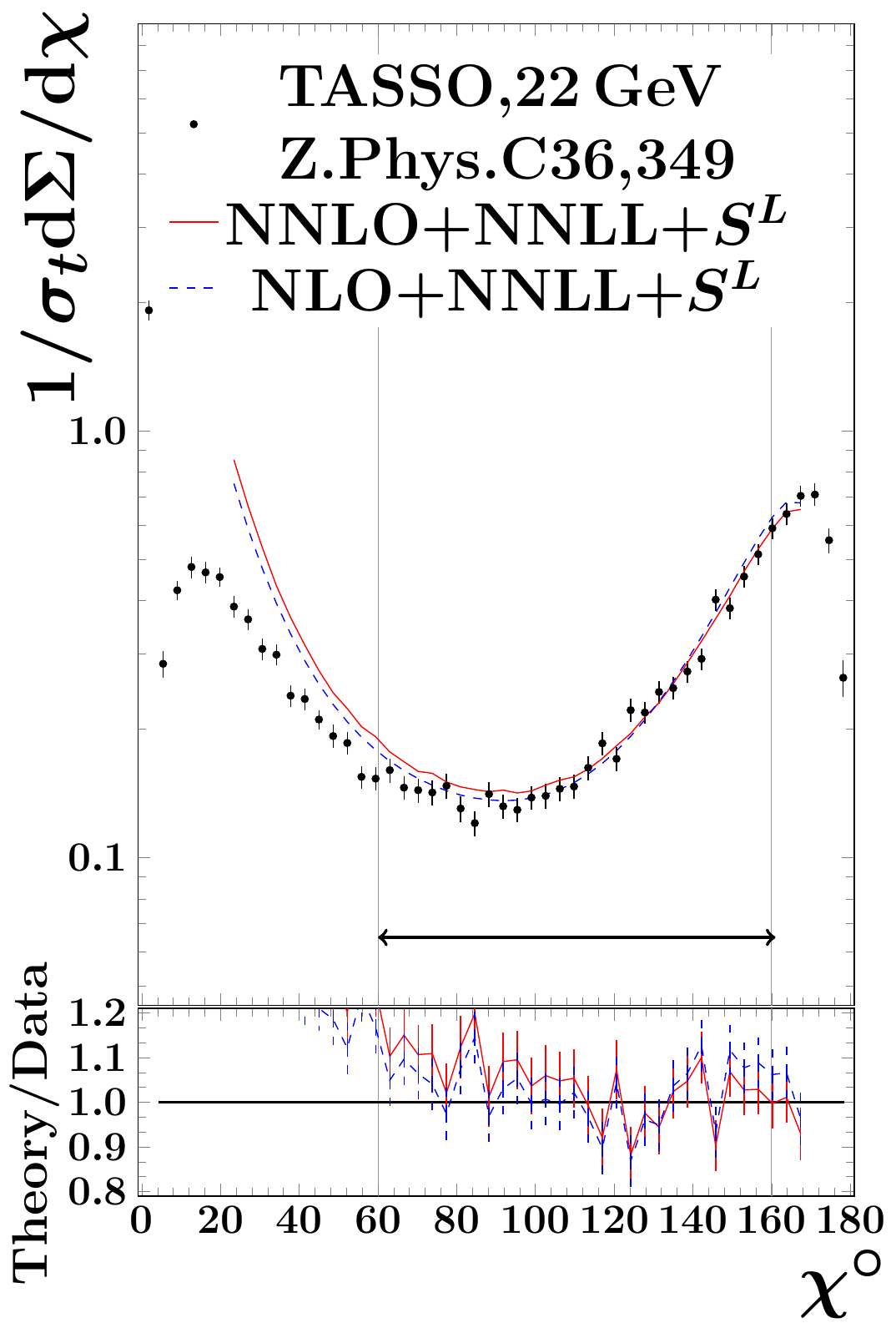}\epjcbreak{}\includegraphics[width=\FIGWONE]{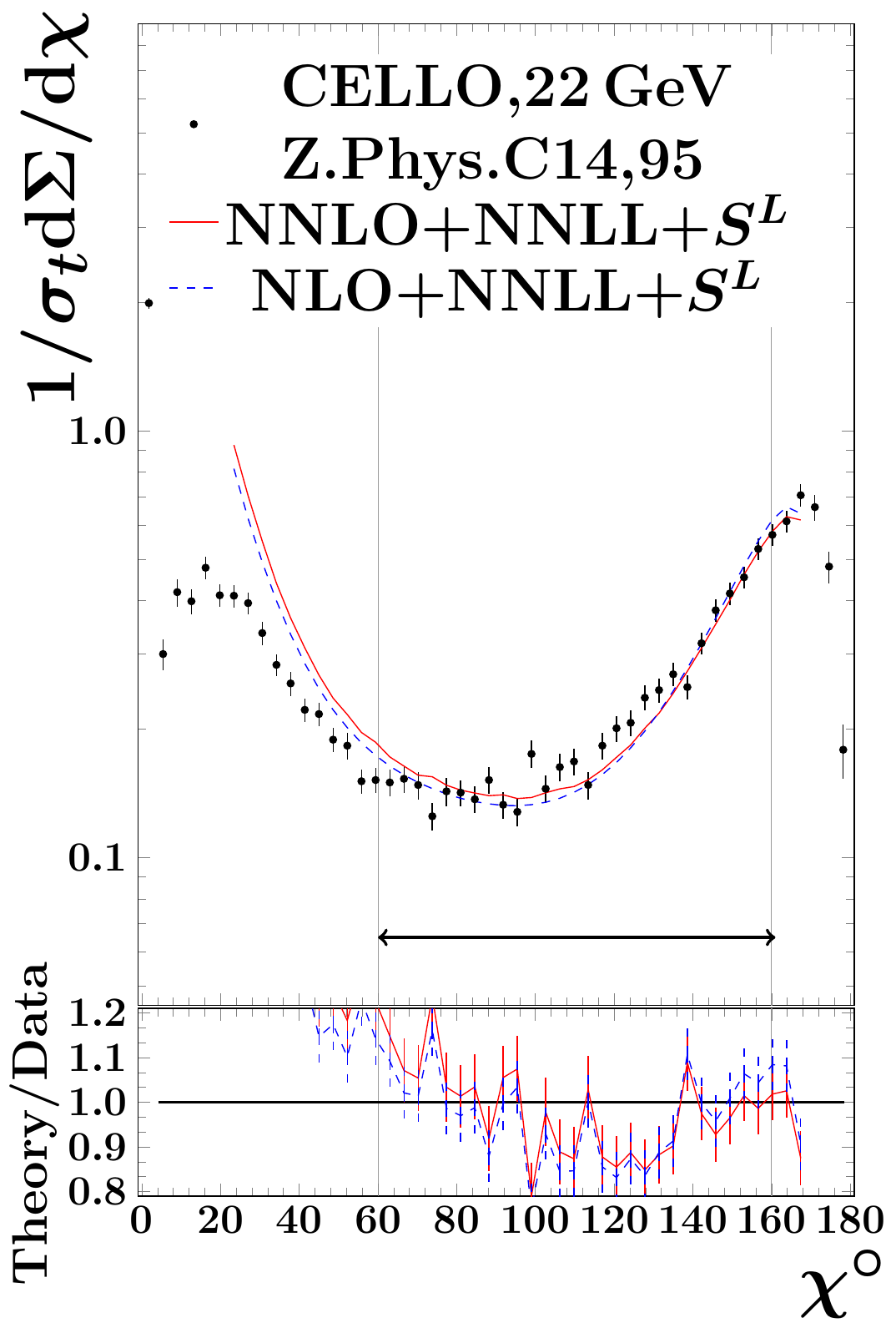}\includegraphics[width=\FIGWONE]{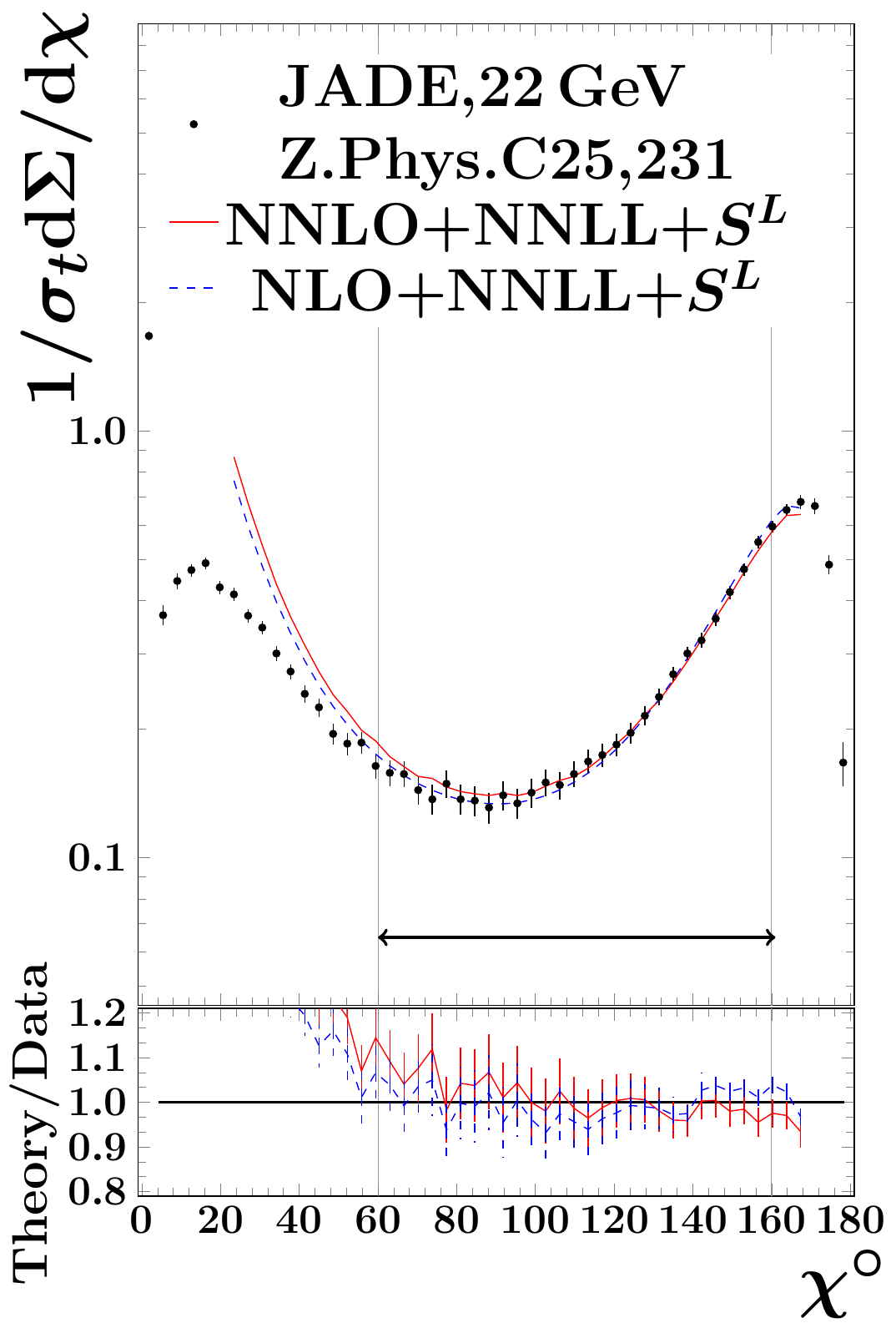}\\
\caption{
Fits of theory predictions to the data for $\sqrt{s}=22-29\GeV$. 
The used fit range is shown with thick line. For the ratio plot 
only the uncertainties of the data are taken into account.
}
\label{fig:result:twonine}
\end{figure}

\begin{figure}[htbp]\centering
\includegraphics[width=\FIGWONE]{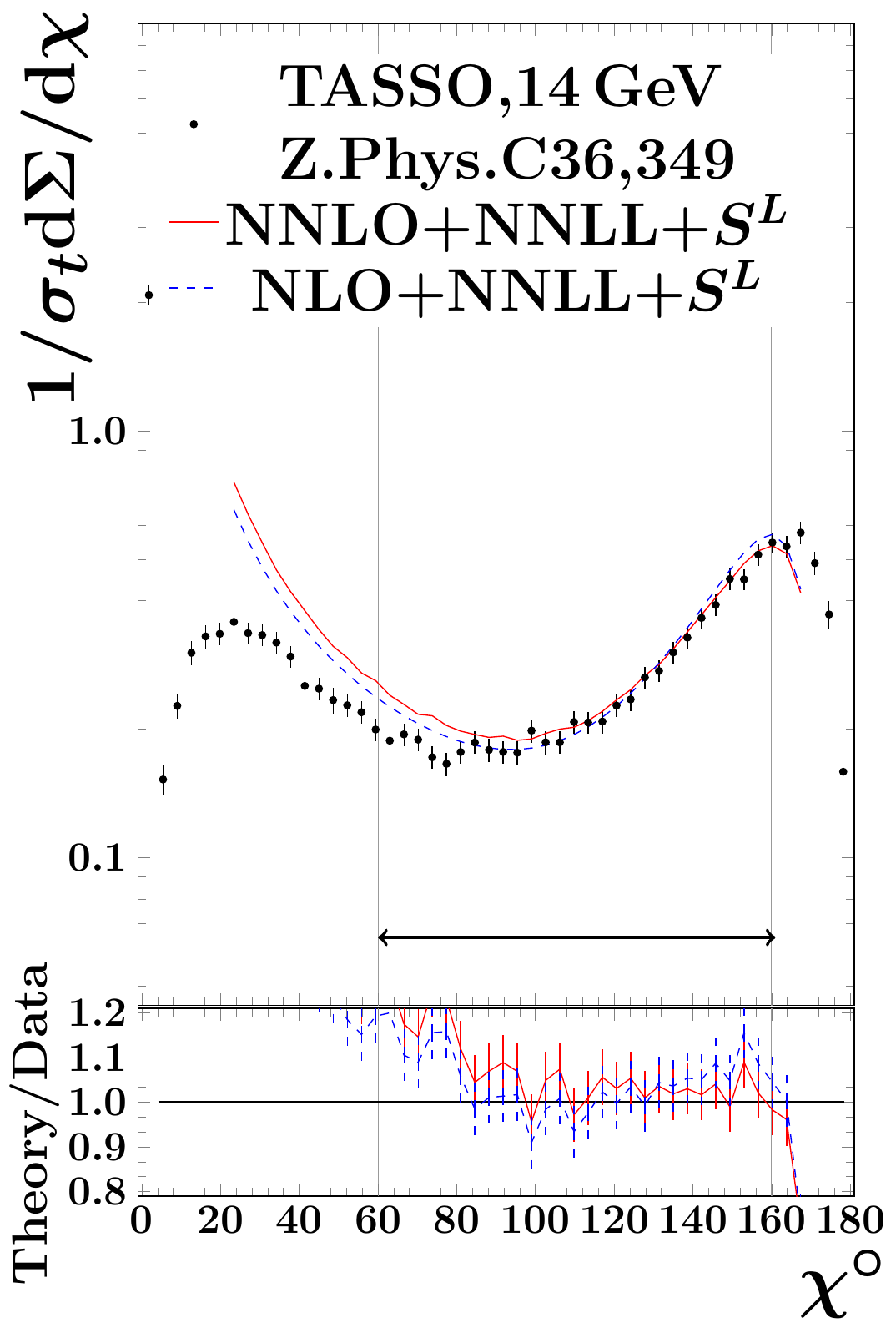}\includegraphics[width=\FIGWONE]{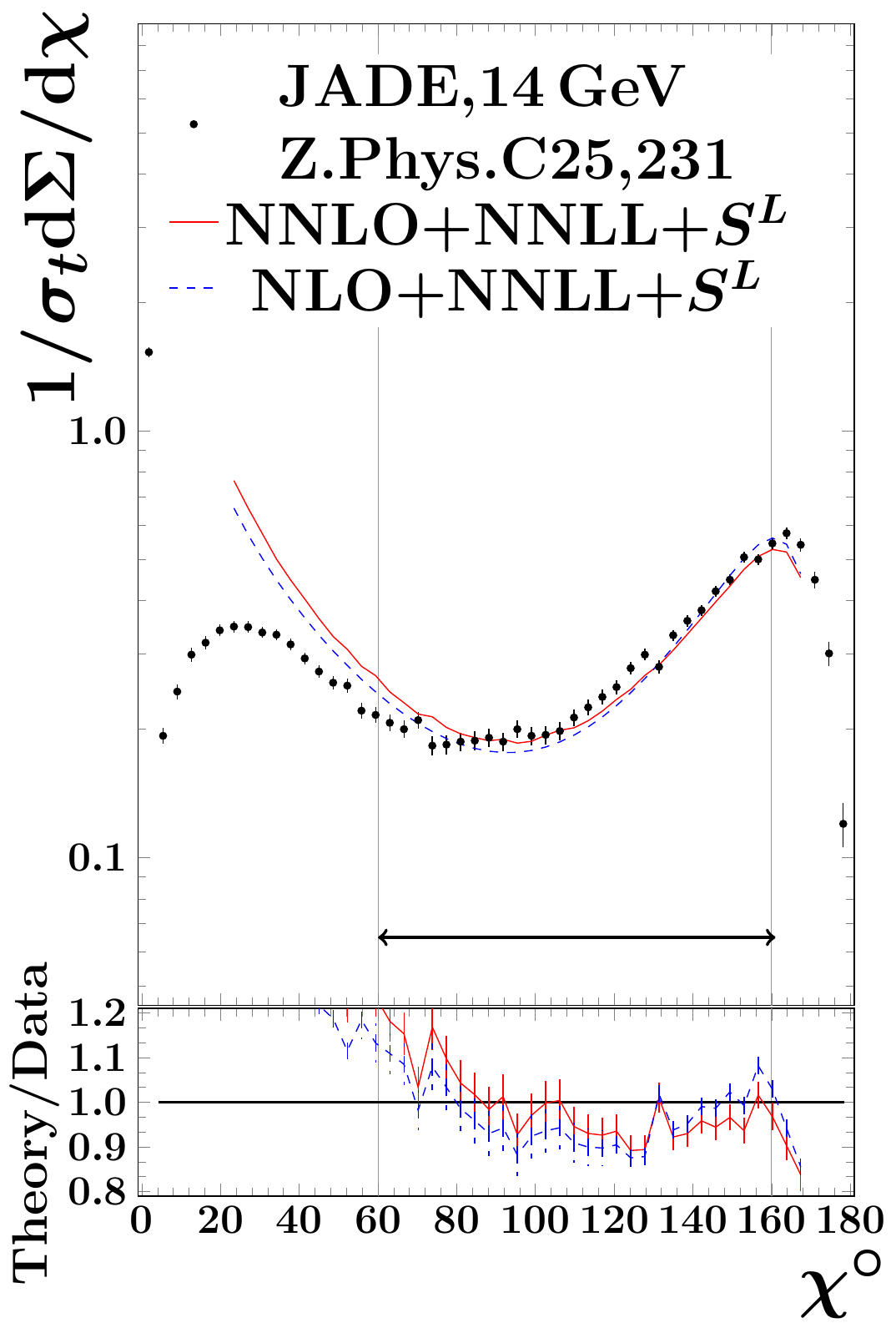}\\
\caption{
Fits of theory predictions to the data for $\sqrt{s}=14\GeV$. 
The used fit range is shown with thick line. For the ratio plot 
only the uncertainties of the data are taken into account.
}
\label{fig:result:onefour}
\end{figure}

The fit ranges were chosen to avoid regions where resummed predictions or 
hadronization correction calculations are not reliable. The selected fit ranges were \fitrangetwo, 
\fitrangethree and \fitrangefour.
The uncertainty on the fit result was estimated with the $\chi^2+1$ criterion 
as implemented in the {\tt MINUIT2} program.
The results of the fits are given in Tab.~\ref{tab:result:full} for the each fit range.
In order to assess the impact of the NNLO corrections, in Tab.~\ref{tab:result:full} we
also present the results obtained using NLO+NNLL predictions. The NNLO corrections are
seen effect the fit in a moderate but non-negligible way.
The obtained values of $\chi^2$ divided by the number of degrees of freedom in the 
fit are of order unity for all cases.
The corresponding distributions obtained from the fits for different $\sqrt{s}$ points are shown in 
Figs.~\ref{fig:result:nineone},~\ref{fig:result:threefour},~\ref{fig:result:twonine}~and~\ref{fig:result:onefour}.

\renewcommand{\arraystretch}{0.9}
\begin{table}[!htbp]\centering
\begin{tabular}{|c||c|c|}\hline
Fit range,$\degree$&  NLO+NNLL                  & NNLO+NNLL              \\
Hadronization      &   $\chi^{2}/ndof$                    & $\chi^{2}/ndof$                  \\\hline\hline
\tabularresult
\end{tabular}
\caption{Results of the fits of the matched predictions at NLO+NNLL and NNLO+NNLL accuracy to 
experimental data.  The given uncertainty is fit uncertainty scaled by $\sqrt{\chi^{2}/ndof}$.
}
\label{tab:result:full}
\end{table}

The systematic uncertainties of the obtained results were estimated with procedures 
used in previous studies~\cite{Jones:2003yv}.
To estimate the bias of the obtained result caused by the absence of higher-order terms in the 
perturbative predictions, the renormalization scale variation procedure was performed.
In this procedure the fits were repeated, with variation of the 
renormalization scale in the range between $x_{R}=1/2$ and $x_{R}=2$.

\begin{figure}[htbp]\centering
\includegraphics[width=0.49\linewidth]{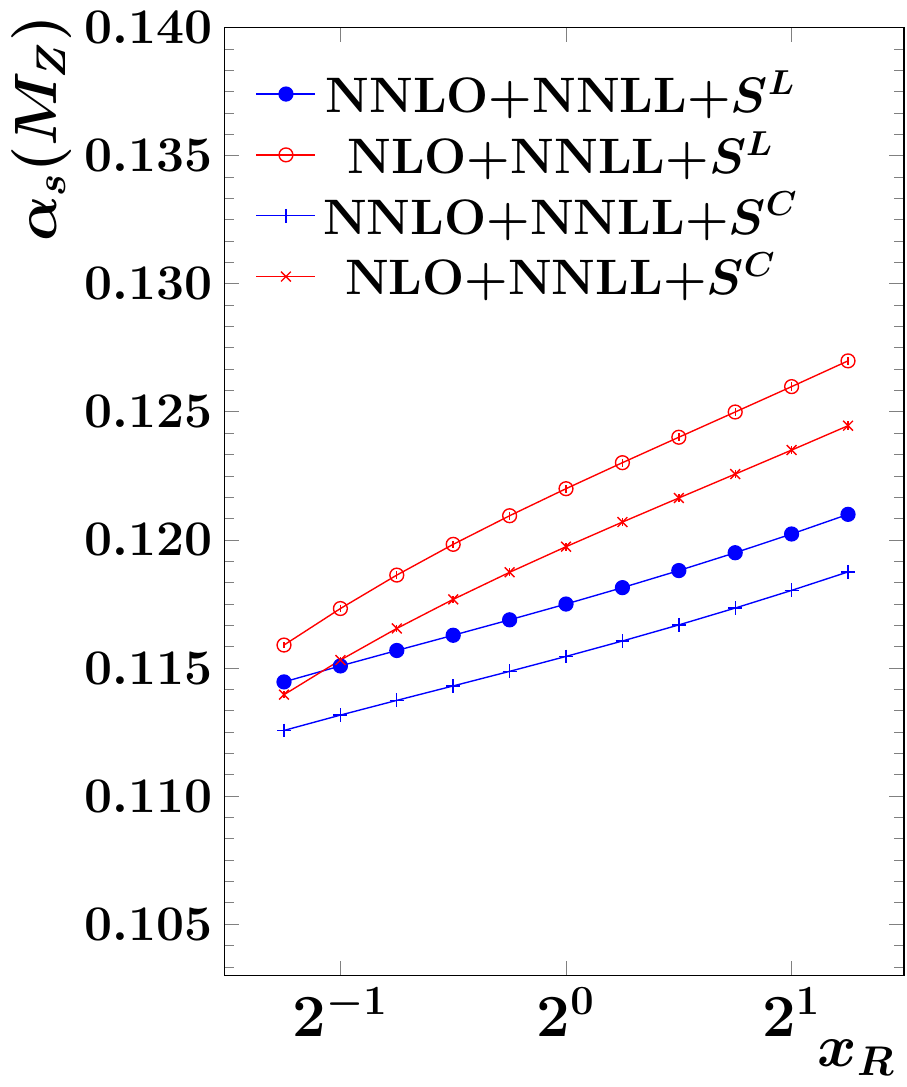}\includegraphics[width=0.49\linewidth]{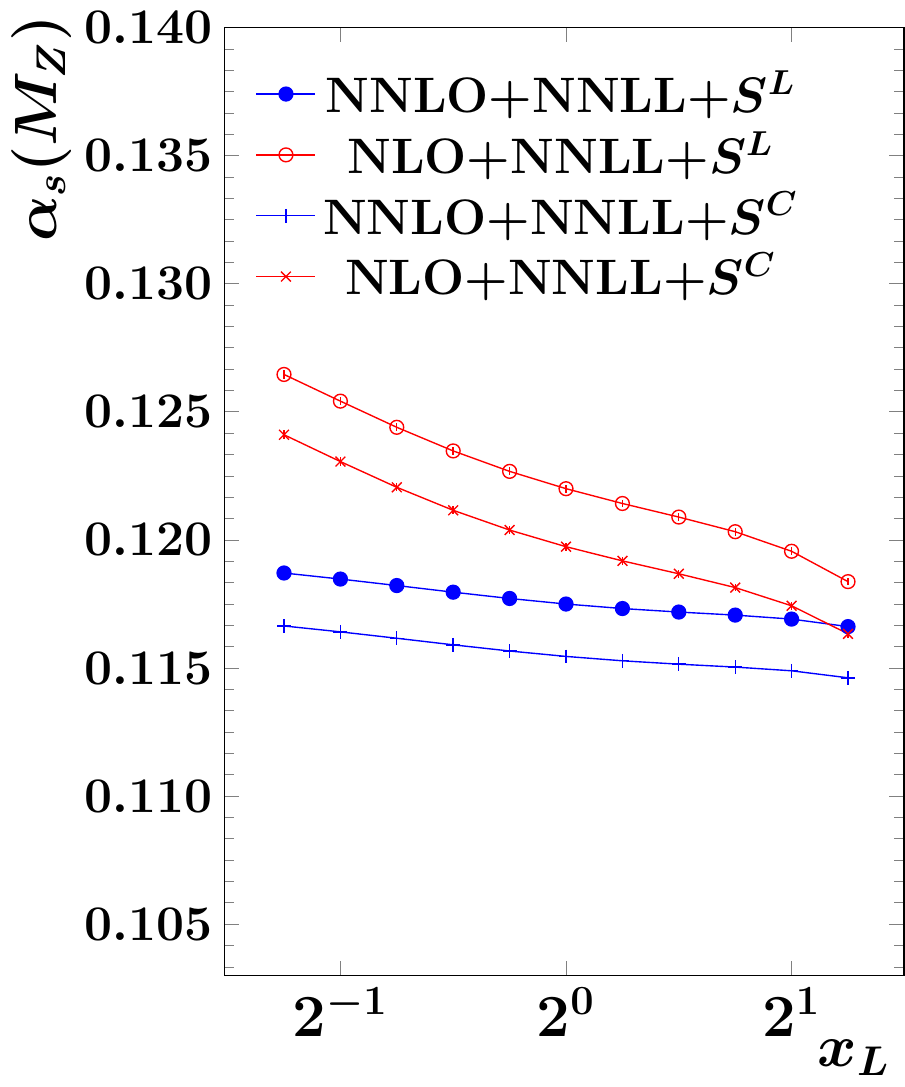}\\
\includegraphics[width=0.49\linewidth]{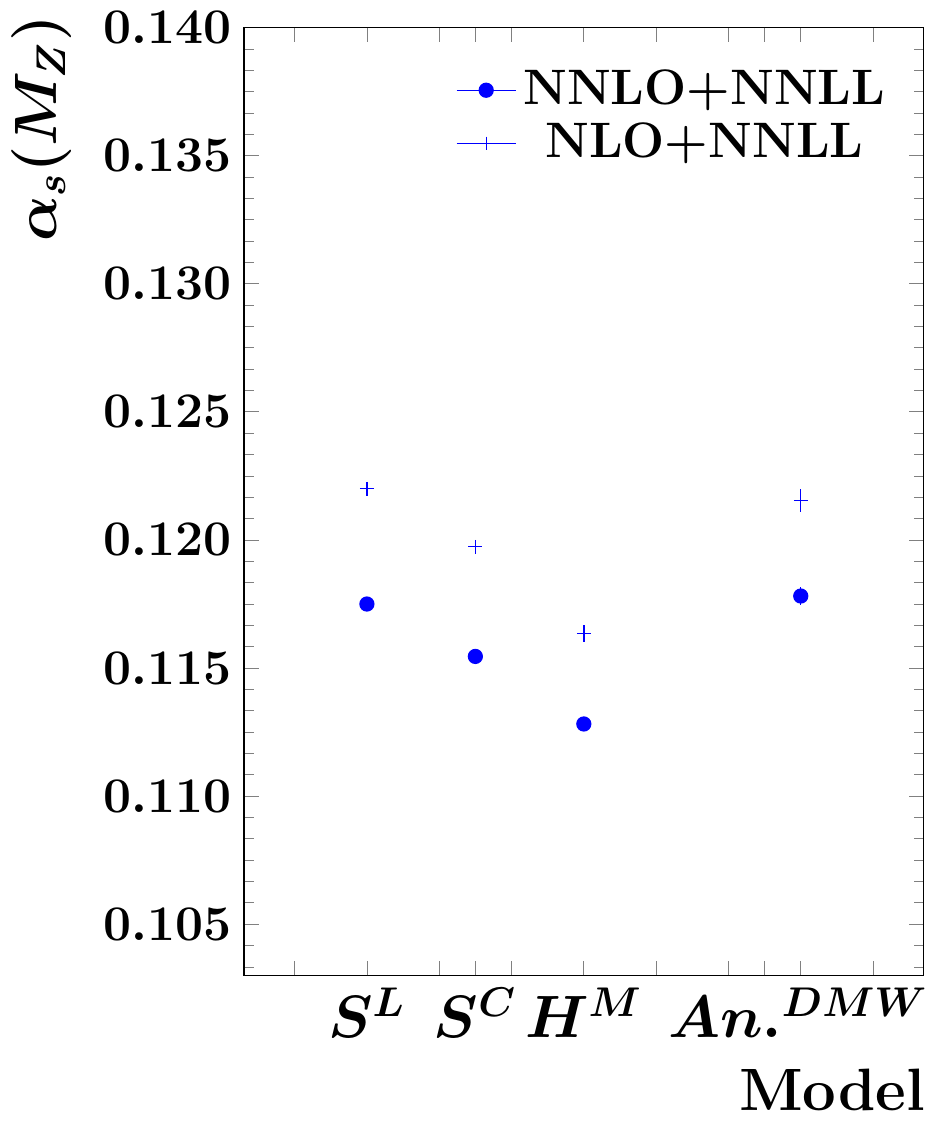}\includegraphics[width=0.49\linewidth]{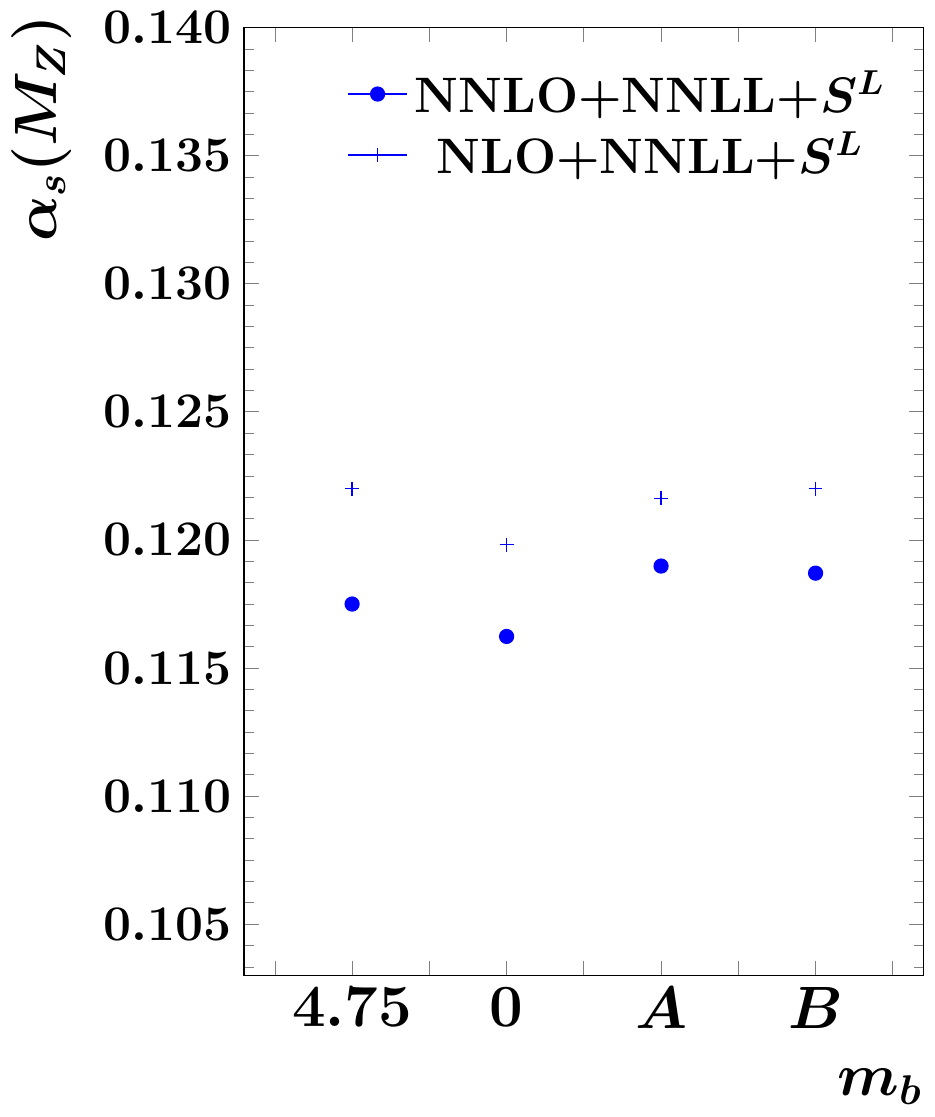}\\
\caption{Dependence of fit results on the renormalization scale (upper left), resummation scale (upper right), 
non-perturbative simulation model (bottom left)   
choice and $b$ mass corrections (bottom right).
The fit range for $S^{L}$, $S^{C}$ and $H^{M}$ setups is \fitrangefour.
The fit range for the $An.^{DMW}$ setup is \fitrangetwo.
}
\label{fig:result:dependence}
\end{figure}

The bias of hadronization model selection is studied 
using the $S^{L}$ and $S^{C}$ setups of hadronization corrections, see results in Fig.~\ref{fig:result:dependence}.
The bias related to the ambiguity of resummation scale choice was estimated by varying $x_{L}$ in the range 
between $x_{L}=1/2$ and $x_{L}=2$. 
To estimate the bias related to the ambiguity of our prescription implementing the unitarity constraint in 
the resummed calculation (see \eqn{eq:unitarity-p}), two values of $p$ were used: $p=1$ and $p=2$.
The difference between results obtained with two options is negligible.
In all cases above the sizes of the biases were estimated numerically as 
half of the difference between the maximal and minimal  $\as$ value 
obtained in the corresponding set of fits.
To estimate the potential bias of the result caused by imperfections of  specific
hadronization model and parton shower model,
the fits were repeated with hadronization corrections obtained with all setups described in
previous subsections. 
The numerical value of the bias  was obtained as half of the difference between the
$\as$ values obtained using non-perturbative corrections 
from Lund and cluster hadronization models implemented in {\tt SHERPA2.2.4}.  
From the Fig.~\ref{fig:result:dependence} it is seen that the estimated biases 
are relatively independent and, therefore combined in the final result as such.

\begin{figure}[htbp]\centering
\includegraphics[width=0.49\linewidth]{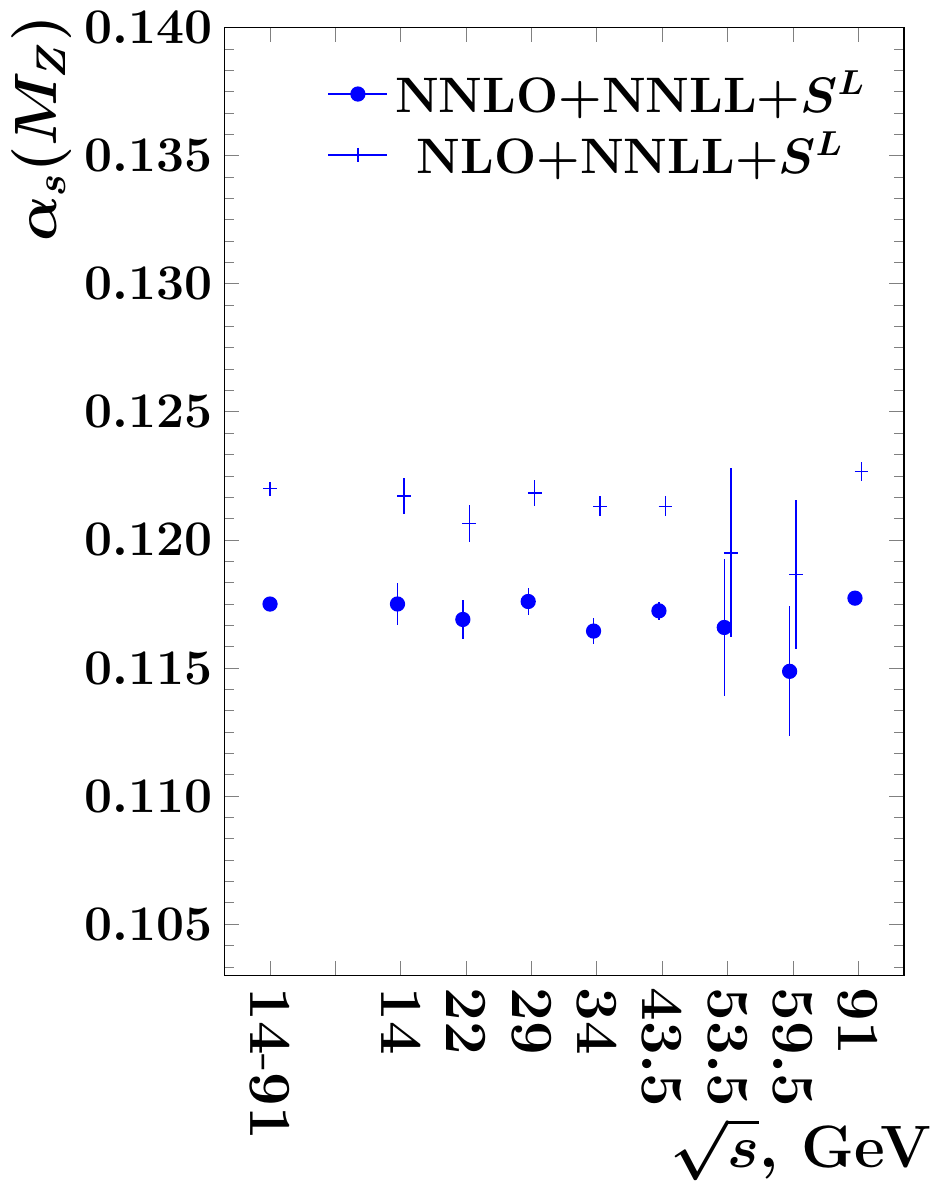}
\includegraphics[width=0.49\linewidth]{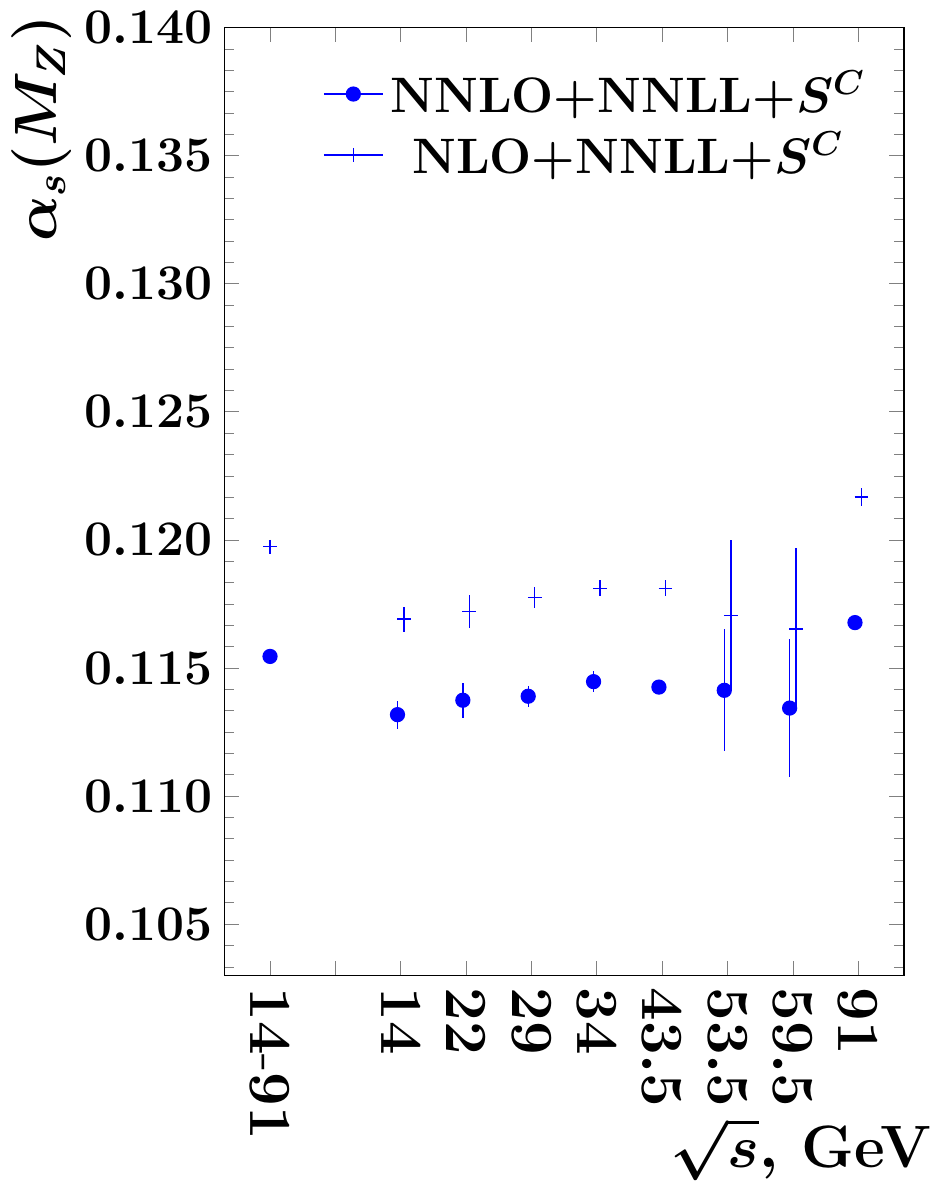}\\
\includegraphics[width=0.49\linewidth]{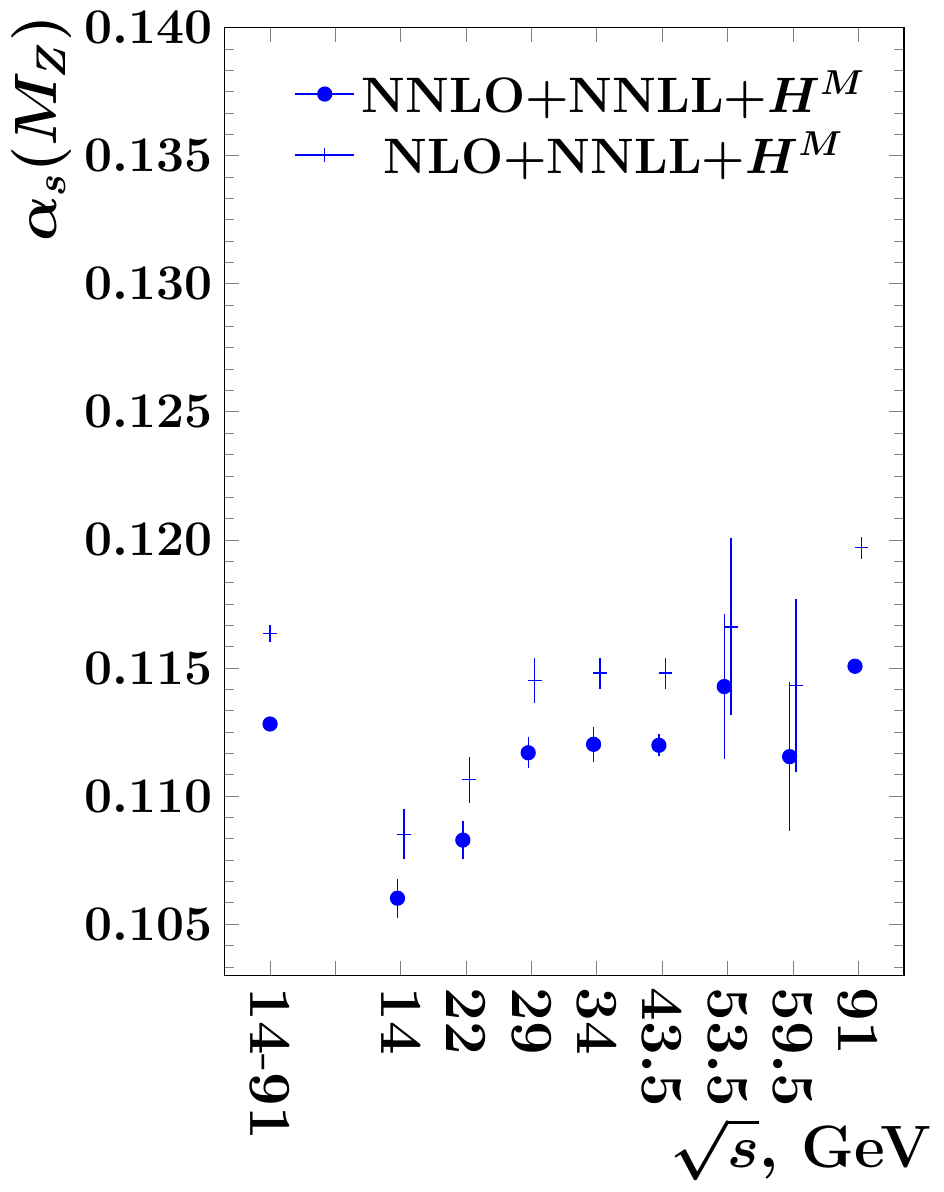}\\
\caption{Dependence of fit results on the used data sets. The fit range for $S^{L}$, $S^{C}$ and $H^{M}$ setups is \fitrangefour.
}
\label{fig:result:q}
\end{figure}

Besides the estimations, several cross-checks of the obtained results were performed.
First, the datasets were grouped according to their energies and fitted separately
for each energy. The results are shown in Fig.~\ref{fig:result:q}. There is no
visible trend for the fitted value of $\as$ with energy in the $S^{L}$ and $S^{C}$ setups.
For the $H^{M}$ setup, the results of the fits are not reliable below $\sqrt{s} < 29\GeV$
due to the sensitivity of this setup to the $b$-quark mass. In addition to the MC hadronization
models the fits were also performed with the analytic hadronization model of Dokshitzer,
Marchesini and Webber (DMW)~\cite{Dokshitzer:1999sh}. In this setup, non-perturbative effects
in EEC were accounted for by multiplying the Sudakov form factor by a correction of the form
$$
S_{\mathrm{NP}} = e^{-\frac12 a_1 b^2} (1 - 2 a_2 b)\,.
$$
Here $a_1$ and $a_2$ are non-perturbative parameters that can be related in the 
dispersive approach to certain moments $\bar{\alpha}_{q,p}$ of the strong coupling $\as$~\cite{Dokshitzer:1999sh}. 
These moments are the fit parameters of the analytic model.
The results obtained from the fits with this setup are listed in Tab.~\ref{tab:result:full}.
They show a high degree of dependence on the selected fit range, but  are close to results obtained
with the Monte Carlo based hadronization corrections in the range \fitrangetwo, see Tab.~\ref{tab:result:full}. 
Hence we conclude that away from the back-to-back region, the analytic model cannot fully account for
hadronization effects.
\FloatBarrier
\section{Results and discussions}
\label{sec:pheno}
In this paper we presented the first combined analysis and extraction of $\as$ at
NNLO+NNLL precision from energy-energy correlation in electron-positron annihilation. 
Moreover, our analysis is the first extraction of the strong coupling based on Monte
Carlo hadronization corrections obtained from NLO Monte Carlo setups at NNLO+NNLL precision. 
For the central value of the final result we quote the results obtained from the fits with the
$S^{L}$ hadronization model in the range \fitrangefour with uncertainties and estimations
of biases obtained as described above. 

At NNLO+NNLL accuracy we obtain the best fit value of
\begin{multline*}
\as(M_{Z})=
\resultNNLO\,.
\end{multline*}
In order to appreciate the impact of NNLO corrections, we also quote the result of the
fit at NLO+NNLL accuracy
\begin{multline*}
\as(M_{Z})=
\resultNLO\,.
\end{multline*}
We see that the inclusion of the NNLO corrections has a moderate but non-negligible effect
on the extracted value of $\as$.

It has been explicitly checked that there are no correlations between estimated biases, 
therefore, the combined values with combined estimations of bias at NNLO+NNLL accuracy are:
$$
\as(M_{Z})=\resultNNLOcomb
$$
while in comparison, for NLO+NNLL accuracy we obtain:
$$
\as(M_{Z})=\resultNLOcomb\,.
$$
The value obtained from the analysis in NNLO+NNLL approximation is in agreement with
the world average as of 2017~\cite{Bethke:2017uli}, however it is  visibly lower than
the results from measurements performed for other $e^{+}e^{-}$ observables using NNLO
perturbative QCD predictions and MC hadronization models~\cite{Bethke:2017uli}.
The estimated uncertainties are dominated by the uncertainty on the theoretical
predictions. The results obtained in this study can be compared to those described
in the original publications with NLO+NNLL precision as well as the results obtained
with analytic hadronization model in the sister paper~\cite{Tulipant:2017ybb}.

\section*{Acknowledgements}
We are grateful to Simon Pl\"{a}tzer and Ludovic Scyboz for fruitful discussions about the 
calculation of NLO predictions with {\tt Herwig7.1.1} and {\tt GoSam}, to Pier Monni for 
stimulating discussions on resummation of event shapes and to Carlo Oleari for providing 
us the {\tt Zbb4} code. Z.T.~was supported by the \'{U}NKP-17-3 New National Excellence 
Program of the Ministry of Human Capacities of Hungary. A.K.~acknowledges financial 
support from the Premium Postdoctoral Fellowship program of the Hungarian Academy of 
Sciences. This work was supported by grant K~125105 of the National Research, Development 
and Innovation Fund in Hungary.

\newpage
{\bibliographystyle{./EECALPHAS}{\raggedright\bibliography{EECALPHAS}}}\vfill\eject
\clearpage
\end{document}